\documentclass[useAMS,usenatbib]{almamemo}
\pdfoutput=1
\usepackage{amsmath}
\usepackage{mathptmx}
\usepackage{graphicx}
\usepackage[colorlinks=true,citecolor=blue]{hyperref}
\usepackage[]{subfig}
\newcommand{\unit}[1]{\mathrm{#1}}
\newcommand{\unitp}[2]{\ensuremath{\mathrm{#1}^{#2}}}
\newcommand{\pycmd}[1]{{\tt #1}}

\volume{587}
\title[]{ALMA Memo \#587\\
Inference of Coefficients for Use in Phase Correction I
}

\author[]{B. Nikolic\\Mullard Radio
  Astronomy Observatory, Cavendish Laboratory, Cambridge CB3 0HE, UK
  \\\url{email:b.nikolic@mrao.cam.ac.uk}
  \\\url{http://www.mrao.cam.ac.uk/~bn204/}}

\date{5 March 2009}
\pagerange{\pageref{firstpage}--\pageref{lastpage}; } \pubyear{2009}
\begin{document}
\label{firstpage}
\maketitle

\begin{abstract}
We present a Bayesian approach to calculating the coefficients that
convert the outputs of ALMA 183\,GHz water-vapour radiometers into
estimates of path fluctuations which can then be used to correct the
observed interferometric visibilities. The key features of the
approach are a simple, thin-layer, three-parameter model of the
atmosphere; using the \emph{absolute\/} measurements from the
radiometers to constrain the model; priors to incorporate physical
constraints and ancillary information; and a Markov Chain Monte Carlo
characterisation of the posterior distribution including full
distributions for the phase correction coefficients. The outcomes of
the procedure are therefore estimates of the coefficients \emph{and\/}
their confidence intervals. We illustrate the technique with
simulations showing some degeneracies that can arise and the
importance of priors in tackling them. We then apply the technique to
an hour-long test observation at the Sub-Millimetre Array and find
that the technique is stable and that, in this case, its performance
is close to optimal. The modelling is described in detail in the
appendices and all of the implementation source code is made publicly
available under the GPL.
\end{abstract}

\section{Introduction}

The performance of millimetre and sub-millimetre wave interferometers
is often limited by the fluctuation of the properties of the Earth's
troposphere along the lines of sight of each of the elements of the
interferometer. If not corrected, these fluctuations lead to a
fluctuating delay to each element and subsequent loss of correlation
(and therefore sensitivity) and a limit on the maximum usable baseline
length. Some recent simulations of the effect of these fluctuations on
ALMA science have been presented by us \citep{ALMANikolic582} and
other authors \citep[e.g.,][]{ALMAAsaki535}.

ALMA plans to correct for these fluctuations by a combination of two
techniques: 
\begin{enumerate}
  \item Fast-switching, that is interleaving science observations with
    observations of near-by phase calibrators that allow antenna phase
    errors to be solved for
  \item Water-vapour radiometry, that is observing the \emph{emission}
    of atmospheric water vapour along the line of sight of each
    element of the array, and inferring and correcting from these
    observations the fluctuating path to each element.
\end{enumerate}
The current plan for ALMA is that fast-switching will be used with a
cycle period of between around 10 and 200\,seconds while fluctuations
on timescales from 1\,second up to the fast-switching timescale will
be corrected by the water vapour radiometry technique. The actual
fast-switching time that will be used will depend on the weather
conditions and the scientific requirements of each project as well as
the achieved accuracy of phase correction with the water-vapour
radiometers.  Furthermore we expect to be able to use the WVRs to
correct for the phase \emph{transfer errors}, that is the errors that
arises due to the different lines of sight to the phase calibration
and science sources.

One of the key requirements for radiometric phase correction is a
prescription for converting the observed sky brightnesses in the
neighbourhood of the 183.3\,GHz water vapour line, as measured by the
WVRs, into a path delay that can be used to correct the observed
visibilities. Any such prescription is complicated by the significant
pressure-broadening and the high cross section of this line (it can be
close to saturated even in the dry conditions of the ALMA site) which
means that the optimum conversion strategy is quite a sensitive
function of the prevalent atmospheric conditions.

In this paper we assume that over the short timescales of interest
(i.e., less than approximately 200\,seconds) the differential path
fluctuation between two telescopes can be predicted by a constant
times the differential fluctuation of the sky brightness between the
same telescopes. The constant of proportionality is the phase
correction ``coefficient'' and we give in this paper an initial
prescription for calculating these coefficients
(Section~\ref{sec:method}). We analyse this prescription with
simulations (Section~\ref{sec:results}) and then apply it to a test
observation at the Sub-Millimetre Array
(Section~\ref{sec:smadataapp}).

\section{Method}
\label{sec:method}

Our aim is to find estimates of, and confidence intervals on, the
coefficients to be used for phase correction. We denote the
coefficients as ${\rm d}L/{\rm d} T_{{\rm B},i}$ where $L$ is the
excess path to the telescope and $T_{{\rm B},i}$ are the sky
brightnesses as measured by the four channels of the WVRs. We use
this notation in general although this differential only makes
mathematical sense when there is a known and invertible function
which connect the cause of fluctuation in $L$ with $T_{{\rm B},i}$. In
this paper we assume that the fluctuation in path are caused by
fluctuations of the total water vapour column \emph{only} so that
the fluctuation water vapour column is also the sole cause of
fluctuations in $T_{{\rm B},i}$.

The approach we take in this paper is to construct a physical model of
the atmosphere with a number of unknown parameters and use some
observables and the basic physical considerations to constrain the
possible values the parameters. Once the distribution of possible
values of model parameters are known, we can use the same model to
compute the distribution of phase correction coefficients. The
physical model employed in these initial studies is extremely simple:
\begin{itemize}
  \item We assume water vapour is the only cause of opacity and path
    fluctuation
  \item We assume all the water is concentrated in a thin layer at a
    single pressure and temperature
  \item We make the plane parallel approximation for computing effects
    due to changes in elevation of the antenna 
\end{itemize}
This means that the model has only three unknown parameters, namely
total water vapour column ($c$), temperature ($T$) and pressure ($P$).

The reason for using such a simple model is that it allows very
extensive computational analysis while containing the basic
ingredients which we know \emph{must\/} influence phase correction. As
we gain experience with simulations and observations at the ALMA site,
we intend to extend this model in directions which show an actual
improvement in the inference of the phase correction coefficients.

For the time being we will assume that the only true observables that
we have available to constrain the model are the four
\emph{absolute\/} sky brightness temperatures measured by the WVRs.
In additional to the direct observables, we have some constraints on
the possible values the model parameters can take; these are called
\emph{priors\/}. In principle the priors are a joint probability
function of all of the model parameters, but in the present paper we
simplify by taking them to be separable into a product of functions of
one parameter only, i.e., $p(cTP)=p(c)p(T)p(P)$.

The method of solution of this problem that we describe here is the
standard Bayesian Markov Chain Monte Carlo approach. As usual, the
inference problem is described by the Bayes equation \citep[see, for
  example][]{Jaynes:PTLS,SiviaD06}:
\begin{equation}
  p(\theta | D ) = \frac{ p(D | \theta) p(\theta)}{p(D)}.
  \label{eq:BayesTheorem}
\end{equation}
where the symbols have following meaning:
\begin{description}
  \item[$\theta$] is the vector of model parameters (in this case
    $\{c,T,P\}$)
  \item[$D$] is the observed data (in this case the sky brightness
    temperatures observed by the WVR)
  \item[$p(\theta)$] is the prior information (in this case
    constraints on model parameters as mentioned above)
  \item[$p(D|\theta)$] is the likelihood, i.e., the probability of
    observing the data we have given some model parameters $\theta$
  \item[$p(D)$] is the so-called Bayesian \emph{evidence}, that is a
    measure of how good our model is in describing the data
  \item[$p(\theta | D )$] is the posterior distribution of model
    parameters
\end{description}

\subsection{Likelihood}

The computation of the likelihood of an observation given model
parameters can be factored into three stages:
\begin{enumerate}
  \item Calculation of the sky brightness temperature as a function of
    frequency
  \item Calculation of the temperatures recorded by the WVRs given a
    sky temperature. Here we model frequency response of the units and
    the coupling to the sky.
  \item Calculating the probability of observed data given the model
    temperatures obtained in the previous step
\end{enumerate}

\begin{figure}
  \subfloat[Precipitable water vapour from 0.6\,mm to 1.3\,mm]{
  \includegraphics[clip,width=0.99\columnwidth]{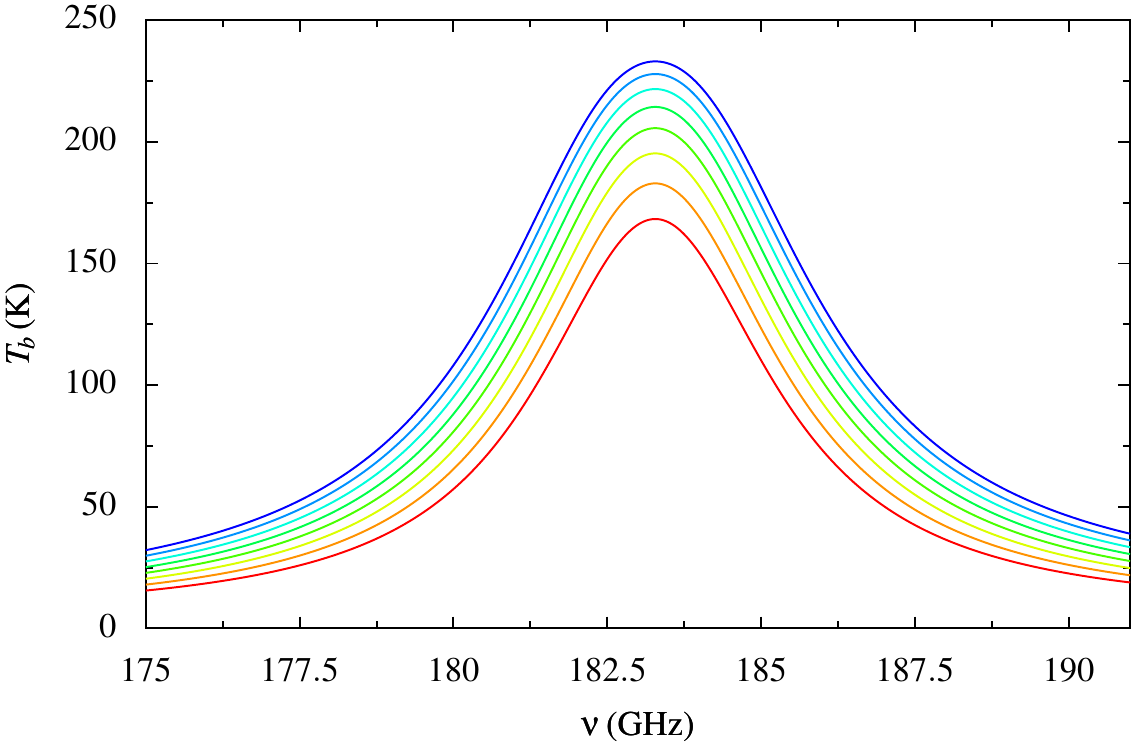}}

  \subfloat[Temperature from 230 to 300\,K]{
  \includegraphics[clip,width=0.99\columnwidth]{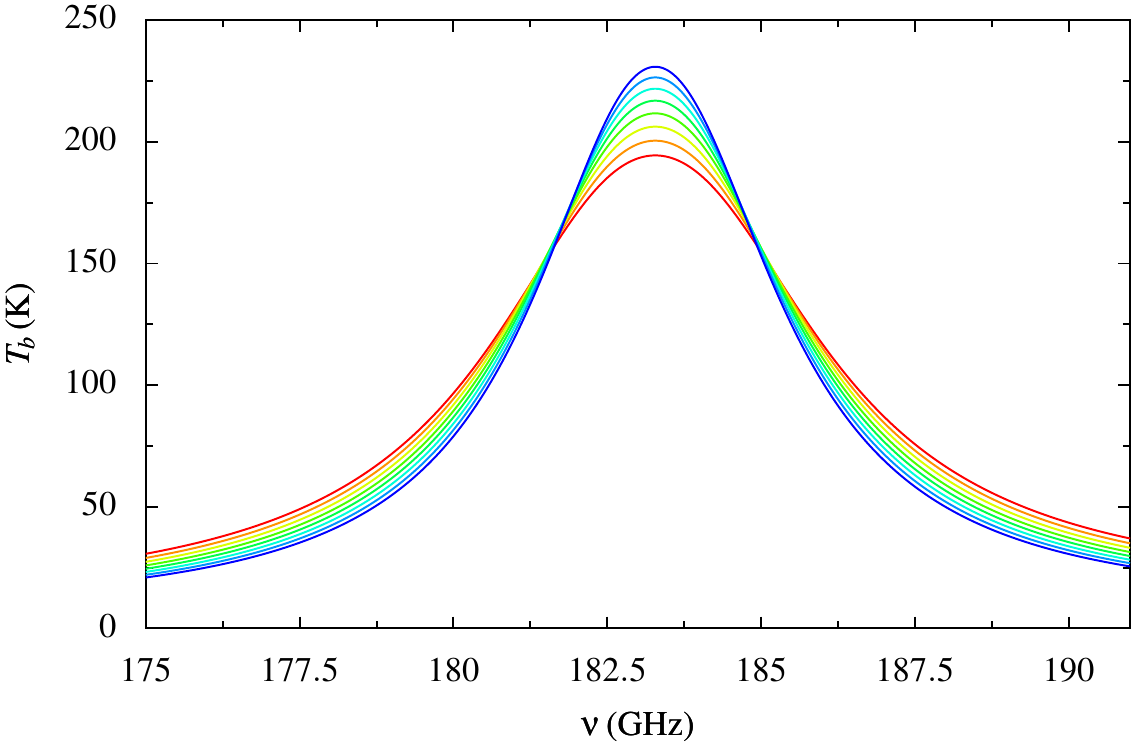}}

  \subfloat[Pressure from 400 to 750\,mBar]{
  \includegraphics[clip,width=0.99\columnwidth]{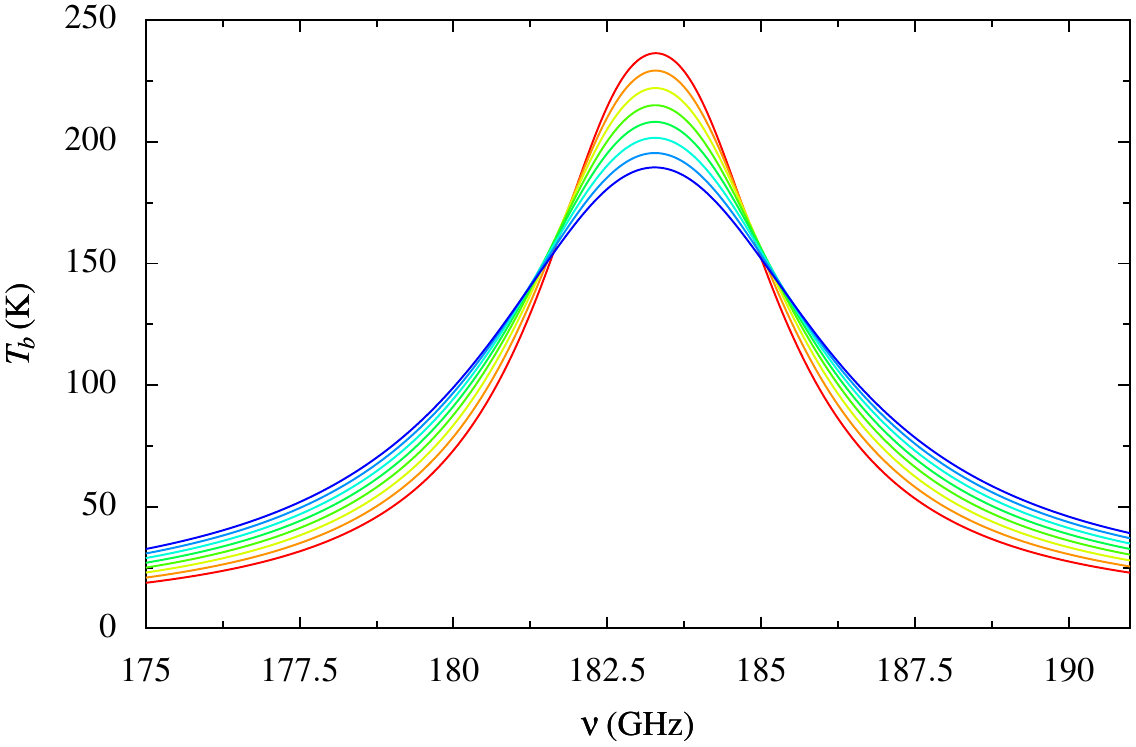}}

  \caption{Variation of the shape of the 183\,GHz water vapour line
    with changes in water vapour column, temperature and pressure}
  \label{fig:waterlineskyb}
\end{figure}

The sky brightness is computed using a simplified model, in which we
assume that the only relevant contributors to the atmospheric opacity
in this band are the water vapour line at 183\,GHz and the
water-vapour pseudo-continuum. We assume the water vapour line has a
Gross line shape and parameters derived from the HITRAN database entry
\citep{HITRAN04} and suitable correction for pressure and temperature
as detailed in Appendix~\ref{sec:atmo-opac-calc}. The parametrisation
that we use for the pseudo-continuum follows closely that in the
program {\tt am} by \cite{PaineAmRev3} and is also given in
Appendix~\ref{sec:atmo-opac-calc}. With the opacity calculation, the
sky brightness can be calculated using simple radiative transfer.  The
resulting sky brightness temperature is shown for a range of
conditions in Figure~\ref{fig:waterlineskyb} illustrating the change
in the shape of the water vapour line with pressure, temperature and
total water vapour column.

The three model parameters are the total water vapour column ($c$),
temperature ($T$) and pressure ($P$). The water vapour column
parameter is taken to refer to the \emph{zenith\/} column; for the
cases when we investigate non-zenith measurements, we assume that this
parameter scales according to the plane-parallel approximation while
the other parameters remain unchanged with a change in elevation of
the telescope.

\begin{figure}
  \subfloat[Prototype radiometers]{
    \includegraphics[clip,width=0.99\columnwidth]{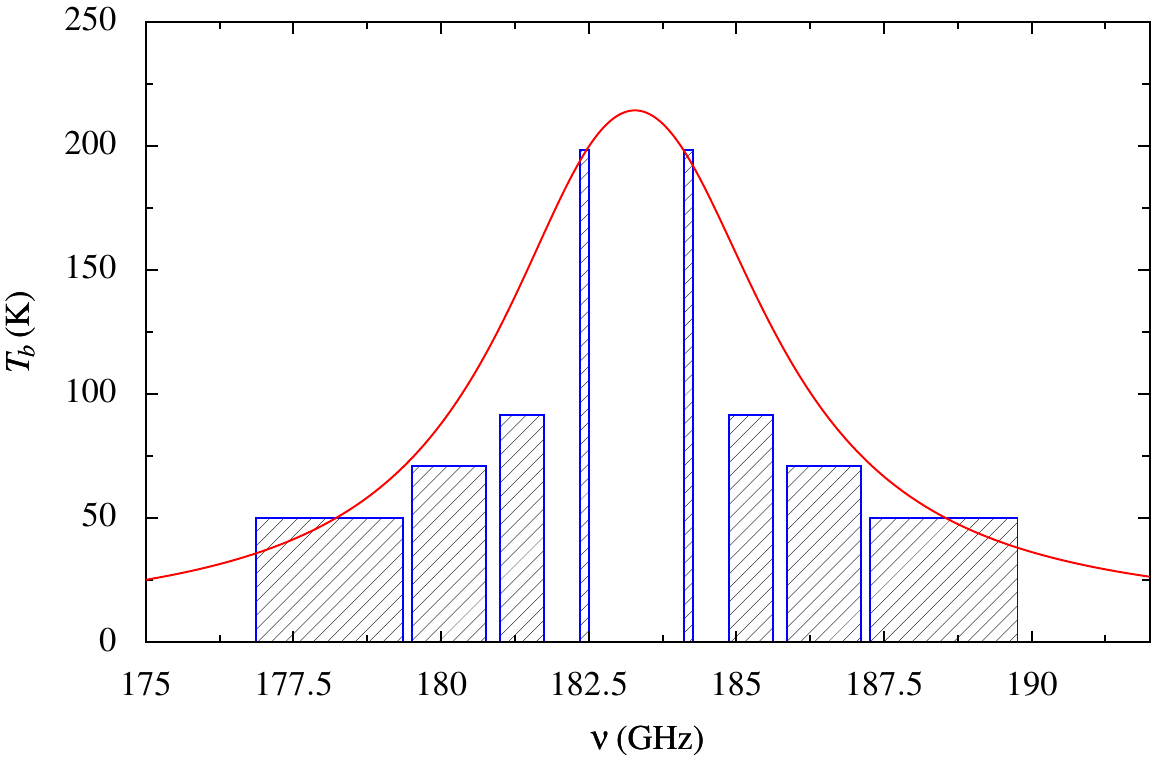}}

  \subfloat[Production radiometers]{
    \includegraphics[clip,width=0.99\columnwidth]{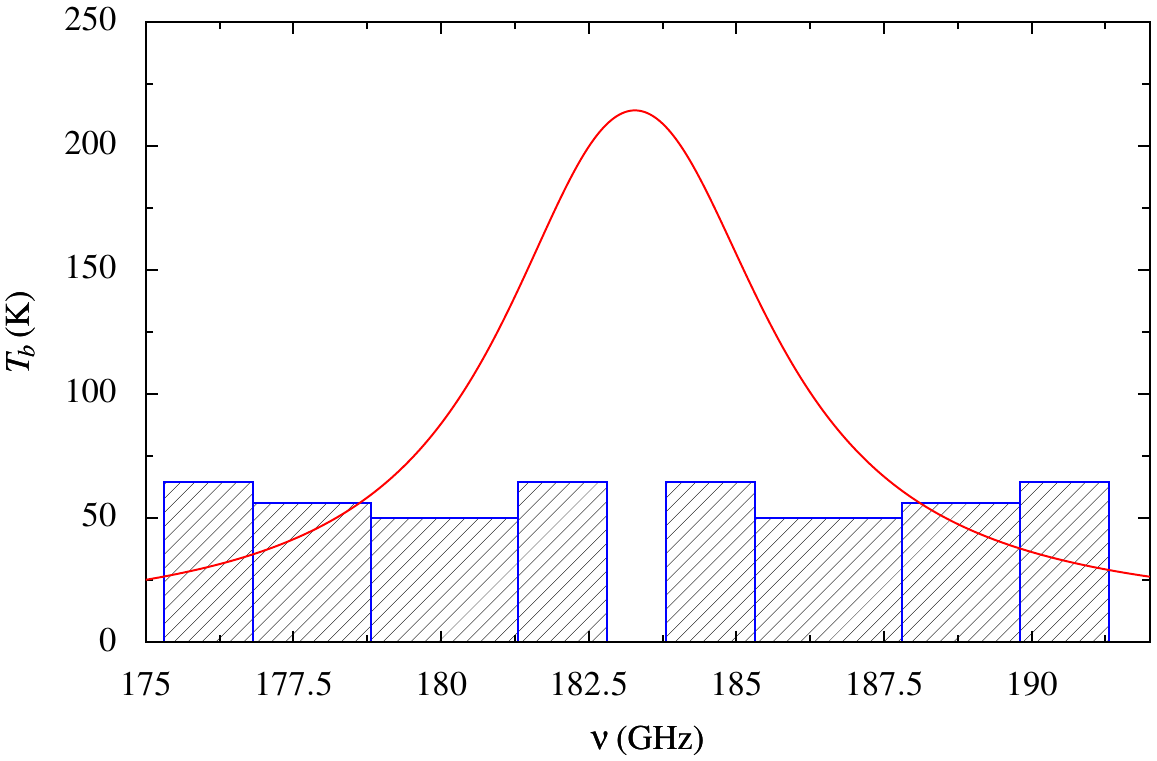}}

  \caption{Filter design centres and bandwidths of the prototype (top)
    and production (bottom) radiometers with the 183\,GHz water vapour
    line also shown (in red). The heights of the rectangles
    representing the filters are inversely proportional to the square
    root of bandwidth, and are therefore an indication of the relative
    sensitivity of the filters.}
  \label{fig:wvrfilters}
\end{figure}

With the sky brightness known, the temperature seen in each of the WVR
filters can be calculated. In the present study we assume the WVRs are
double-sideband (like the production units on ALMA will be) and that
their filter set corresponds to the prototype filter set. The reason
for using the prototype filter set definition is that later we will
make an analysis of a sample test data set collected with the
prototype radiometers at the SMA.  The prototype filter set is shown
in Figure~\ref{fig:wvrfilters}, which also shows for contrast the
filter set of the production ALMA WVRs.

In this study we assume the filters are perfectly sharp and at their
design centre frequencies and bandwidths. The average sky brightness
across each filter sideband is calculated using the five point
Gauss-Legendre Quadrature \citep[e.g.,][]{abramowitz+stegun} which
provides reasonable (but, since it is \emph{not} an adaptive
algorithm, a non-uniform) accuracy and is extremely efficient since
the sky brightness needs to be calculated at only five frequencies per
filter sideband.

The second effect which needs to be taken into account at this stage
is the non-perfect coupling of the radiometer in the sky. Our initial
analyses of sky-dip measurements during this test campaign (to be
published subsequently) suggest that the coupling was about 0.91 and
that the termination temperature of the parts of the beam that did not
reach the sky was about 265\,K. We therefore use these values in the
analysis of the data from the SMA later
(Section~\ref{sec:smadataapp}); however for the simulations shown in
Section~\ref{sec:results} for simplicity we assume perfect coupling. 

The results at stage of computation are the four temperatures,
$T_{{\rm B},i}$, seen by the WVR as functions of
$\theta\equiv\{c,T,P\}$. Given these parameters $\theta$, what is the
probability of \emph{observing\/} a set of temperatures $T^*_{{\rm
    B},i}$? This is the probability that we denote by the likelihood
function $L$ and it is governed by the instrumental effects within the
radiometer.

For the present analysis, the dominant source of error is the
uncertainty in the absolute calibration of the radiometers. The
production radiometers specification require this error to be smaller
than 2\,K at all times. If the underlying errors were
normally-distributed (which may be a reasonable approximation) that
would require an underlying $\sigma$ of less than 1\,K.  We do not
however know at this time precisely what the final distribution will
be and further it is likely that the calibration errors will be
correlated between the filter outputs.

Nevertheless, for simplicity we presently take the errors to be
normally and independently distributed so that the likelihood function
takes the well known form:
\begin{equation}
 \log p(D| \theta) = \log{L} = - \sum_{i} \left( \frac{ T^*_{{\rm
       B},i} - T_{{\rm B},i}}{\sigma_{T,i}} \right)^{2}
\end{equation}
where we take $\sigma_{T,i}=1\,K$. When we better understand the
calibration uncertainties in calibration of the WVR units, it will be
important to incorporate this information into the above equation. (By
comparison the \emph{thermal\/} noise in one second of integrating
time will be below 0.1\,K.)

\subsection{Priors}

Besides the sky brightness observed by the WVRs, we have (and will
have) some constraints on the model parameters that are the results of
physical considerations, or derived from independent past
observations. For the purposes of this paper we will consider three
different priors, shown in Table~\ref{tab:priors}, with the aim of
illustrating the effects they have on the final results. The priors
are specific to the Mauna Kea site rather than ALMA site since they
will also be used for analysis of testing data collected at the Mauna
Kea. The conclusions derived from them are expected, however, to
transfer directly to the ALMA site.

\begin{table*}

\begin{tabular}{llccc}
  1 &
  Basic & 
  $  p(c) = \left\{
  \begin{array}{cc}
    1 &   0\,\unit{mm}< c< 5\,\unit{mm}\\
    0 &   {\rm otherwise}
  \end{array}  \right.  $
  & $ p(T) = \left\{
    \begin{array}{cc}
       1 &   200\,\unit{K}<T<320\,\unit{K}\\
       0 &   {\rm otherwise}
    \end{array}
  \right.    $ 
  &$  p(P) = \left\{
    \begin{array}{cc}
       1 &   100\,\unit{mB}<P<650\,\unit{mB}\\
       0 &   {\rm otherwise}.
    \end{array}
  \right.$\\
  2 &
  Reasonable& 
  $  p(c) = \left\{
  \begin{array}{cc}
    1 &   0\,\unit{mm}< c< 5\,\unit{mm}\\
    0 &   {\rm otherwise}  
  \end{array}
  \right.  $
  & $ p(T) = \left\{
    \begin{array}{cc}
       1 &   260\,\unit{K}<T<280\,\unit{K}\\
       0 &   {\rm otherwise}
    \end{array}
  \right.    $ 
  &$  p(P) = \left\{
    \begin{array}{cc}
       1 &   530\,\unit{mB}<P<610\,\unit{mB}\\
       0 &   {\rm otherwise}.
    \end{array}
  \right.$
  \\
  3&
  Pressure constraint& 
  $  p(c) = \left\{
  \begin{array}{cc}
    1 &   0\,\unit{mm}< c< 5\,\unit{mm}\\
    0 &   {\rm otherwise}  
  \end{array}
  \right.  $
  & $ p(T) = \left\{
    \begin{array}{cc}
       1 &   260\,\unit{K}<T<280\,\unit{K}\\
       0 &   {\rm otherwise}
    \end{array}
  \right.    $ 
  &$  p(P) = \left\{
    \begin{array}{cc}
       1 &   570\,\unit{mB}<P<590\,\unit{mB}\\
       0 &   {\rm otherwise}.
    \end{array}
  \right.$
\end{tabular}
\caption{Priors used for the simulations and the analysis of SMA data}
\label{tab:priors}
\end{table*}

The first prior we consider (number 1 in Table~\ref{tab:priors}) is an
extremely relaxed prior which puts very loose constraints on the
pressure and temperature of the water vapour layer:
\begin{enumerate}
  \item Uniform probability that the pressure is between 100 and
    650\,mBar. Since the mean pressure at the peak of Mauna Kea is
    605\,mBar this prior only assumes that the water vapour is not
    extremely high in the atmosphere
  \item Uniform probability that the temperature is between 200 and
    320\,K. Clearly this is a much wider range than typical
    tropospheric temperatures at altitudes where there is significant
    water vapour.
  \item Uniform probability that the zenith water vapour column is
    between 0 and 5\,mm. This uninformative constraint is used for all
    of the other priors too.
\end{enumerate}
Since this prior is less informational then the constraints we will
have even without any ancillary measurements at the site, it is used
to illustrate the degeneracies present if no priors at all are
present.

The second prior we consider (number 2 in Table~\ref{tab:priors}) is
designed to be representative of information on water vapour have
might have with basic understanding of the site but without any
sophisticated ancillary measurements. In this prior, we assume we know
the temperature of the water vapour layer within 20\,K and the
pressure of the layer to within 80\,mBar. This prior is used to
analyse the test data from Mauna Kea.

Finally, the third prior we consider (number 3 in
Table~\ref{tab:priors}) is used as an illustration of the improvement
in accuracy that can be obtained with a tight prior on one of the
parameters. In this case we still assume that we know the temperature
to within 20\,K but that we know the pressure of the water vapour
layer to within 20\,mBar instead of 80\,mBar assumed in prior 2. Such
a constraint on pressure of the water vapour layer might be derived
from, for example, determination of its height using one of the
techniques described below.

\subsection{Markov Chain Monte Carlo}

With the expressions for the likelihood and the priors that we have
described above, we have the necessary information to compute the
Bayes equation (Eq. \ref{eq:BayesTheorem}). In general, this is
computationally expensive \citep[see for example,][]{MackayBayesain}
because of the large volume of parameter space that must be
characterised in order to determine the maximum of $p(D | \theta)
p(\theta)$ and the numerical value $p(D) = \int {\rm d}\theta p(D |
\theta) p(\theta)$ .

The approach we take is standard Markov Chain Monte Carlo (MCMC) using
the \cite{1953Metropolis} algorithm \citep[for a tutorial review, see
  also][]{NealR1993}. In this approach a chain of points in the
parameter space is calculated such that the next point in the chain is
found by proposing a new point by random displacement from the current
point and calculating the relative likelihood of the two points. If
the new point is more likely it is accepted onto the chain; if it is
\emph{less} likely, it accepted with probability determined by the
ratio of the likelihoods of the current and new points. The proposal
distributions we use in this paper are Gaussian with
$\sigma_{c}=0.001\,\unit{mm}$, $\sigma_{T}=0.2\,\unit{K}$ and
$\sigma_{P}=0.5\,\unit{mBar}$. We use the implementation of the MCMC
algorithm in the BNMin1 library \citep{BNMin1}.

By construction of the Metropolis algorithm, the density of points of
the chain in a volume of parameter space is an estimate of the
posterior probability $p(\theta | D )$. This straight-forward approach
does not however allow estimation of $p(D)$ on its own, and so it is
not possible to analyse the relative benefits of different
models. There are however techniques available for estimation of
$p(D)$ and we intend to implement these in the future to allow proper
model comparison.

\subsection{Marginalisation and calculating the ${\rm d}L/{\rm d} T_{{\rm B}}$}
\label{sec:margcalcdldt}

We use the information contained in the Markov Chains in two ways: we
marginalise and histogram the points to make estimates of the model
parameters; and we, for each point in the chain, calculate the phase
correction coefficients ${\rm d}L/{\rm d} T_{{\rm B},i}$. The
marginalisation of Markov Chains in directions parallel to parameter
axes is trivially accomplished by simply ignoring those
parameters. Subsequent computation of histograms is also easily done
(we use the \pycmd{numpy.histogram} routine for this).

Computation of the coefficients ${\rm d}L/{\rm d} T_{{\rm
      B},i}$ is more involved as new physics must be introduced to
compute the delay introduced by the water vapour layer. In the
interest of simplicity, we split this calculation into two parts:
calculation of the non-dispersive and the dispersive path delay. We
compute the non-dispersive path delay using the Smith-Weintraub
equation as described in Appendix~\ref{sec:non-dispersive} taking into
account the temperature and the pressure of the water vapour. As we
are interested here in the effects of water vapour \emph{only}, we do
not here consider the effect a change in temperature will have on the
refractive index of the dry air.

The dispersive delay calculation is more complicated as well as
dependent on the observing frequency (unlike the rest of the
discussion presented in this memo). However, at the frequencies of
relevance to the SMA test data presented later, the dispersive effect
is relatively small, i.e., around 5\%.  We therefore only calculate an
adjustment using the ATM by \cite{PARDOATM} which is then used to
scale the non-dispersive path. The details of this calculation are
presented in Appendix~\ref{sec:dispersive} and the conclusion is that
we scale up the non-dispersive phase coefficients by a factor of 1.05
to take into account the dispersion at 230\,GHz.

With the path delay calculated we calculate the coefficients
${\rm d}L/{\rm d} T_{{\rm B},i}$ by making a small
perturbation to the parameter $c$, i.e., the quantity of water vapour,
and computing the differential as a finite difference.

\section{Simulation}
\label{sec:results}

In this section we present analysis of a simulated single observation
of the sky brightness with a WVR. The goals for this section are to
illustrate the outputs of the technique we have described above, to
show the effects and the importance of the priors, and to illustrate
the approximate accuracy with which it will be possible to predict the
correction coefficients ${\rm d}L/{\rm d} T_{{\rm B},i}$ from
the inputs consisting of the sky brightness only.

The atmosphere from which we simulate our data point has 1\,mm of
water vapour toward the zenith at a temperature of 270\,K and pressure
of 580\,mBar. We assume the observation is toward the zenith. As
described above, we simulate the sky brightness measured by the filter
set of prototype WVRs. We find in this case that the simulated
temperatures are: 194.8, 142.6, 90.7 and 47.5\,K for the inner to
outer channels respectively. These simulated sky temperatures are then
used for subsequent inference as the observable temperatures
($T^*_{{\rm B},i}$).

\begin{figure*}
  \begin{tabular}{ccc}
    \includegraphics[clip,width=0.33\linewidth]{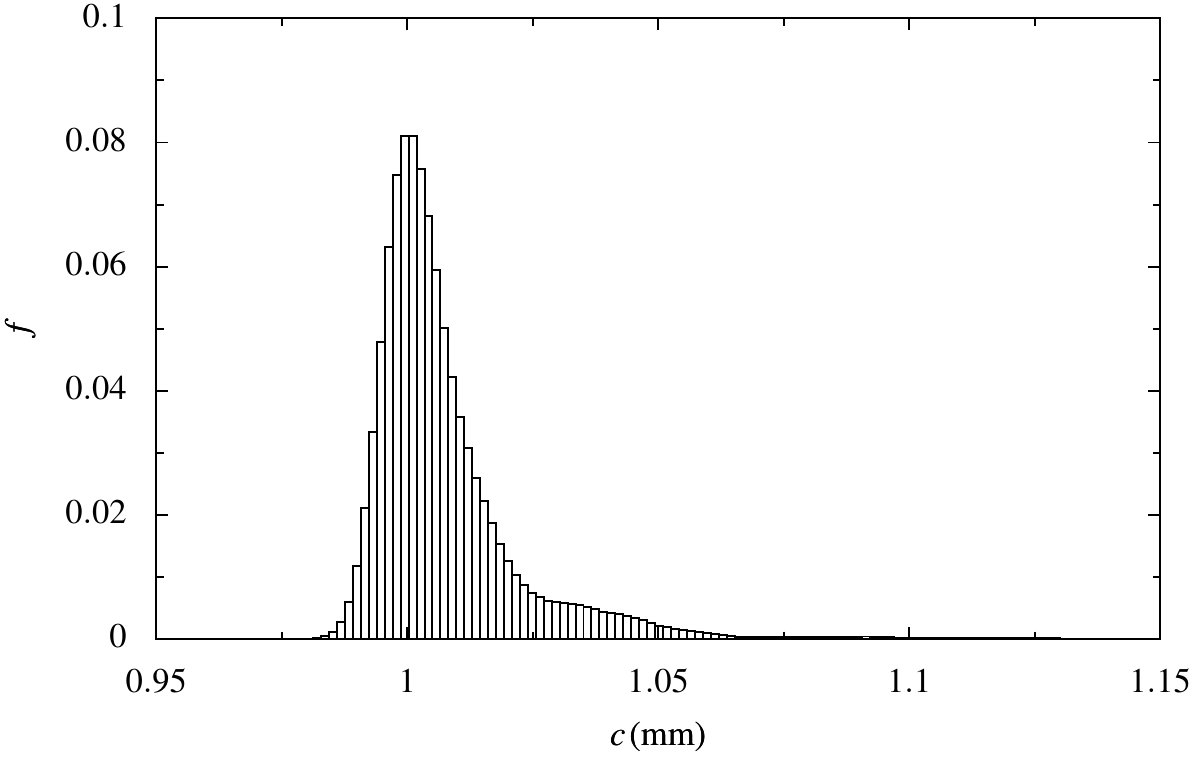}&
    \includegraphics[clip,width=0.33\linewidth]{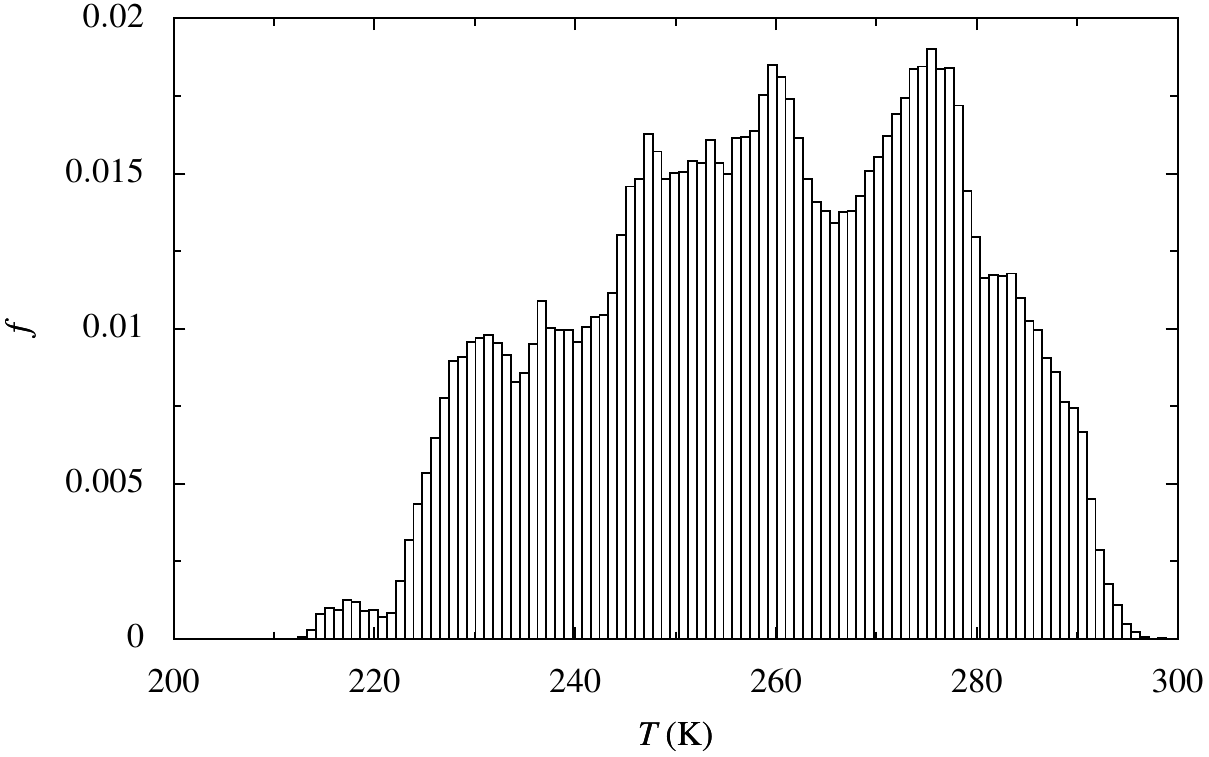}&
    \includegraphics[clip,width=0.33\linewidth]{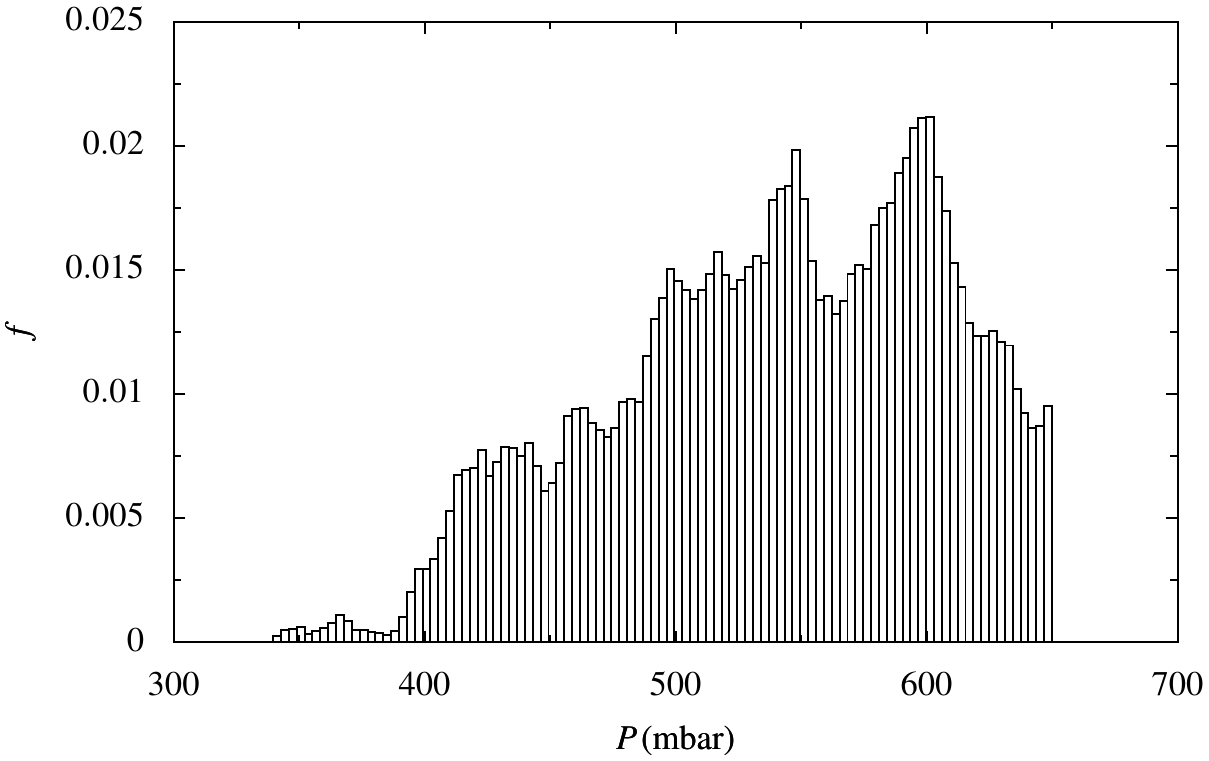}\\
    \includegraphics[clip,width=0.33\linewidth]{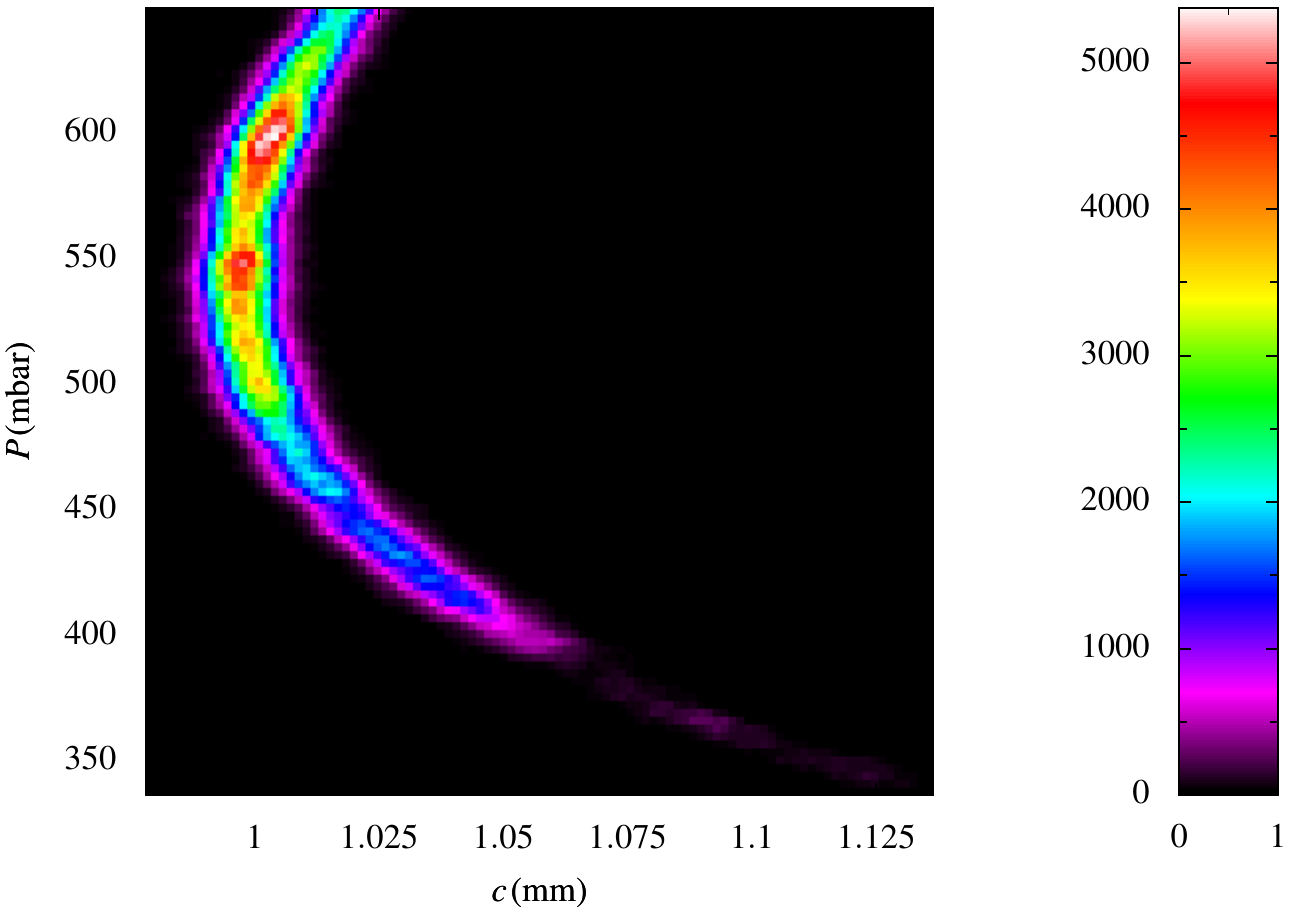}&
    \includegraphics[clip,width=0.33\linewidth]{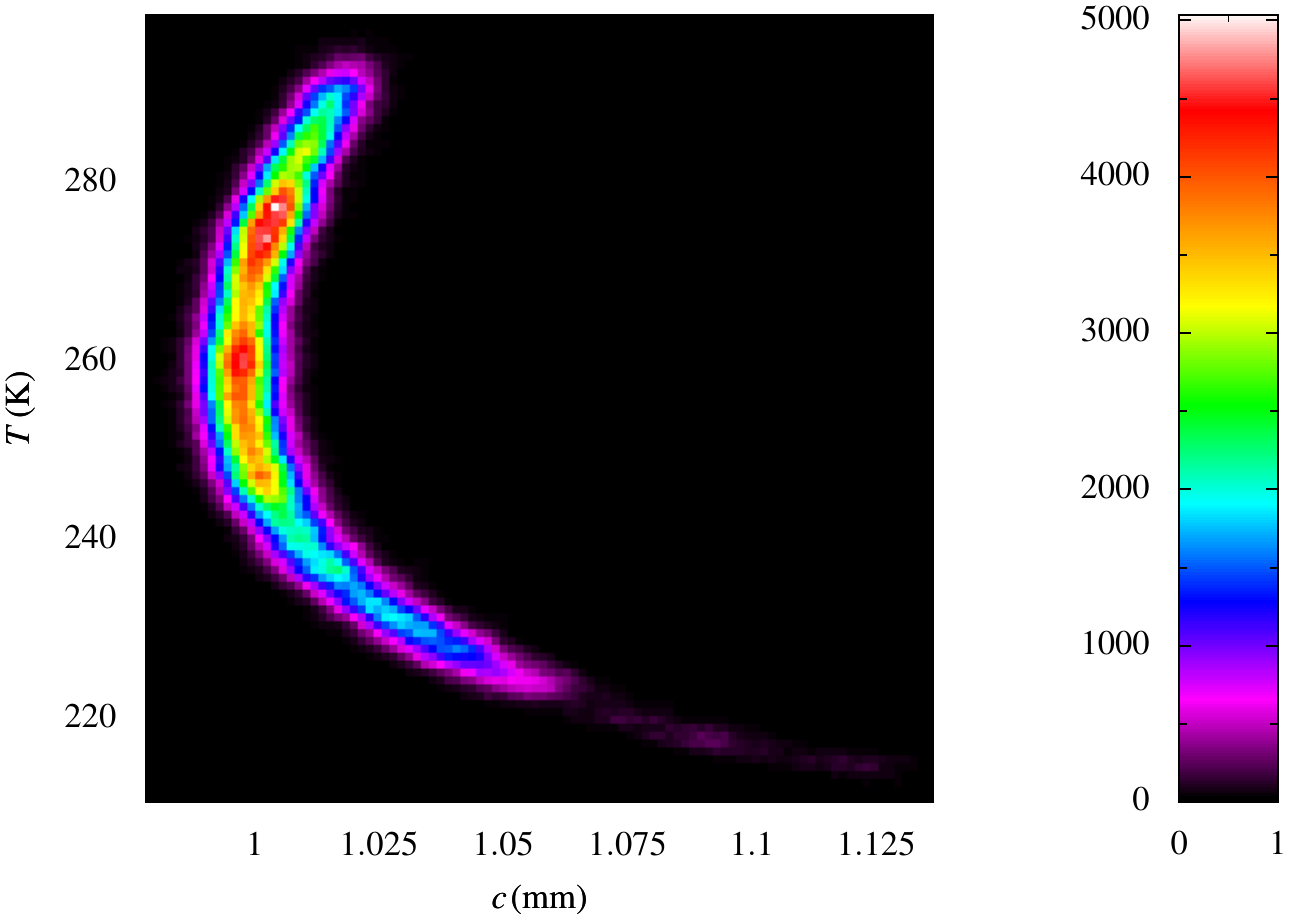}&
    \includegraphics[clip,width=0.33\linewidth]{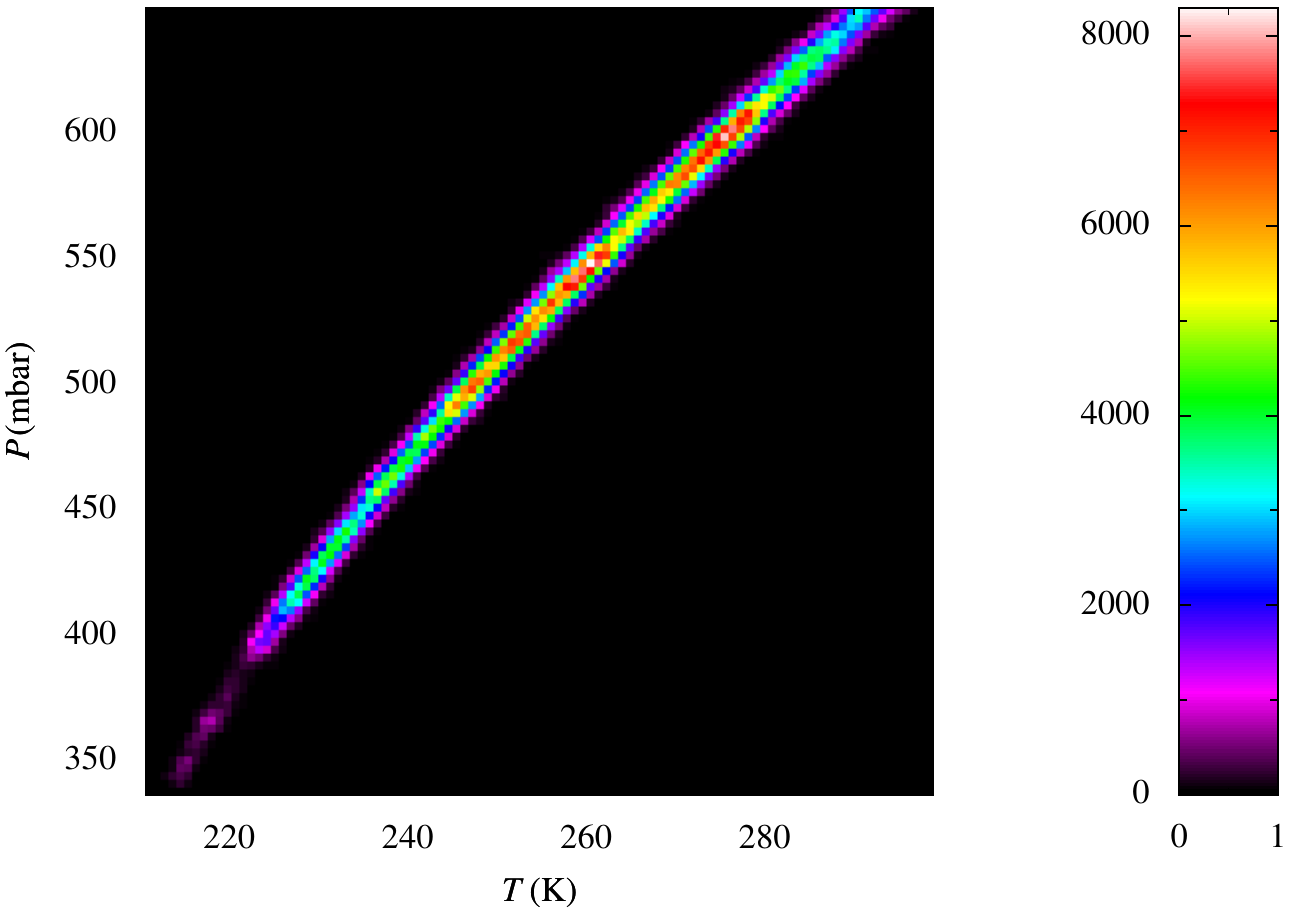}
  \end{tabular}
  
  \caption{Posterior distribution of model parameters from a retrieval
    of simulated measurement at \{$c=1.0\,\unit{mm}$,
    $T=270\,\unit{K}$, $P=580\,\unit{mBar}$\} with very weak priors
    (row 1 of Table~\ref{tab:priors}). The top row shows the
    marginalised distributions of each of the model parameters, while
    the bottom row shows the joint distribution of combinations of two
    parameters with the third marginalised.}
  \label{fig:abssimretr}
\end{figure*}

The first inference we present is with the weak priors from row 1 of
Table~\ref{tab:priors}. Recall that in this case we assume weak
constraints on the model parameters, including allowing the pressure
to be higher than the ground level pressure; hence, this case is an
illustration of the inference when essentially no prior information is
supplied. The posterior distribution of the model parameters for this
case is shown in Figure~\ref{fig:abssimretr}. Like in the other
figures of the model parameter posterior distribution, we present
these results as the marginalised posterior distributions of each of
$c$, $T$ and $P$; and as the joint distribution of $c$ vs $P$, $c$ vs
$T$ and $c$ vs $T$ with the remaining (third) parameter marginalised.

The marginalised probabilities in the top row of
Figure~\ref{fig:abssimretr} show that in this case we can place
relatively weak constraint on the amount of water vapour (to about
5\%) and very poor constraints on both the pressure and the
temperature of the water vapour. It can also be clearly noted that the
posterior distributions are not well approximated by a Gaussian
distribution; for example, the posterior distribution $p(c)$ shows a
long tail toward higher water columns. 

The reason for the poor inference can be understood from the second
row of Figure~\ref{fig:abssimretr} which shows the joint distributions
of each combination of the model parameters. Considering first the
plot of the joint probability $p(TP)$ in the right panel of the lower
row, Figure~\ref{fig:abssimretr}, we can see that the retrieval of the
pressure and temperature are almost exactly degenerate.  In other
words if a certain combination of pressure and temperature explain the
observed data well, then a higher temperature and a proportionally
higher pressure describes the observation also sufficiently well.

This degeneracy between pressure and temperature also affects the
accuracy of the retrieval of the water vapour column as shown in the
left and middle panels of the lower row of
Figure~\ref{fig:abssimretr}. There we see that the extreme values of
pressure and temperature that are permissible due to the degeneracy
give rise to a tail of likelihood to higher values of the retrieved
water vapour column.

\begin{figure*}
  \begin{tabular}{ccc}
    \includegraphics[clip,width=0.33\linewidth]{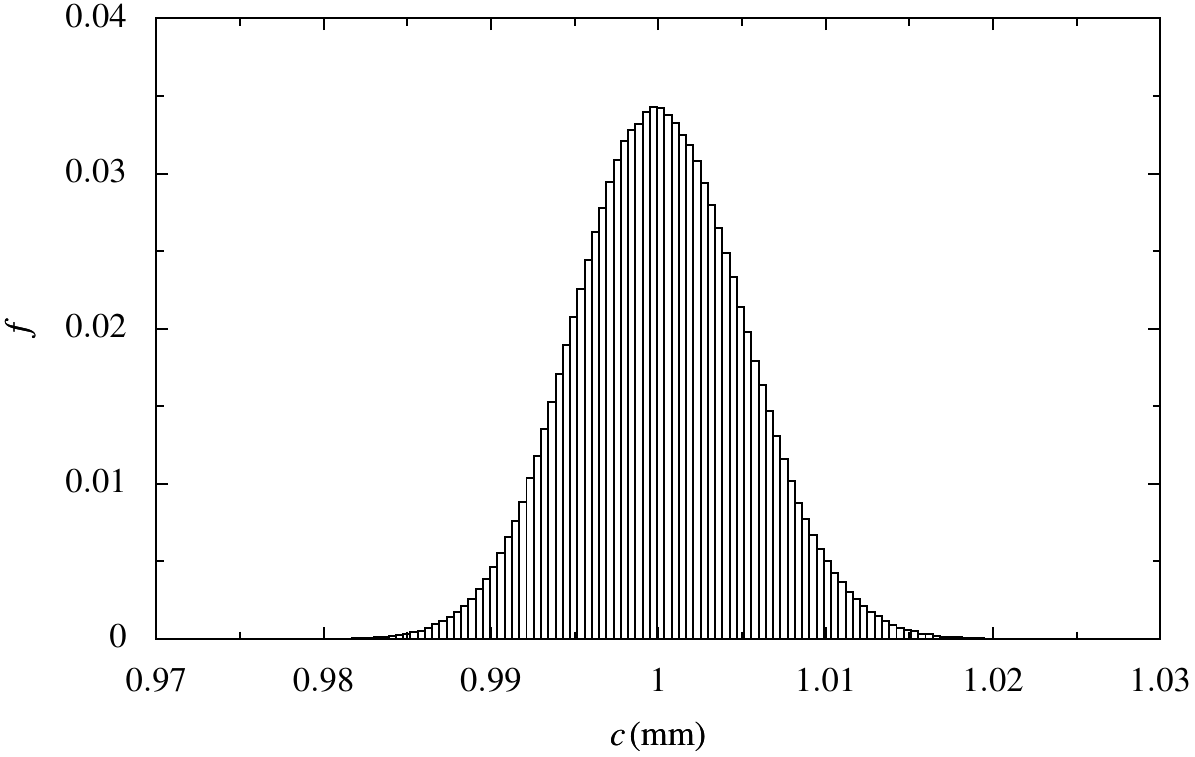}&
    \includegraphics[clip,width=0.33\linewidth]{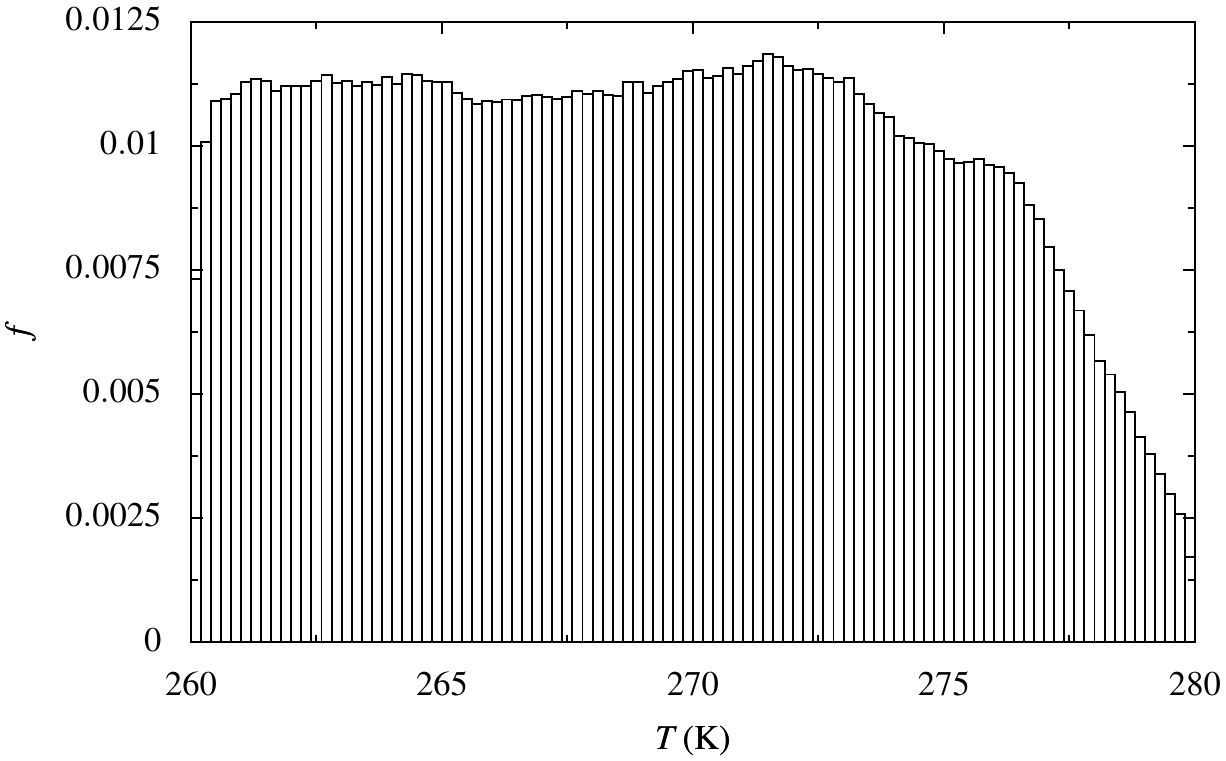}&
    \includegraphics[clip,width=0.33\linewidth]{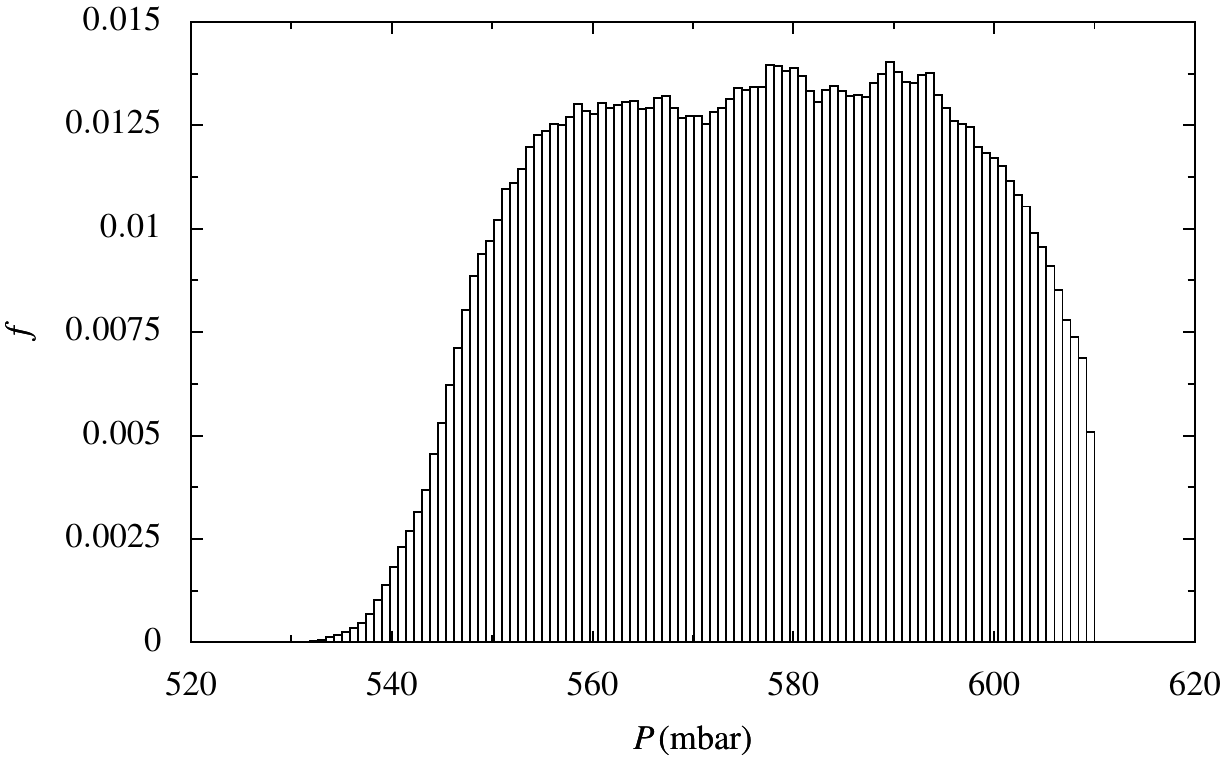}\\
    \includegraphics[clip,width=0.33\linewidth]{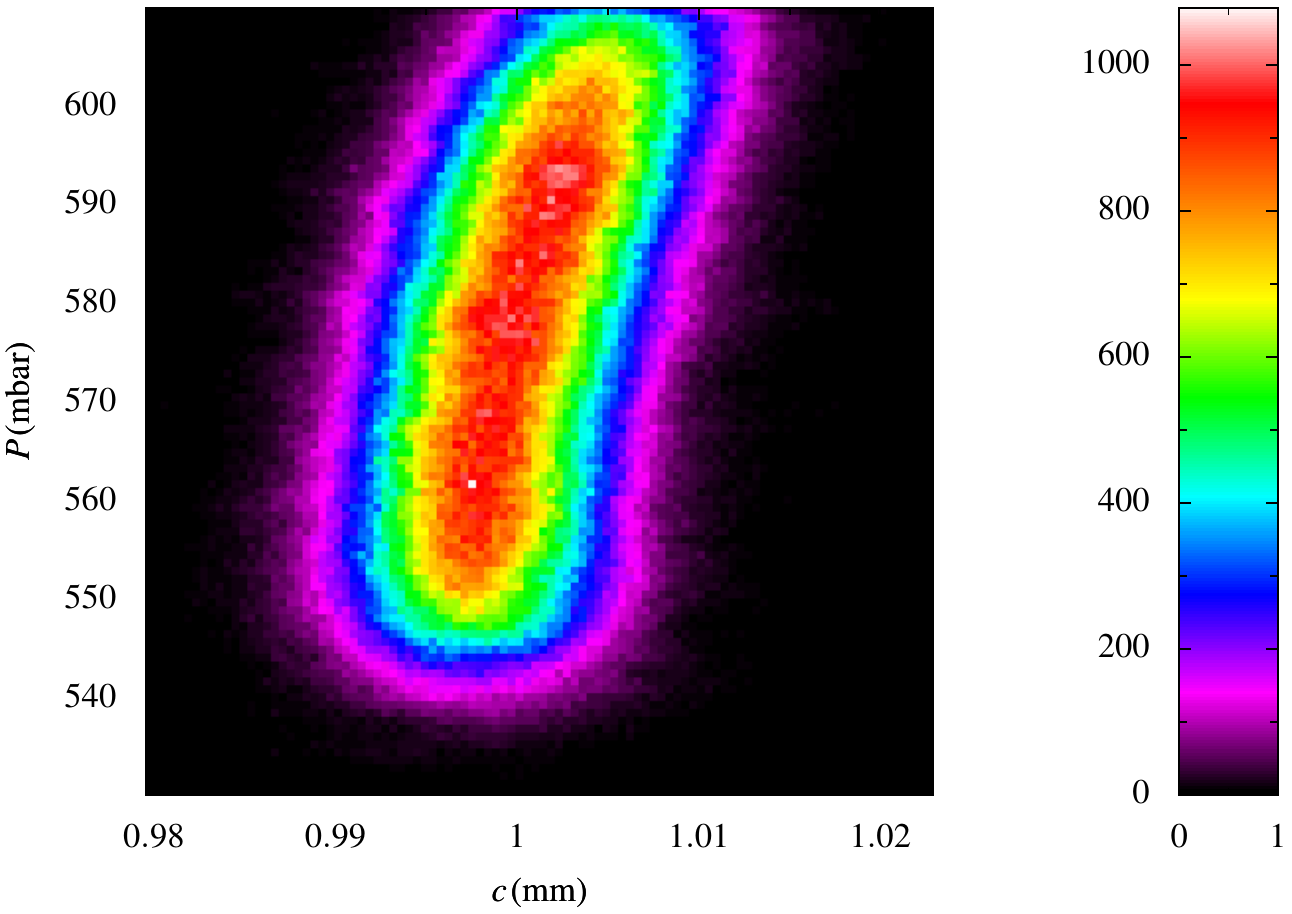}&
    \includegraphics[clip,width=0.33\linewidth]{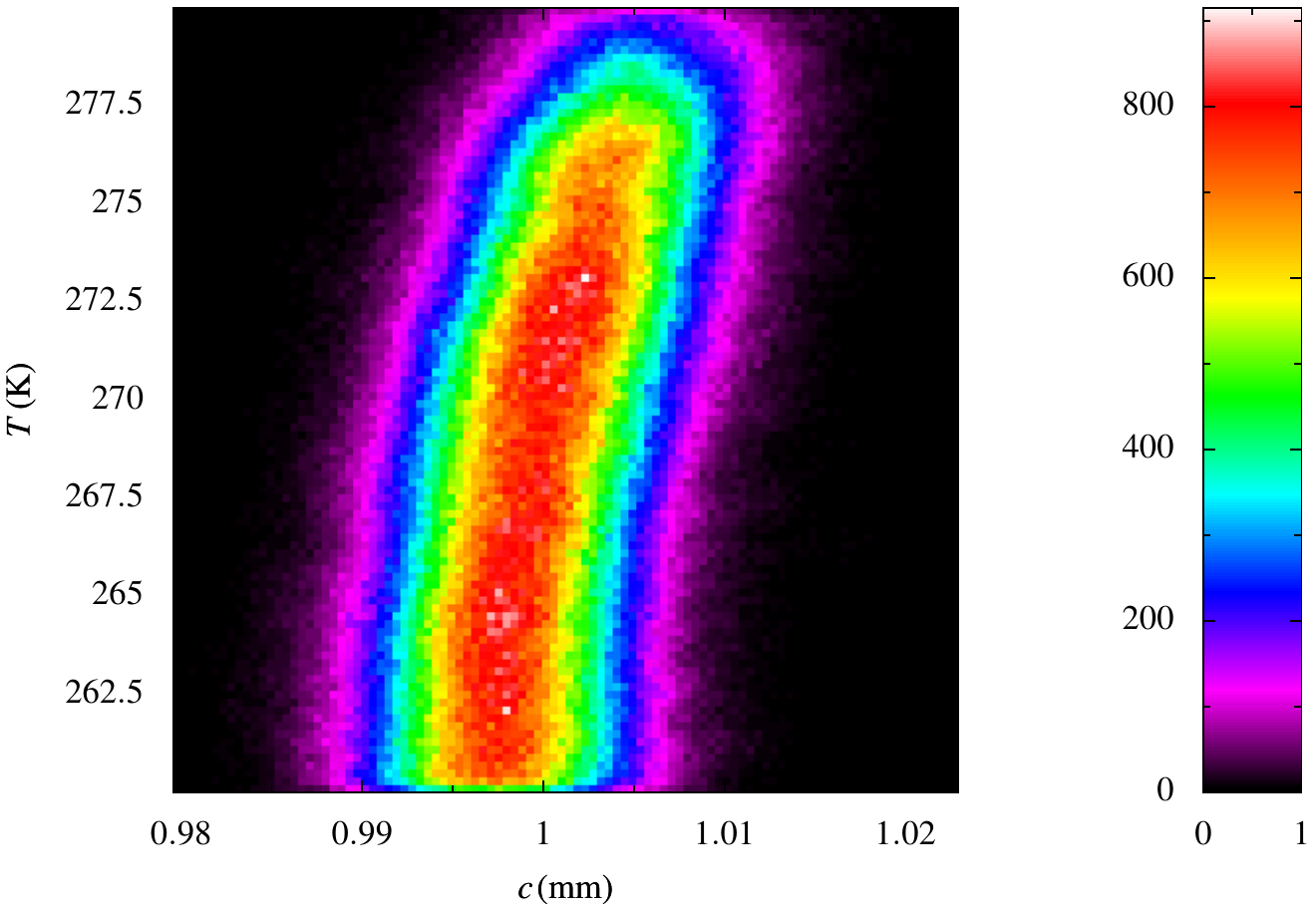}&
    \includegraphics[clip,width=0.33\linewidth]{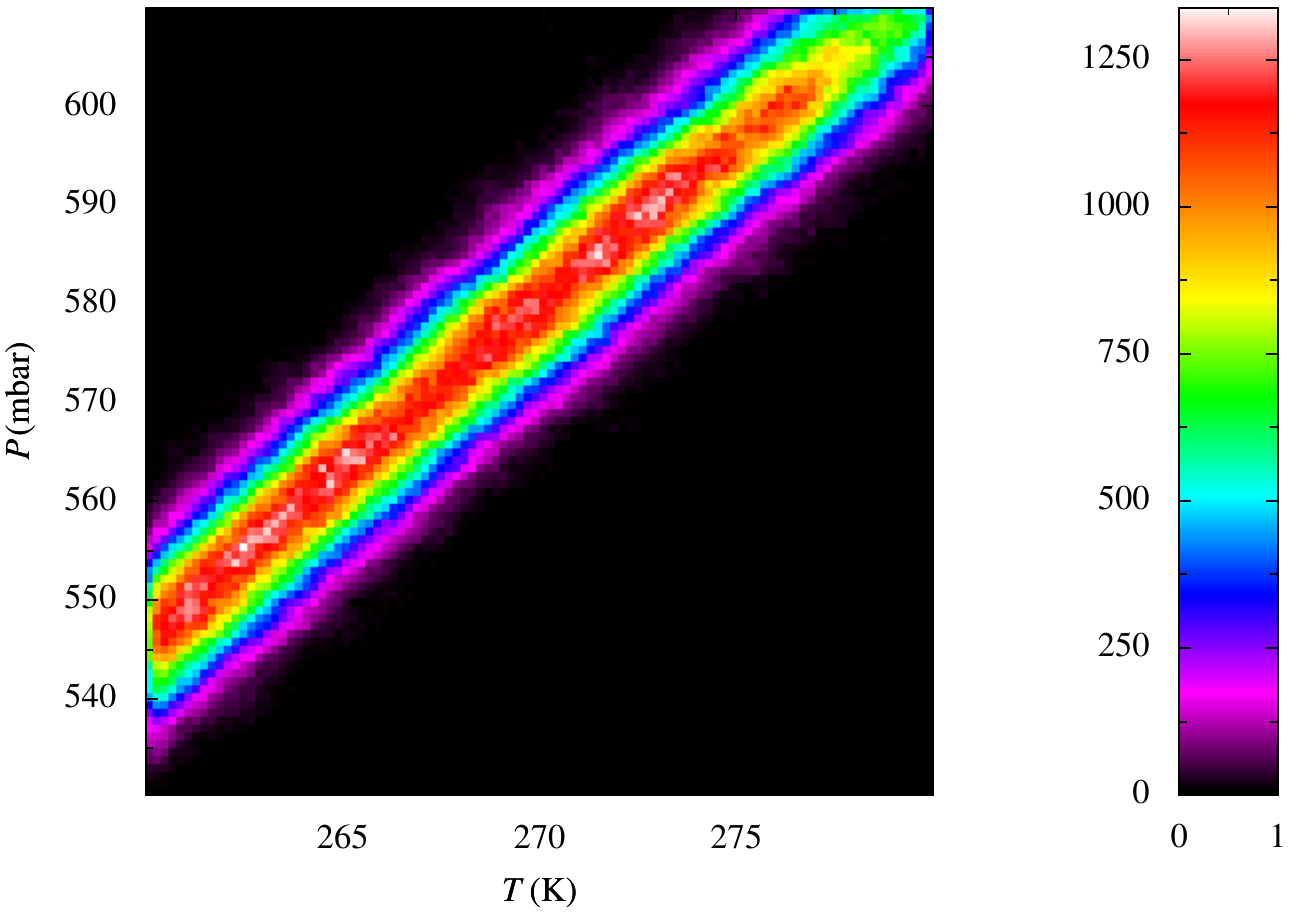}
  \end{tabular}
  \caption{Posterior distribution of model parameters derived from
    four absolute sky brightness temperatures only, i.e., like
    Figure~\ref{fig:abssimretr}, but now including reasonable flat
    priors on all three of the $c$, $T$ and $P$ model parameters.}
  \label{fig:abssimprior}
\end{figure*}

The next result we describe combines the same simulated data point
with priors in the second row of Table~\ref{tab:priors}. These are the
priors that we might have without significant ancillary information,
i.e., that we know the temperature of the water vapour layer to within
20\,K and its pressure to within 80\,mBar. 

The posterior distribution from this inference is shown in
Figure~\ref{fig:abssimprior}. It can be seen from the top row of this
figure that these results are qualitatively different from the
inference with very non-informative priors. In this case, the
inference of the water vapour is well approximated by Gaussian with a
full-width-half-maximum of about 0.012\,mm and the entire distribution
is within 0.02\,mm of the model value. The inferences of the
temperature and pressure are still poor however; in fact, it can be
seen that their distributions fill almost entirely the space allowed
by their priors, indicating that the priors in this case are providing
important information.

The lower row of Figure~\ref{fig:abssimprior} again provides an
explanation for the marginalised distributions of the model
parameters.  The pressure-temperature joint distribution again shows
the degeneracy which explains the poor retrieval of each of those
model parameters individually. The water column-pressure and water
column-temperature distributions still show spreads but with two
important differences: 
\begin{enumerate}
  \item The priors mean that the range over which pressure and
    temperature can vary is much smaller, therefore leading to smaller
    errors in the retrieved water column
  \item The joint distributions are highly elongated along the
    vertical axes, which means a relatively large change in temperate
    or pressure is required to cause an error in the retrieved water
    column
\end{enumerate}
The condition (ii) is somewhat specific to the simulated point in
parameter space, and will not be as true for a general combination of
filter centres/widths and observed sky brightnesses.

\begin{figure*}
  \subfloat[Marginalised distributions of the phase correction coefficients]{
  \begin{tabular}{cc}
    \includegraphics[clip,width=0.45\linewidth]{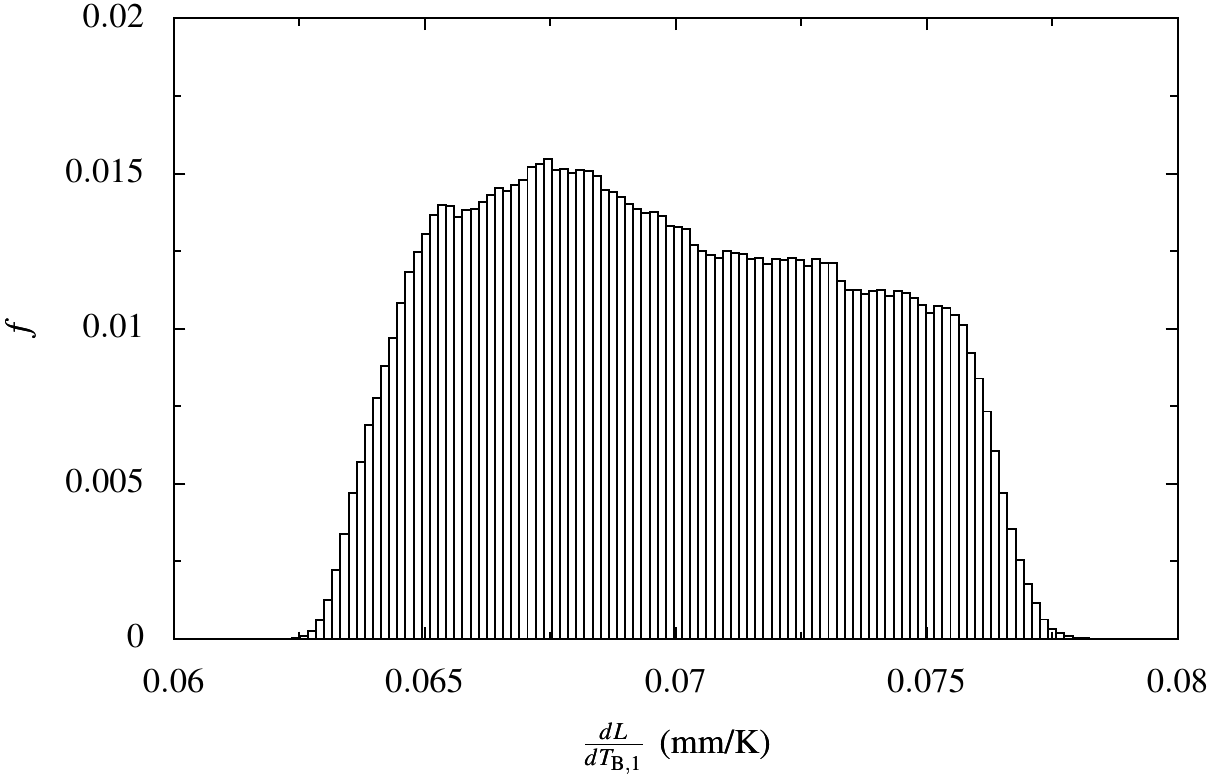}&
    \includegraphics[clip,width=0.45\linewidth]{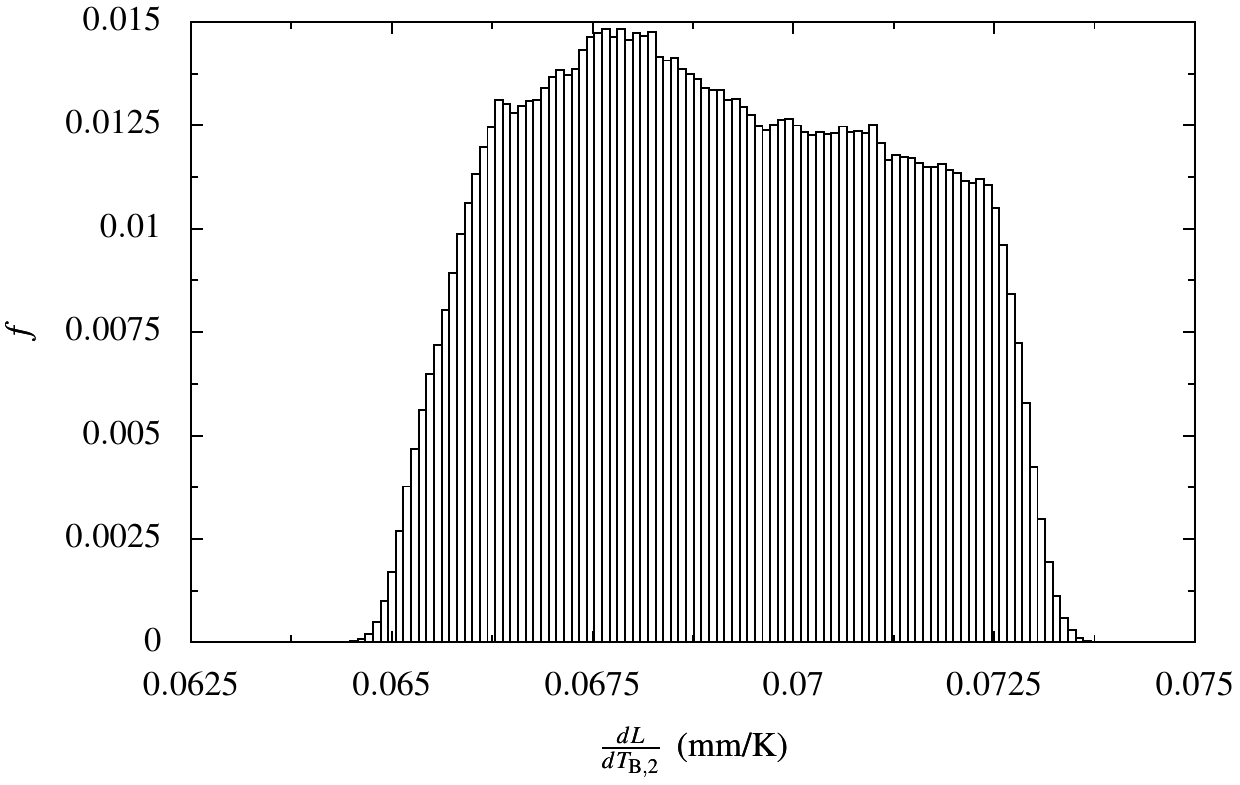}\\
    \includegraphics[clip,width=0.45\linewidth]{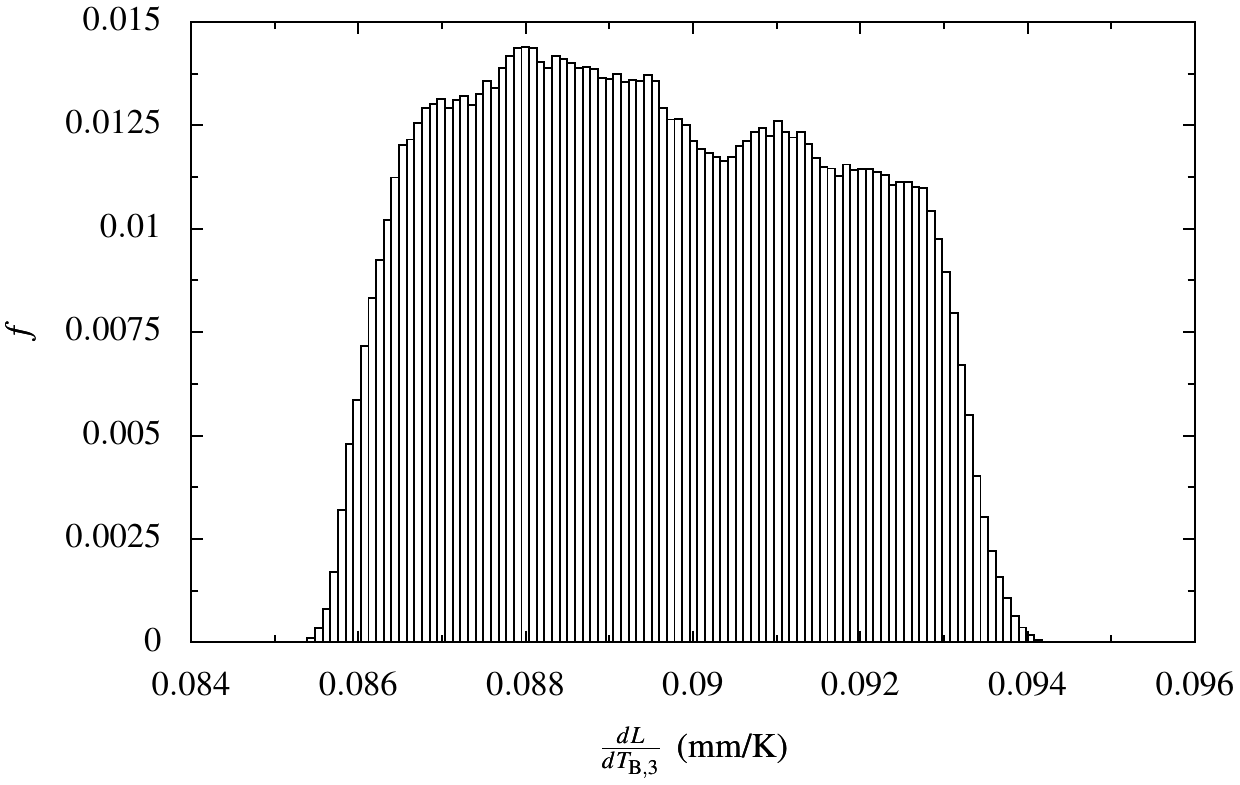}&
    \includegraphics[clip,width=0.45\linewidth]{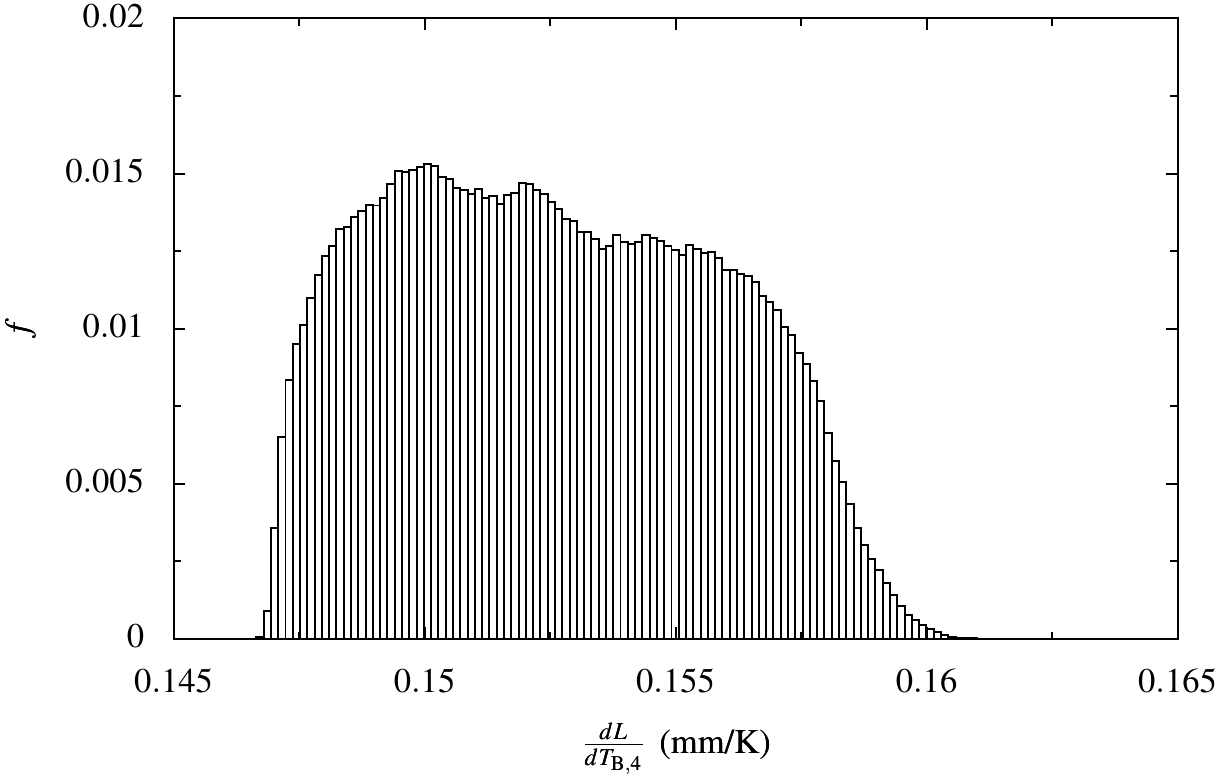}
  \end{tabular}}

  \subfloat[Join distributions of the phase correction coefficiens]{
  \begin{tabular}{ccc}
    \includegraphics[clip,width=0.33\linewidth]{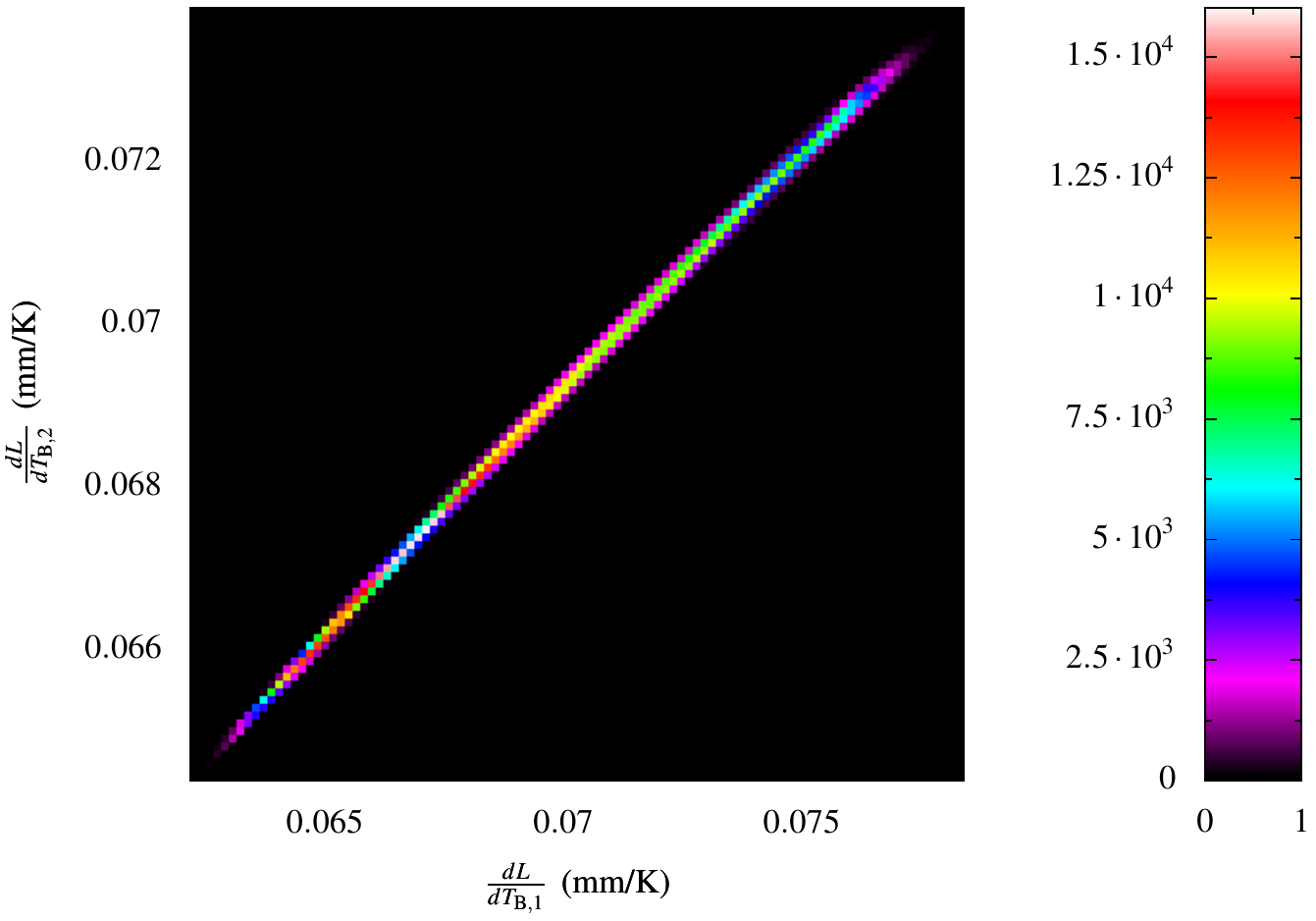}
    &
    \includegraphics[clip,width=0.33\linewidth]{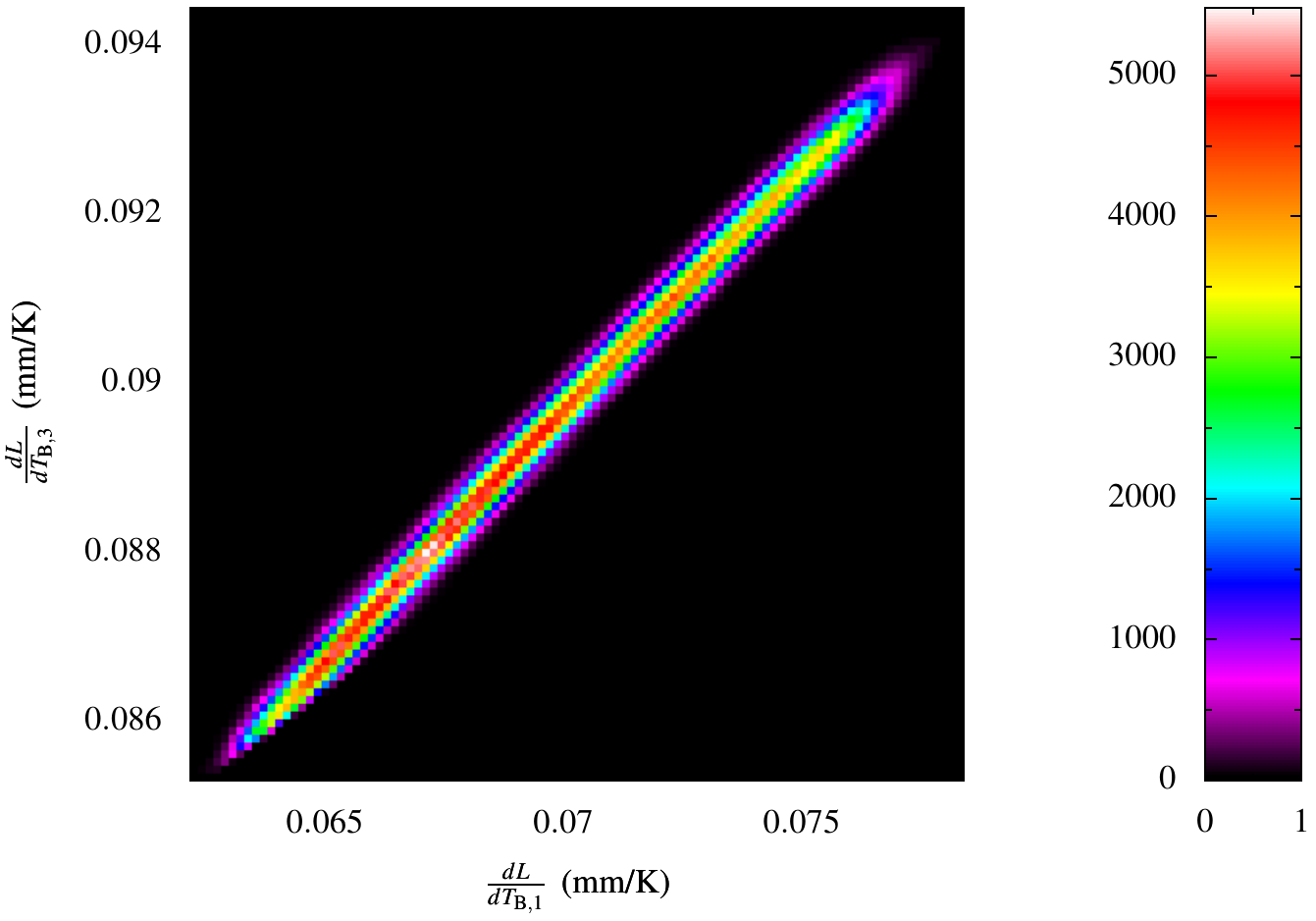}
    &
    \includegraphics[clip,width=0.33\linewidth]{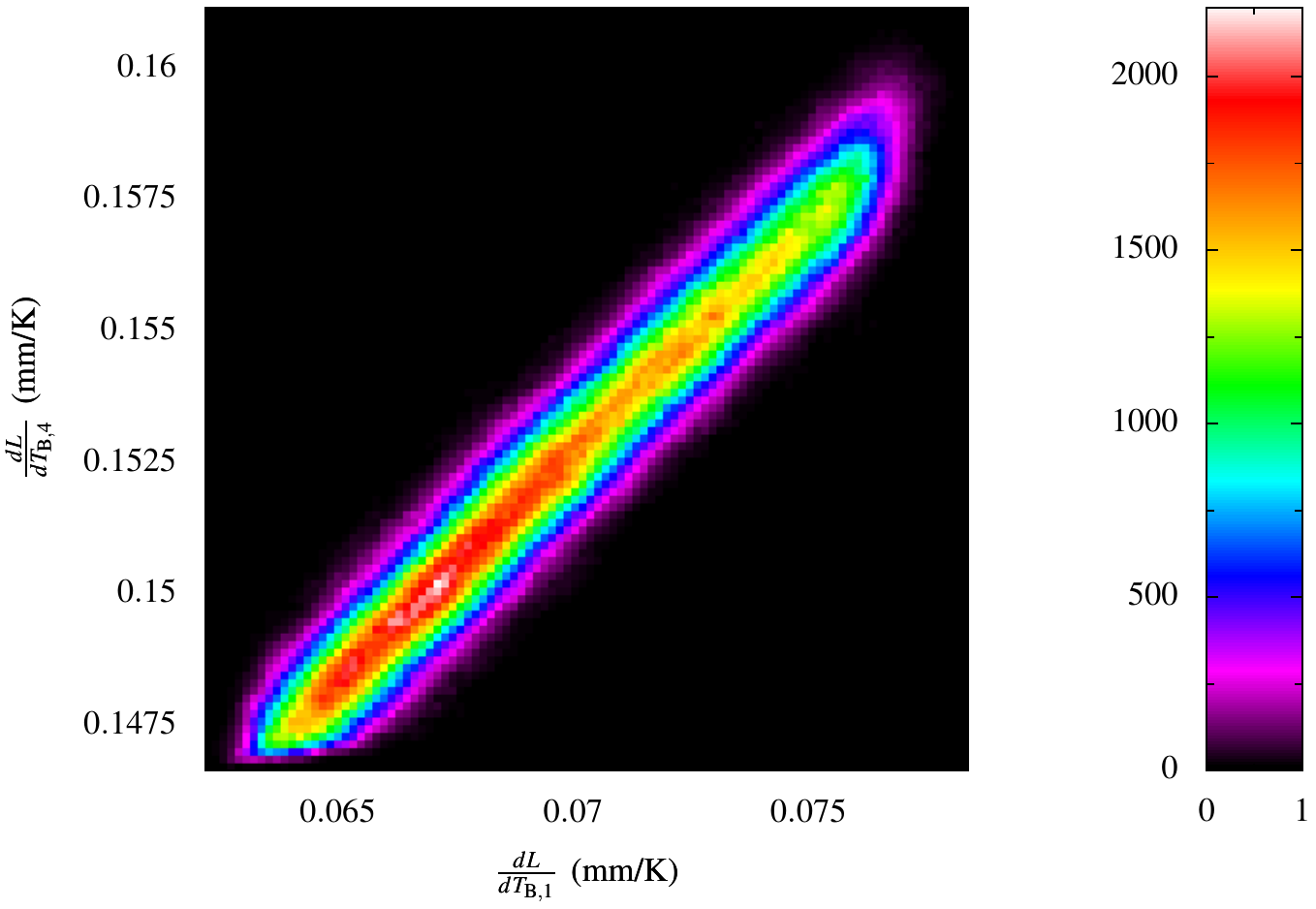}\\
    \includegraphics[clip,width=0.33\linewidth]{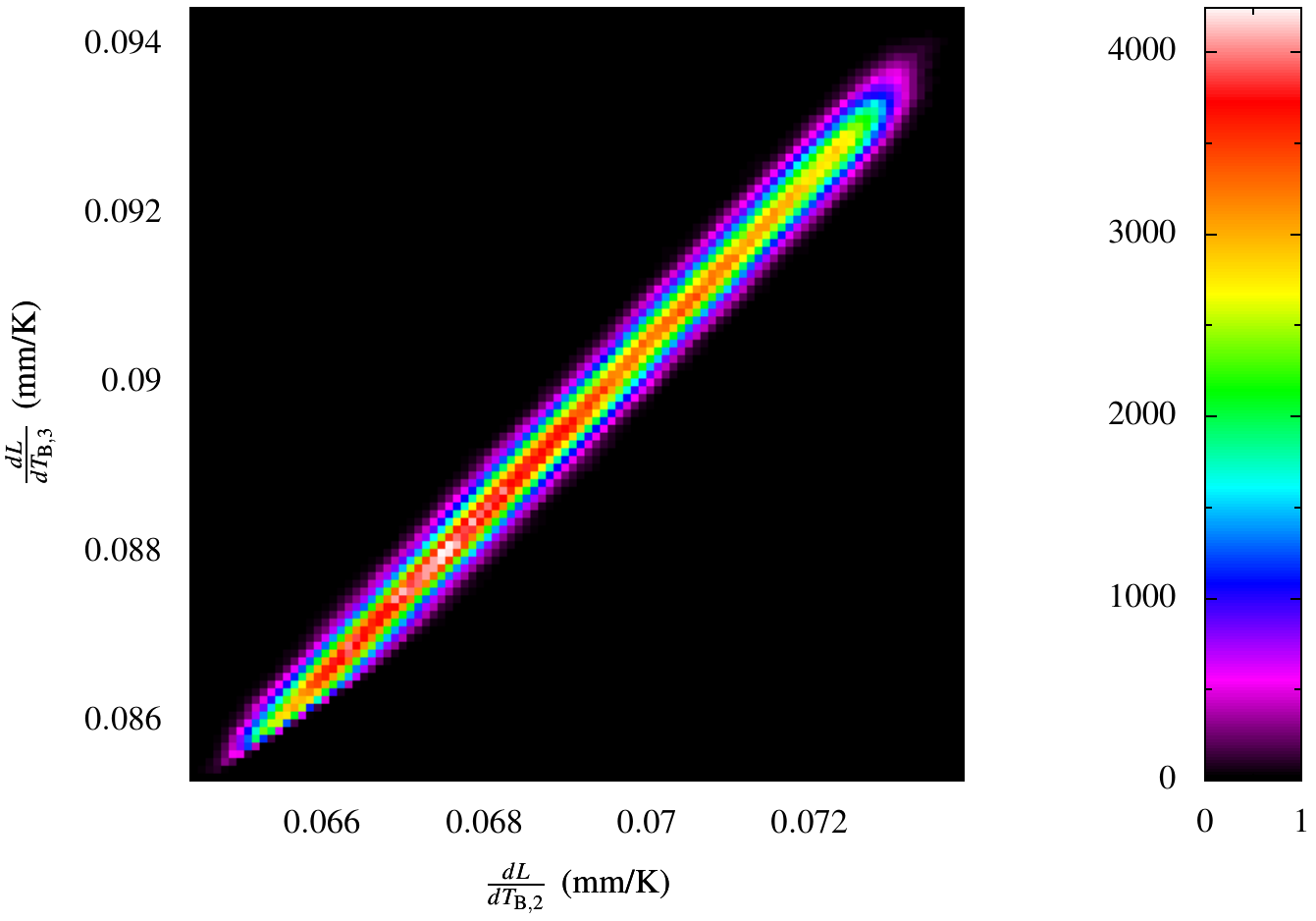}
    &
    \includegraphics[clip,width=0.33\linewidth]{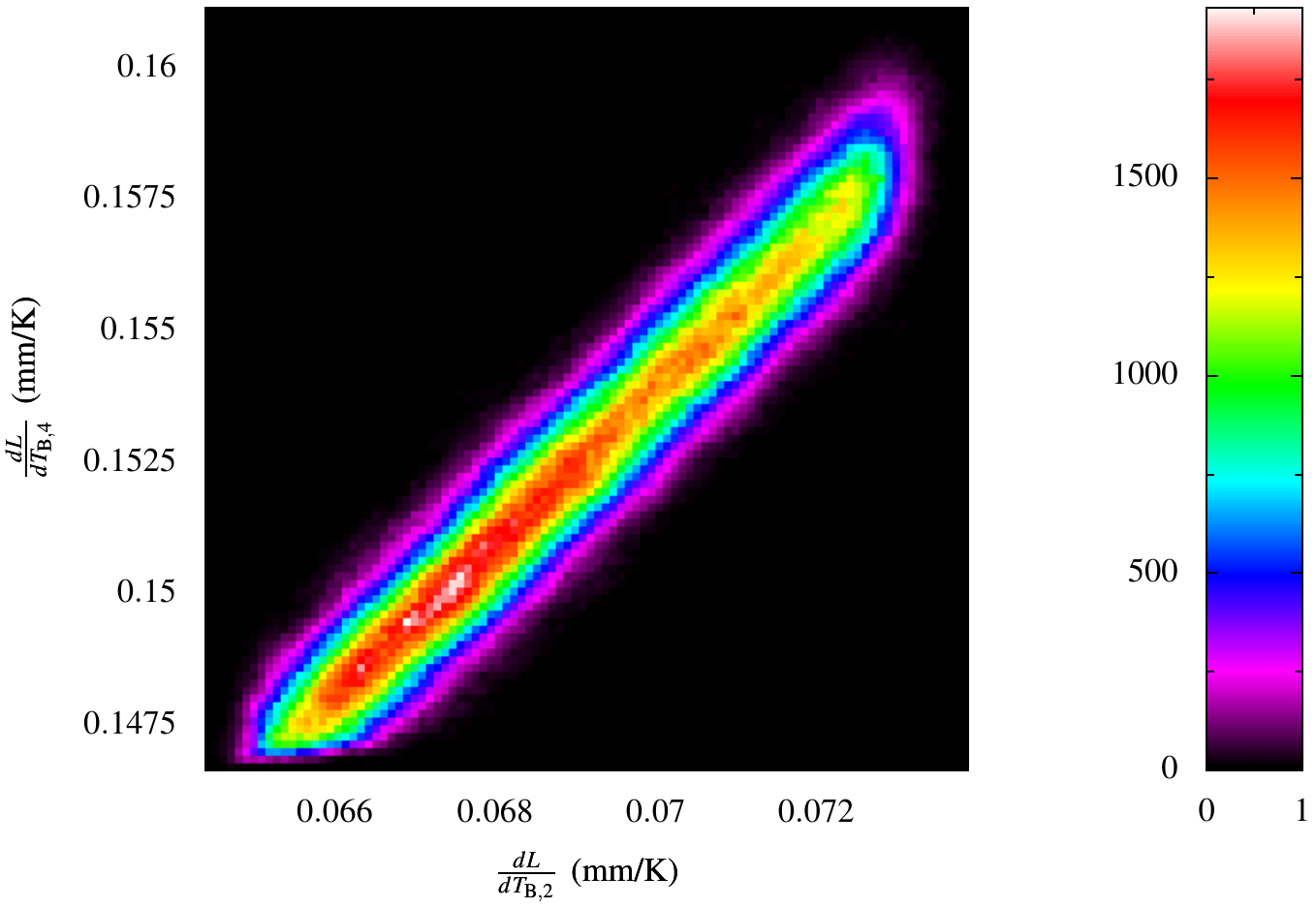}
    &
    \includegraphics[clip,width=0.33\linewidth]{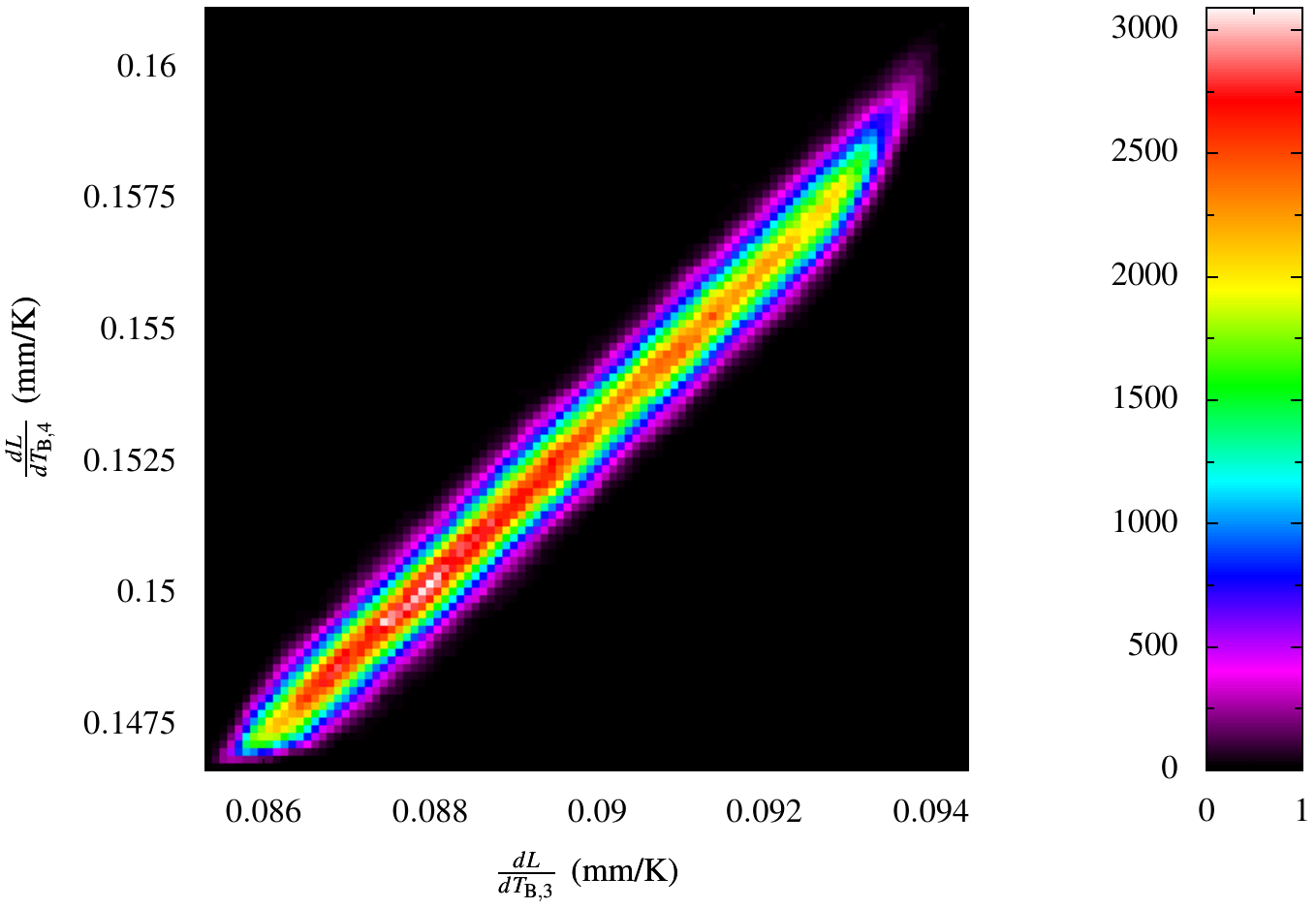}
  \end{tabular}}

  \caption{Posterior distributions of the coefficients ${\rm
        d}L/{\rm d} T_{{\rm B},i}$ used to convert brightness
    fluctuations into path fluctuations for the model parameters
    posterior shown in Figure~\ref{fig:abssimprior}.}
  \label{fig:abssimprior-dtdl}
\end{figure*}

Since the inference shown in Figure~\ref{fig:abssimprior} is
constrained to a reasonable volume of the parameter space we can
compute the ${\rm d}L/{\rm d} T_{{\rm B},i}$ to find how well
we can predict the phase correction coefficients. The results are
shown in Figure~\ref{fig:abssimprior-dtdl} as the marginalised
distribution of each of the coefficients (upper part of the figure)
and also the joint distribution of each pair of coefficients (lower
part). It can be seen in the upper part of this figure that the
posterior distributions are non-Gaussian, in fact almost square, and
of width of about 10\%. This large spread is probably dominated by the
uncertainty in the retrieved temperature of the water vapour which
influences its refractive index (see
Equation~\ref{eq:nondisp-simple}). 

The joint distributions plotted in the lower part of
Figure~\ref{fig:abssimprior-dtdl} show that in this case the errors on
inference of the coefficients are very highly correlated, with similar
correlation for each pair of the channels. This means that it is
unlikely that making use of all of the radiometer channels
\emph{simultaneously\/} would reduce the error in phase correction due
to the uncertainty in the inference of the coefficients. Under
different conditions and perhaps with additional observational data,
this may be however be possible.

\begin{figure*}
  \begin{tabular}{ccc}
    \includegraphics[clip,width=0.33\linewidth]{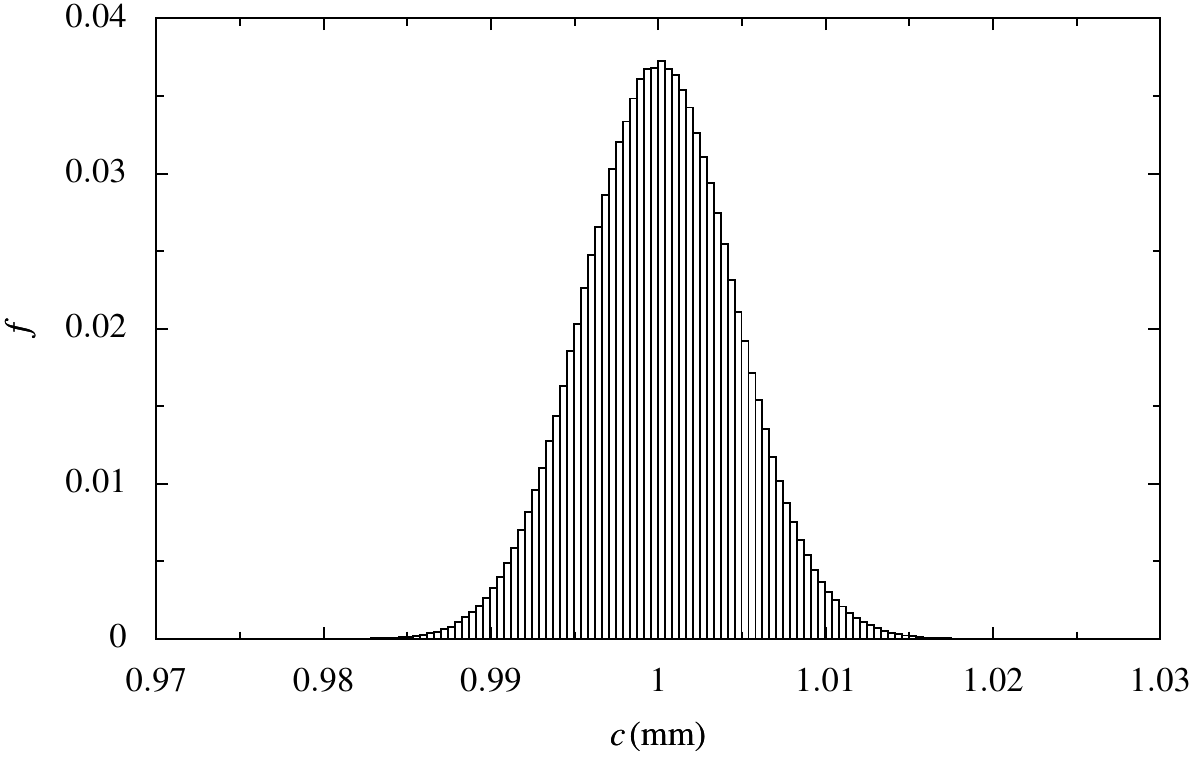}&
    \includegraphics[clip,width=0.33\linewidth]{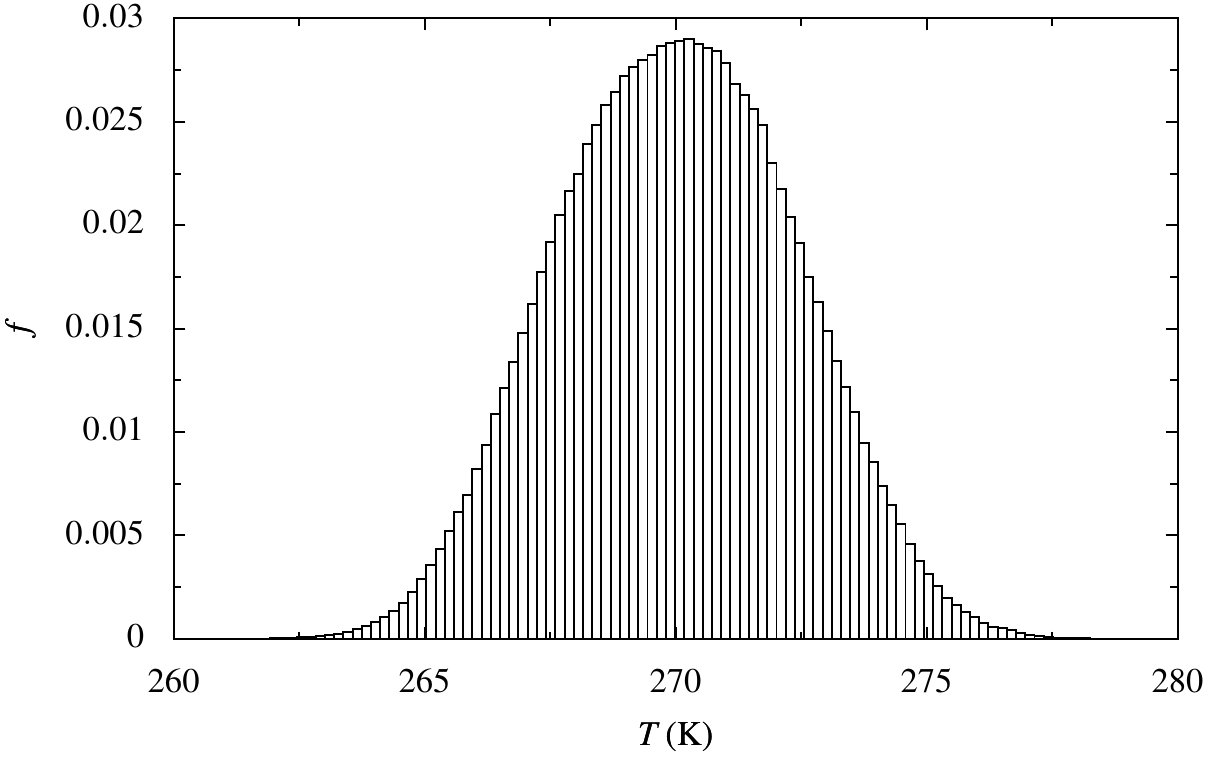}&
    \includegraphics[clip,width=0.33\linewidth]{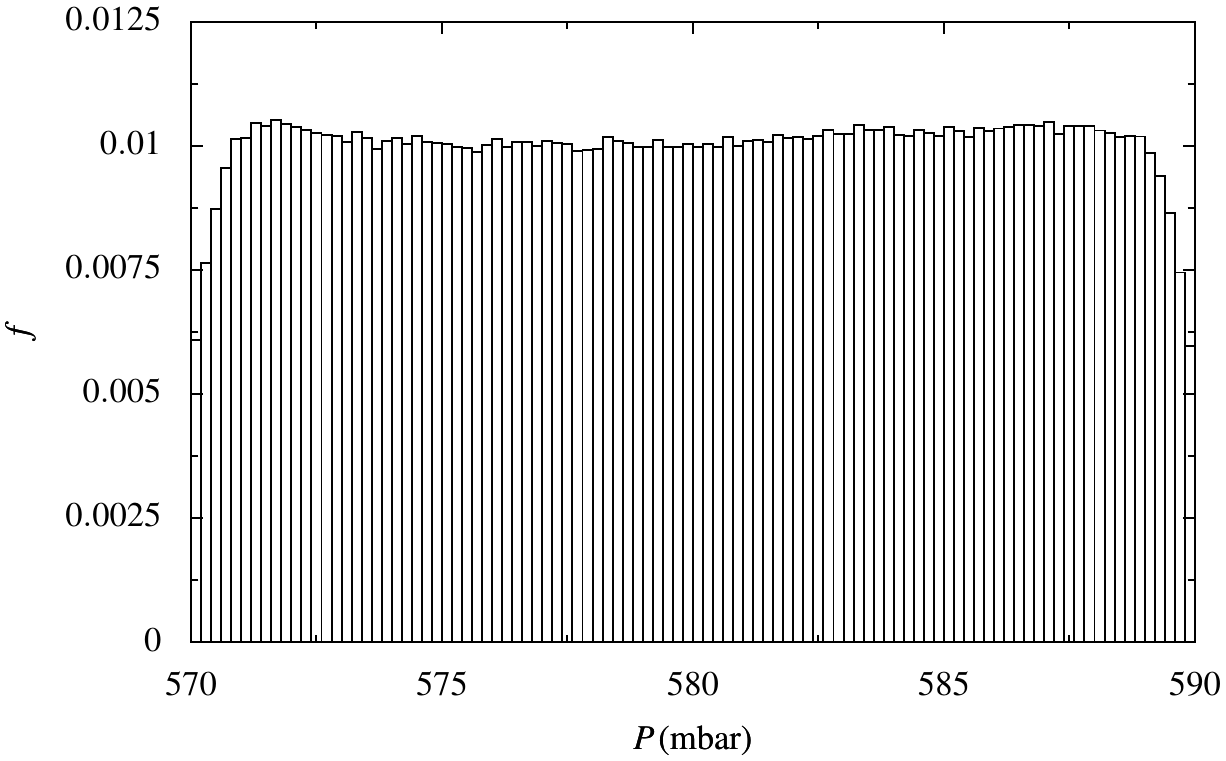}\\
    \includegraphics[clip,width=0.33\linewidth]{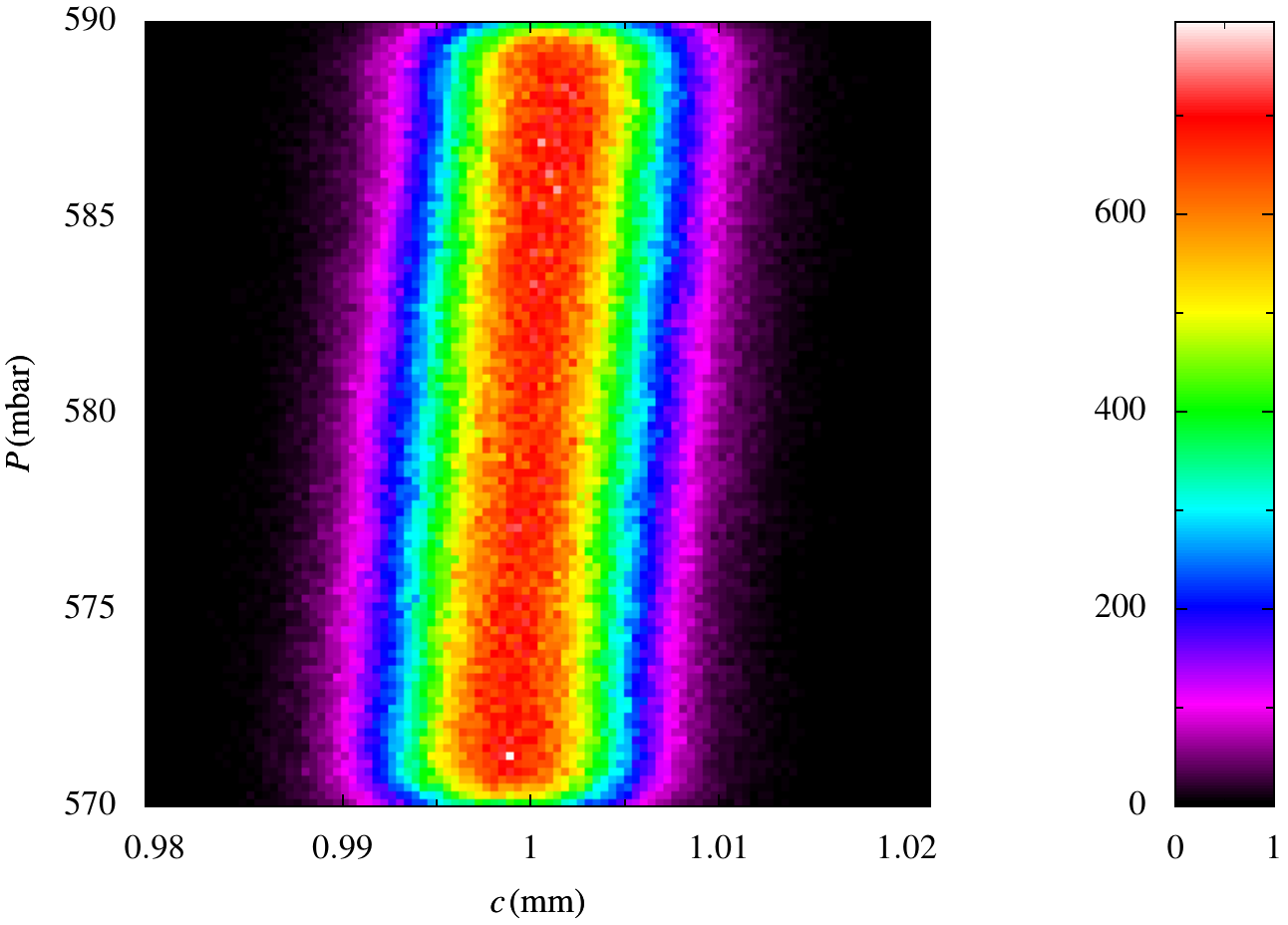}&
    \includegraphics[clip,width=0.33\linewidth]{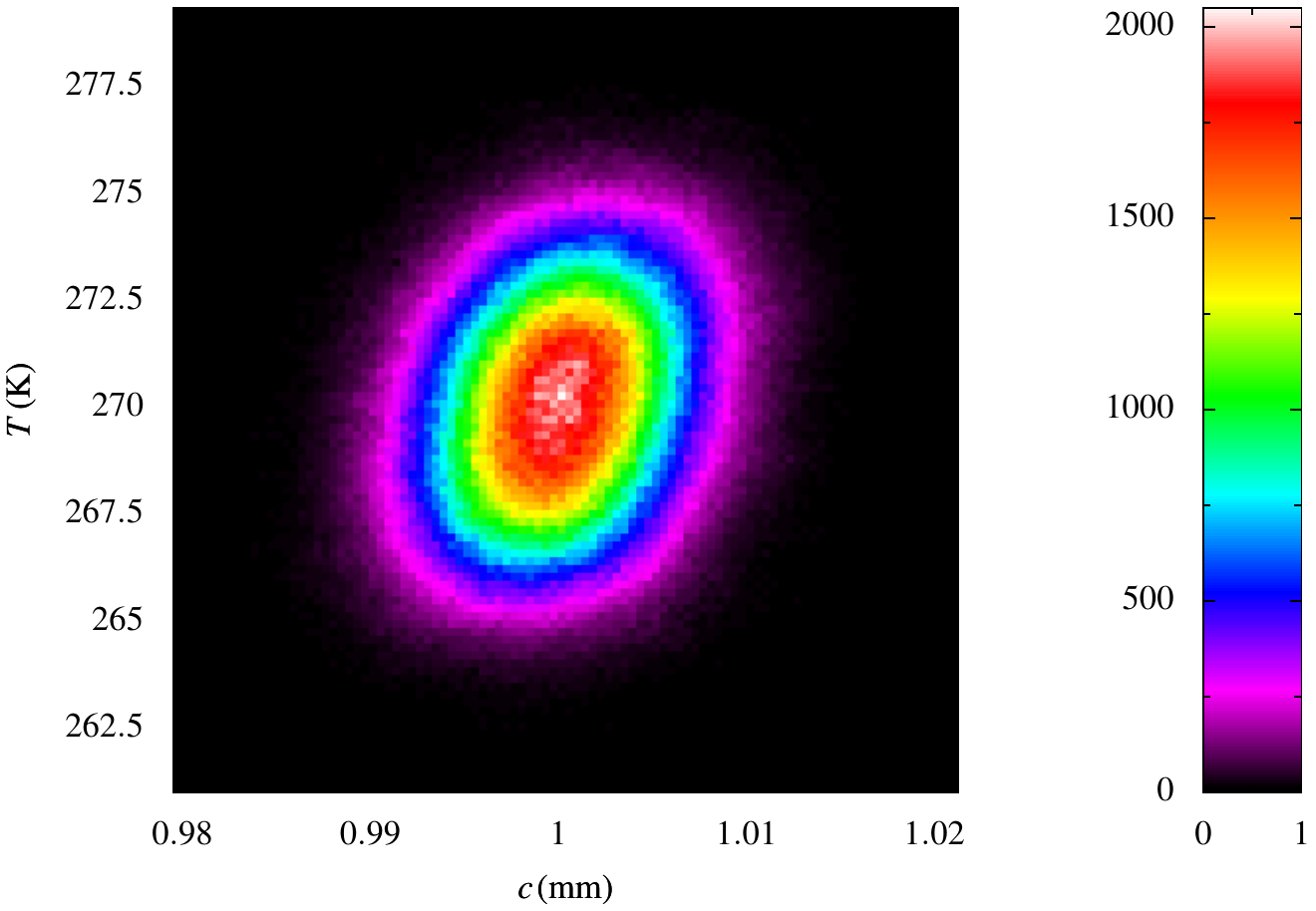}&
    \includegraphics[clip,width=0.33\linewidth]{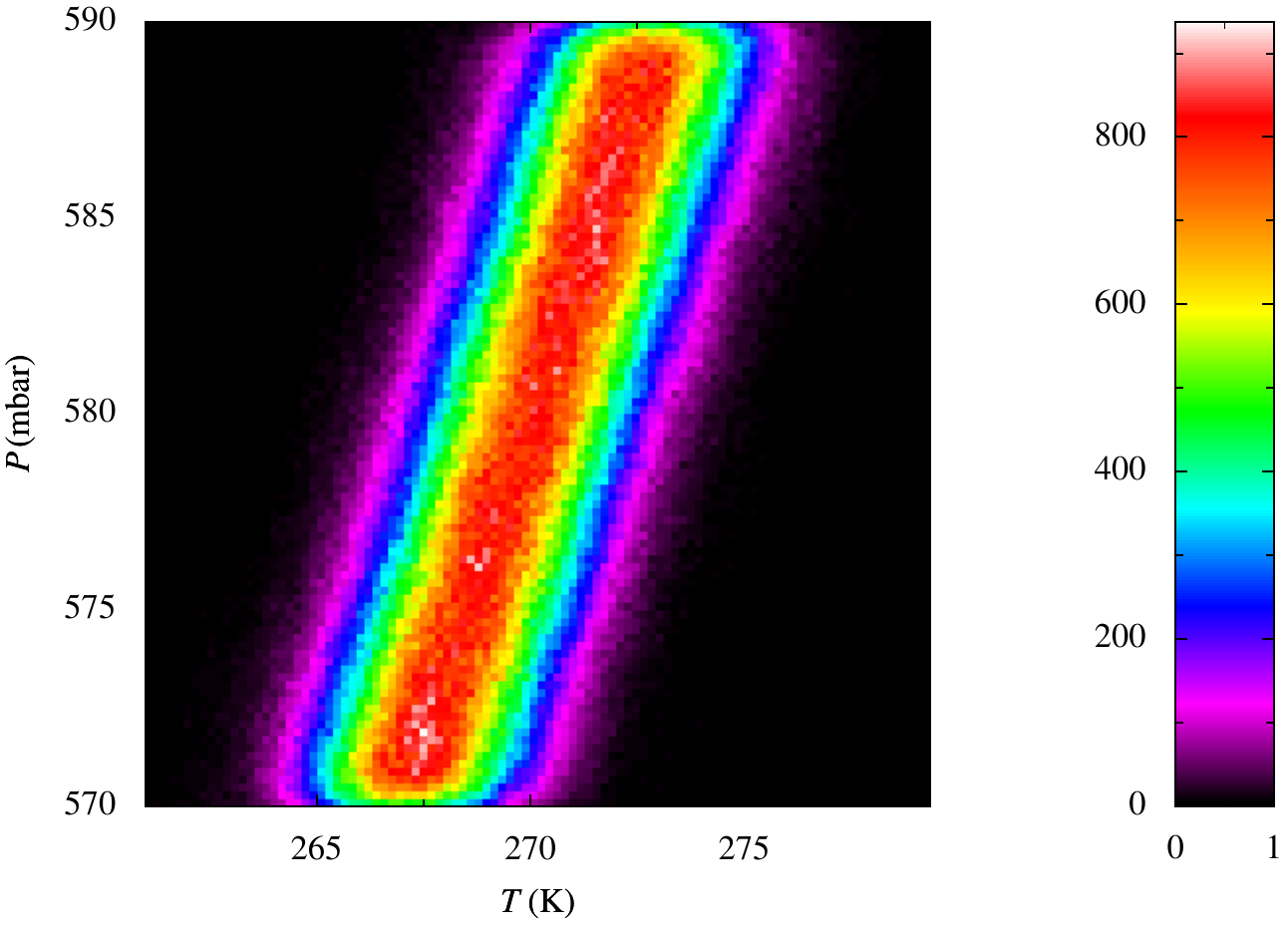}
  \end{tabular}
  \caption{Model parameter posterior distribution with tight prior on
    pressure.}
  \label{fig:abssimpriortight}
\end{figure*}

The last result that we show in this section is an inference with a
tight prior on the pressure (or, equivalently, the height) of the
water vapour layer, i.e., $570\,\unit{mBar}<P<590\,\unit{mBar}$, but
with the prior on temperature as before. The posterior distribution of
the model parameters for this case is shown in
Figure~\ref{fig:abssimpriortight}.

We again find that the results of the inference are qualitatively
different, primarily in that the posterior distribution of the
temperature of the water vapour is now approximately Gaussian and with
a full-width-half-maximum of about 8\,K, significantly less than its
prior range of 20\,K. The posterior distribution for the pressure
however is, as expected, completely dominated by the prior and simply
flat over its prior range. The implication is that a tighter
\emph{a-priori} range of the pressure allows a much better inference
of the temperature of the water vapour layer.

\begin{figure*}
  \begin{tabular}{cc}
    \includegraphics[clip,width=0.45\linewidth]{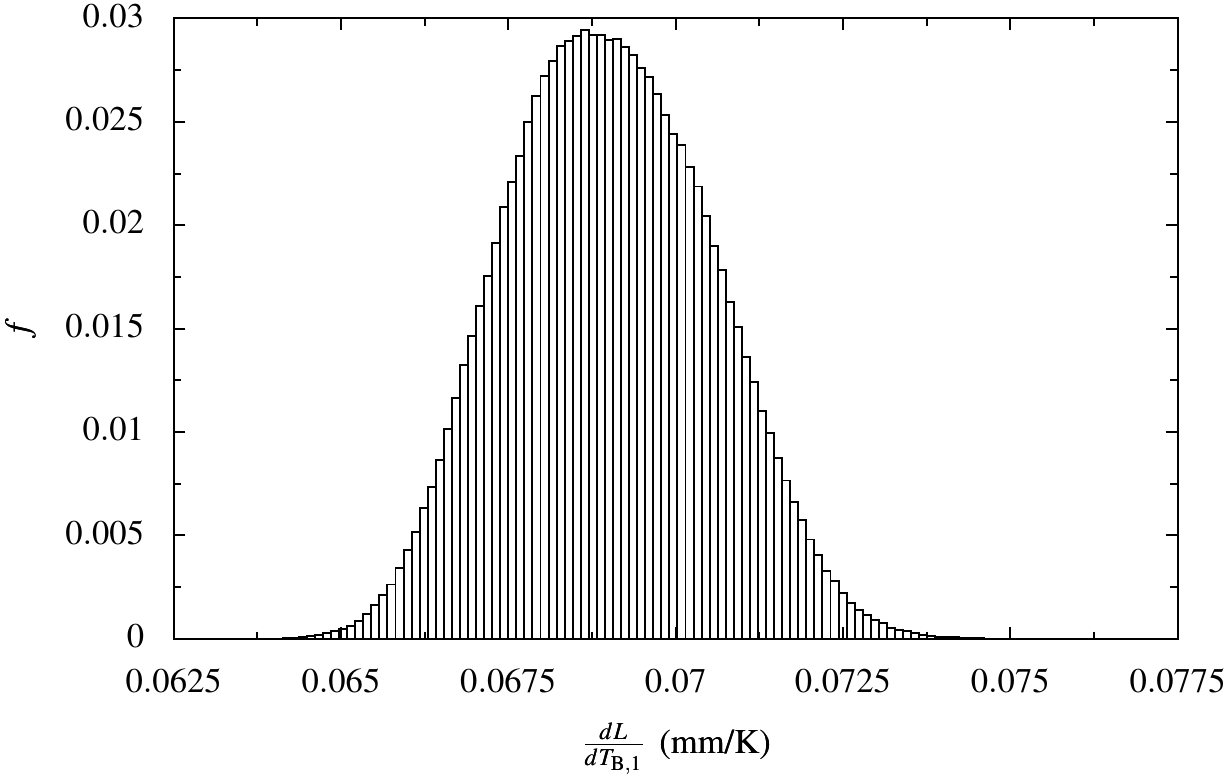}&
    \includegraphics[clip,width=0.45\linewidth]{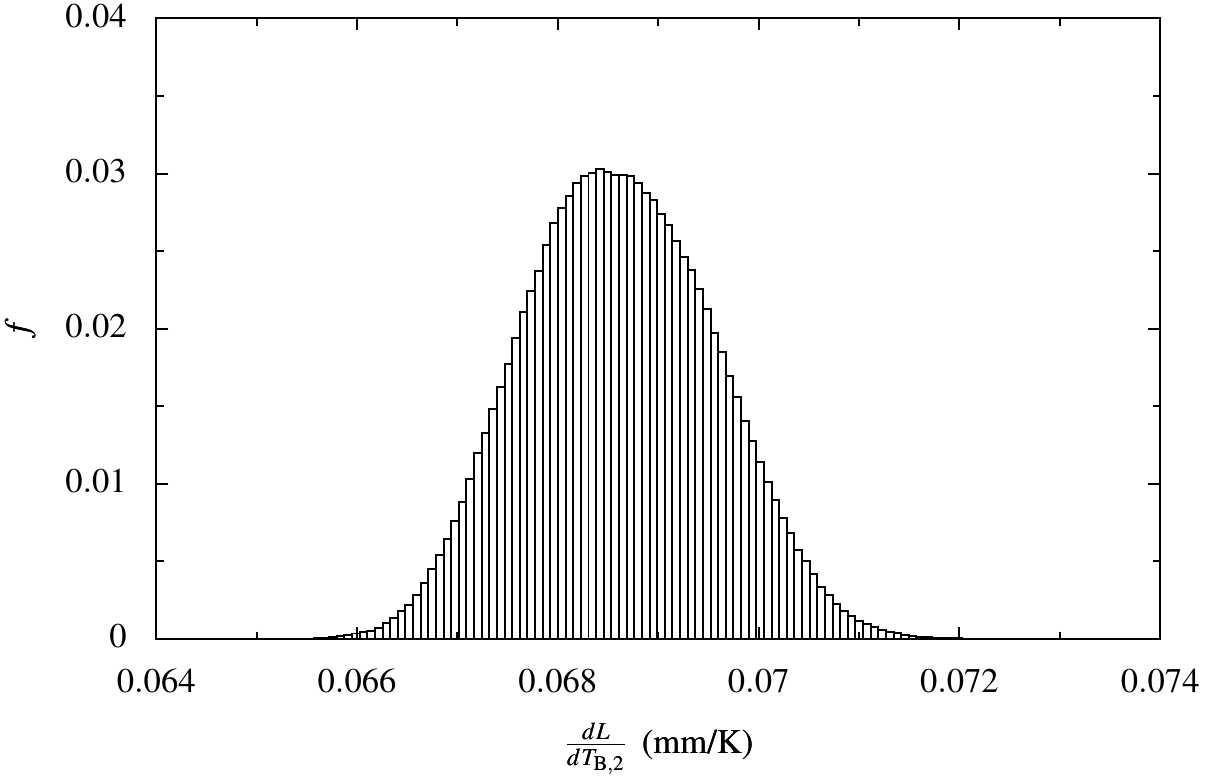}\\
    \includegraphics[clip,width=0.45\linewidth]{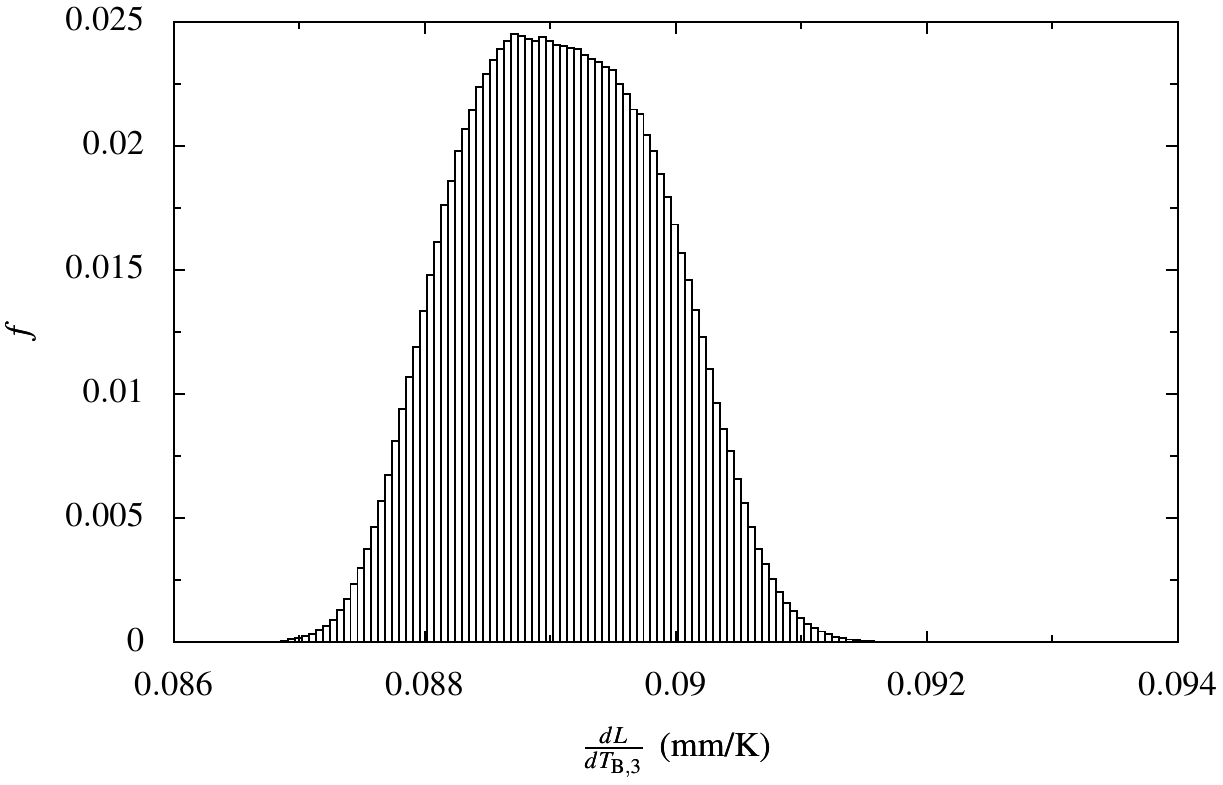}&
    \includegraphics[clip,width=0.45\linewidth]{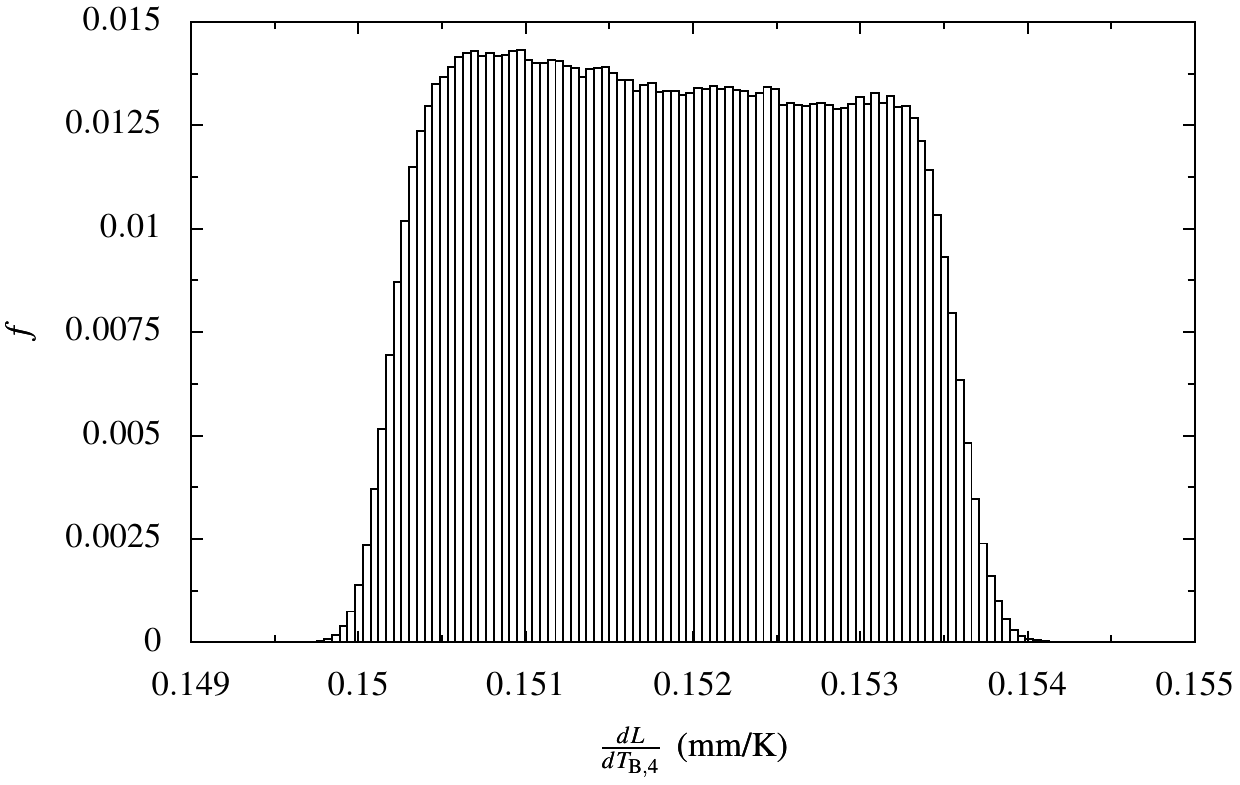}
  \end{tabular}
  \caption{The posterior distribution of ${\rm d}L/{\rm d}
      T_{{\rm B},i}$ corresponding to
    Figure~\ref{fig:abssimpriortight}.}
  \label{fig:fig:abssimpriortight-dldt}
\end{figure*}

The posterior distributions of the ${\rm d}L/{\rm d} T_{{\rm
      B},i}$ coefficients for this posterior distribution are shown
in Figure~\ref{fig:fig:abssimpriortight-dldt}. The qualitatively
different inference of the temperature has an important impact on the
quality of the inference of these coefficients too, as can be seen
from the approximately Gaussian shape of coefficients for channels 1
to 3 and by their significantly tighter ranges. For example. the
distribution of coefficient for channel 2 for the inference with prior
2, (Figure~\ref{fig:abssimprior-dtdl}) is essentially flat with a
range of about 0.008\,mm/K. With a tight prior on the pressure
however, the distribution of this coefficient is close to Gaussian
with a full-width-half-maximum of 0.003\,mm/K, clearly substantially
better.

\subsection{Discussion}

One of the main aims of this section was to illustrate the outputs of
an analysis based on the methods described in
Section~\ref{sec:method}, i.e., viewing the problem of estimating the
phase correction coefficients as Bayesian inference problem. 

The main outcome of such an analysis are the posterior distributions
for the phase correction coefficients such as presented in
Figures~\ref{fig:abssimprior-dtdl} and
\ref{fig:fig:abssimpriortight-dldt}. The posterior distributions allow
us both to pick a specific coefficient to use for each channel of the
radiometer \emph{and} give us confidence intervals for the accuracy of
those coefficients. Obtaining such confidence intervals is important
since the combination of dynamic scheduling, wide range of ALMA
configurations and a large number of projects will mean that each
science observation is likely to be in conditions which are just
marginal for its requirements. Therefore, if phase correction doesn't
work as well as expected, it is likely to seriously impact the science
aims.

We should note however that the confidence intervals calculated using
the model we have specified and therefore do not properly capture the
probability that simply our model is not very good. Within the
Bayesian framework this can be done through the evidence, or $p(D)$,
value and we intend to implement this functionality in the near
future.

Beside the distributions of the phase correction coefficients, the
outcome of the analysis is also the full joint distribution of the
model parameters, which in this case are the water vapour column, and
its pressure and temperature. It is the availability of this full
posterior distribution that makes reliable estimates of the phase
coefficients possible. This becomes particularly important when more
parameters are added to the problem, as will no doubt be necessary in
our case. With more parameters it becomes less and less feasible to
pick a single representative point in the parameter space to calculate
the phase correction coefficients at and making use of the full
distribution becomes increasingly important.

The model we have been using in this paper is fairly simplistic in
that it assumes that the \emph{only\/} observables we have are the
four absolute sky brightness temperatures observed by the
radiometers. What we find is that if we have no further constraints at
all, we can estimate the water vapour column to an accuracy of about
5\%. We can not however constrain the temperature of the water vapour
at all because it is almost exactly degenerate with the pressure and
therefore we can not make any estimate of the phase correction
coefficients (since they depend on the temperature, see
Equation~\ref{eq:nondisp-simple}).

This shortcoming can be improved on by adding even a fairly loose
constraint on the pressure and temperature of the water, such as can
be derived from historical distributions of water vapour in the
atmosphere and its temperature. Such loose constraints should be able
to provide estimates of the phase correction coefficients in the 10\%
range. Even tighter \emph{a-priori} constraints can provide further
improvements as shown in Figures~\ref{fig:abssimpriortight}
and~\ref{fig:fig:abssimpriortight-dldt}.

With the model as simple as the one we have presented here and no data
from the site of ALMA itself it would be somewhat premature to discuss
in detail already the improvements in accuracy particular constraints
can make. We however note that we \emph{will\/} have a substantial
amount of ancillary information such as:
\begin{itemize}
  \item Ground level air temperature, pressure and relative humidity
    and a number of points at the site
  \item Inference of the vertical temperature profile of the
    atmosphere at the centre of the array from a commercial O$_2$ line
    sounder
  \item Library of vertical profiles of atmospheric parameters from
    past radiosonde launches
  \item Meso-scale meteorological forecasts 
  \item Inference of atmospheric parameters from specialised telescope
    observations such as sky-dips and crossed beam observations
\end{itemize}
Inferences from these measurements can be used as \emph{a-priori}
probabilities on model parameters, or, indeed, in some cases, as
further observables which are analysed simultaneously with the sky
brightness measurement.

It should be noted however that further and better prior information
on the temperature and pressure parameters will leave two other
sources of uncertainty: errors due to the inaccuracies in the model
that we use and limitations in estimating the water vapour column
which arises due to calibration accuracy of the WVRs. The best way of
tackling these uncertainties may turn out to be to use the
\emph{observed} correlation between changes in the excess path to the
telescopes ($\delta L$) and the fluctuations of the temperature
observed by the radiometers ($\delta T_{{\rm B},i}$) as an additional
observable that can constrain the models. We will discuss this
approach in the next memo in this series.

Finally, we now consider ways in which the model presented above may
be improved. Firstly, the model may fairly easily be re-parametrised
in terms of the height of the water vapour layer, the temperature
lapse rate of the atmosphere and the ground level pressures and
temperatures instead of the current parametrisation in terms of
pressure and temperature of the layer itself. This would allow us to
easily take into account the measured ground-level temperature and
pressure and the further information that we will have on the
temperature lapse rate (through historical radiosonde measurements and
oxygen line profiling) and water vapour height (through historical
radiosonde data and specialised telescope scans). The second
improvement is to consider the effect of a thick layer of water
vapour, perhaps with an exponential fall off in water vapour content
as a function of height.

Assessing the benefits of such improvements to the model will require
computation of the evidence value and, ideally, test data from the
ALMA site.

\section{Application to test data collected at SMA}
\label{sec:smadataapp}

In this section we analyse an observation taken with the two prototype
ALMA water vapour radiometers \citep{ALMAHills352} at the
Submillimetre Array (SMA) on Mauna Kea. 

The observation was made on 17 February 2006 and consisted of an
approximately one hour long track of a bright quasar. The
interferometric visibility between the two SMA antennas with the
water-vapour radiometers was recorded together with the sky
brightnesses seen by the radiometers.  This observation was part of
the testing campaign of the prototype radiometers at the SMA which
will be described in detail in a separate paper.

The effective LO frequency of the observation was 235\,GHz and both
the upper and the lower sideband were recorded. In the present
analysis we use only the upper side band with the on-sky frequency of
240\,GHz.  The water-vapour radiometers were read out with an
integration time of 1 second while the interferometer was read out at
2.5\,seconds. In order to match the two data sets, the radiometer data
ware re-sampled to 2.5\,second time resolution, and a small adjustment
to the time-stamps was made to correct for a known timing drift
problem.  The radiometer data were recorded in already calibrated
format and required no further processing. The interferometric data
were converted to a text format by M.~Reid of the SMA. Subsequently
the variation in the observed visibility was transformed to the path
fluctuation between the two antennas.

\begin{figure*}
  \begin{tabular}{cc}
    \includegraphics[clip,width=0.45\linewidth]{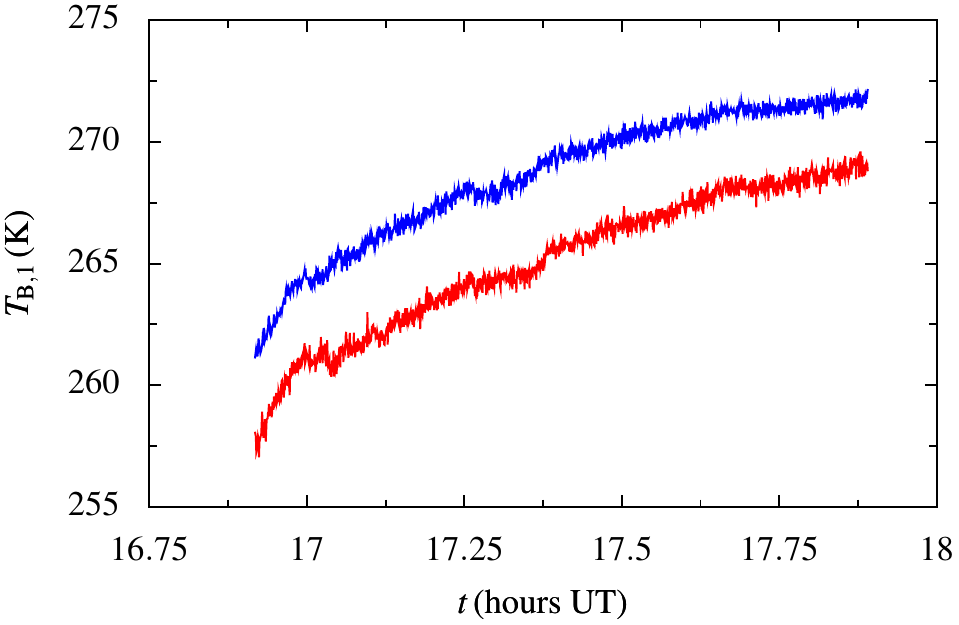}&
    \includegraphics[clip,width=0.45\linewidth]{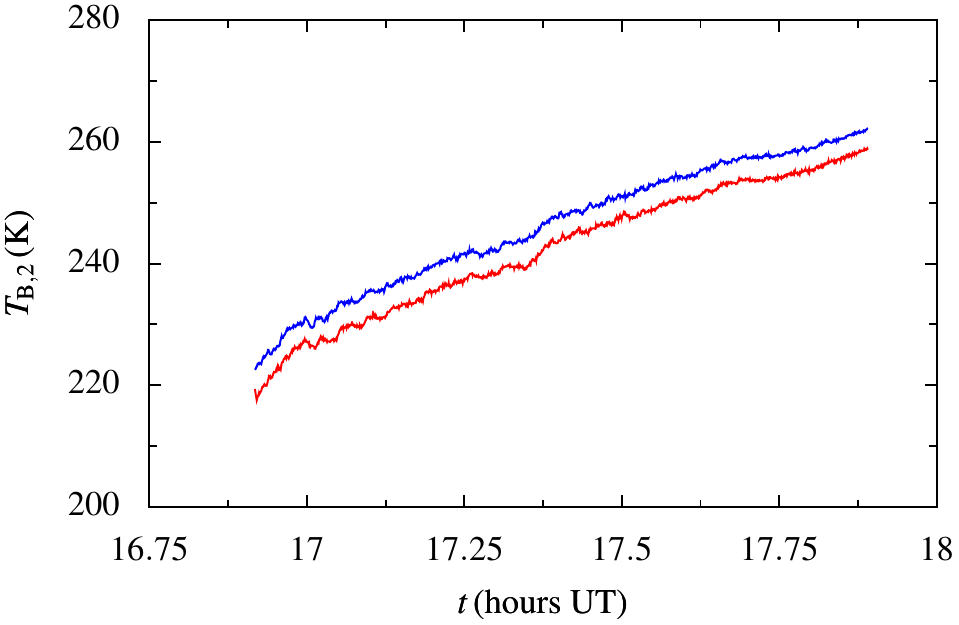}\\
    \includegraphics[clip,width=0.45\linewidth]{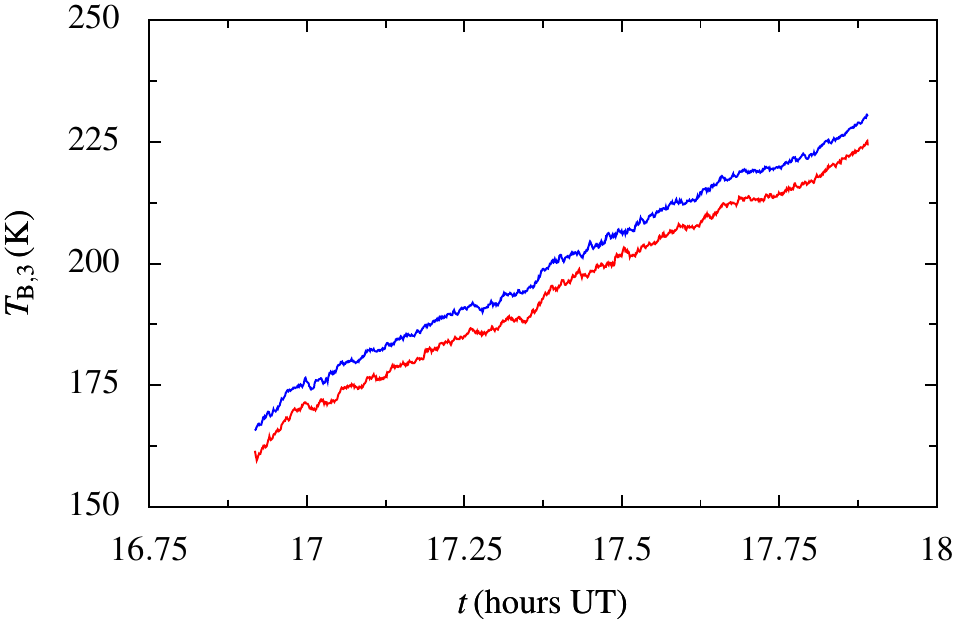}&
    \includegraphics[clip,width=0.45\linewidth]{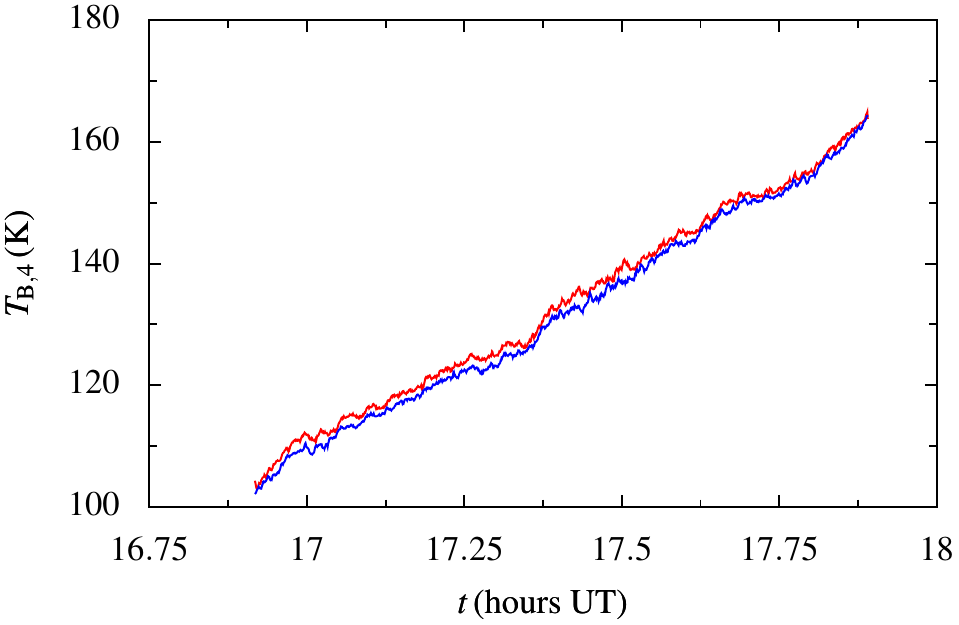}
  \end{tabular}
  \caption{The sky brightness temperatures observed by the two
    prototype WVRs while at tests at the SMA, which was tracking a
    quasar at the time. Blue is one of the radiometers and red is the
    other.  The four panels represent the four radiometer channels.}
  \label{fig:radiodata}
\end{figure*}

The total sky brightness temperatures observed by four channels of two
radiometers during this observation are shown in
Figure~\ref{fig:radiodata}. It can be seen that observed brightness is
increasing during the observation which is a consequence of the
decreasing elevation of the source and therefore increasing airmass as
the observation progressed. It can also be seen that the innermost two
channels (channels 1 and 2, top row of Figure~\ref{fig:radiodata})
were almost saturated, indicating significant water vapour along the
line of sight. The overall levels of the blue and red curves in this
plot allow us to make inference about the atmospheric conditions at
the time of the observation while the relative fluctuations between
the two curves, once multiplied by the phase correction coefficients,
should correlate closely with the path fluctuation measured by the
interferometric observation.

\begin{figure}
  \includegraphics[clip,width=\columnwidth]{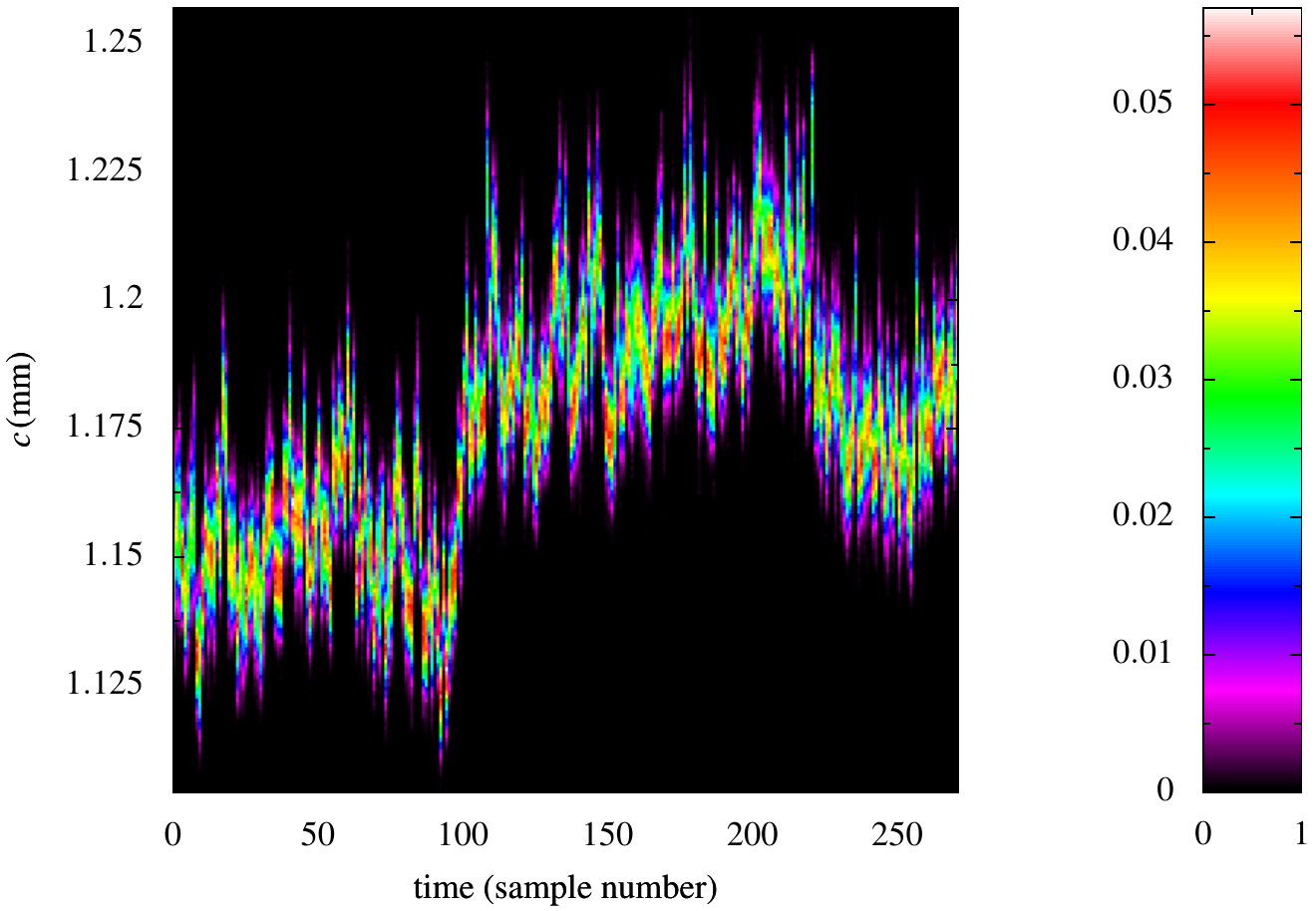}  

  \caption{Retrieval of the \emph{zenith\/} water vapour column as a
    function of time from the data recorded by one of the radiometers
    during the February 17th observation. The priors used in the
    retrieval are the priors from row 2 of Table~\ref{tab:priors}. The
    retrieval was calculated once every 25\,seconds for the hour-long
    observation.}
  \label{fig:retseq}
\end{figure}

A retrieval of atmospheric parameters can be made from every one
second integration of each of the two radiometers without significant
loss in accuracy since, as mentioned before, the thermal noise in one
second is much smaller than the expected calibration error of the
radiometers. In practice however we expect retrievals will be made
rather less often since it is likely that variation in atmospheric
conditions will be fairly slow at the ALMA site. In this study we will
present retrievals at three different time-resolutions:
\begin{enumerate}
  \item Calculation of the marginalised zenith water vapour at 300
    time points during the observation
  \item A detailed analysis of a single data point in the middle of
    the observation
  \item Calculation of the marginalised phase correction coefficients
    at three points of observation 
\end{enumerate}

We first present a sequence of 300 retrievals from sky brightness
temperatures measured at intervals spaced by 25 seconds during the
observation. As with the other retrievals in this section, these were
made with the priors as shown in row 2 of Table \ref{tab:priors}. We
plotted these retrievals in Figure~\ref{fig:retseq} by marginalising
them to obtain the posterior distribution of the \emph{zenith} water
vapour column and plotting these histograms in colour scale as a
function of time.  Therefore in this plot, time runs in the horizontal
direction, the zenith water vapour column parameter along the vertical
and the colour represents the posterior probability.

The first point to note is that because we retrieve for the zenith
water vapour column, we do not expect to see a large increase in the
parameter $c$ as the air-mass increases during the observation. The
overall range of the water vapour measured in the retrievals is about
5\% indicating that at least approximately the plane-parallel
approximation holds and the referencing to zenith column is
a reasonable approach.  

Secondly it can be noted that although fluctuations in the water
vapour column can be seen, adjacent retrievals (which are separated by
25 seconds in time) generally show similar values indicating a high
degree of \emph{stability\/} in the retrieved posterior
distribution. The magnitude of the fluctuations seen in the retrieved
water column is about 50\,\micron\ of water vapour which corresponds
approximately to a 300\,\micron\ of path fluctuation (using the rough
conversion of $1\,\unit{mm}\,\unit{water}\approx
6\,\unit{mm}\,\unit{path}$).  As we will present later, e.g.,
Figure~\ref{fig:smaradiointfdata}, this roughly corresponds to the
fluctuations seen in the path by the interferometric measurements.
Therefore, over the hour of the observation and a large change in
airmass there is no obvious evidence of instability in the water
vapour column retrieval.

\begin{figure*}
  \begin{tabular}{ccc}
    \includegraphics[clip,width=0.33\linewidth]{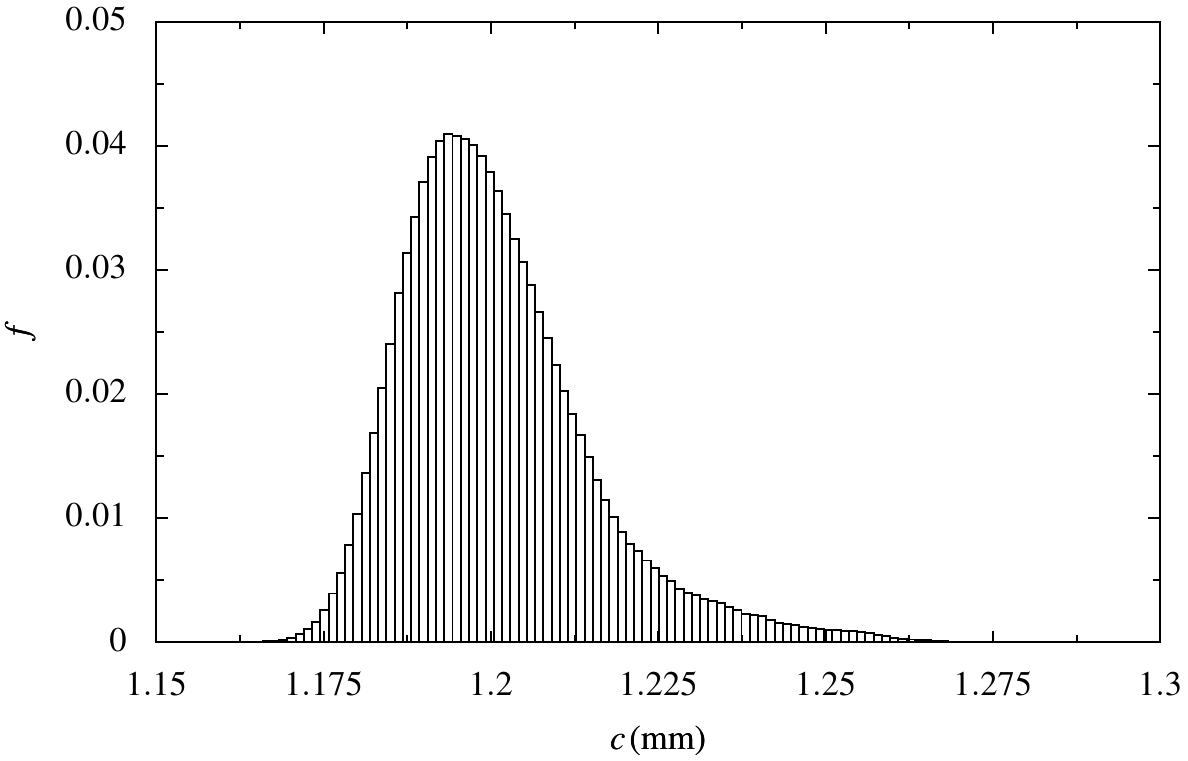}&
    \includegraphics[clip,width=0.33\linewidth]{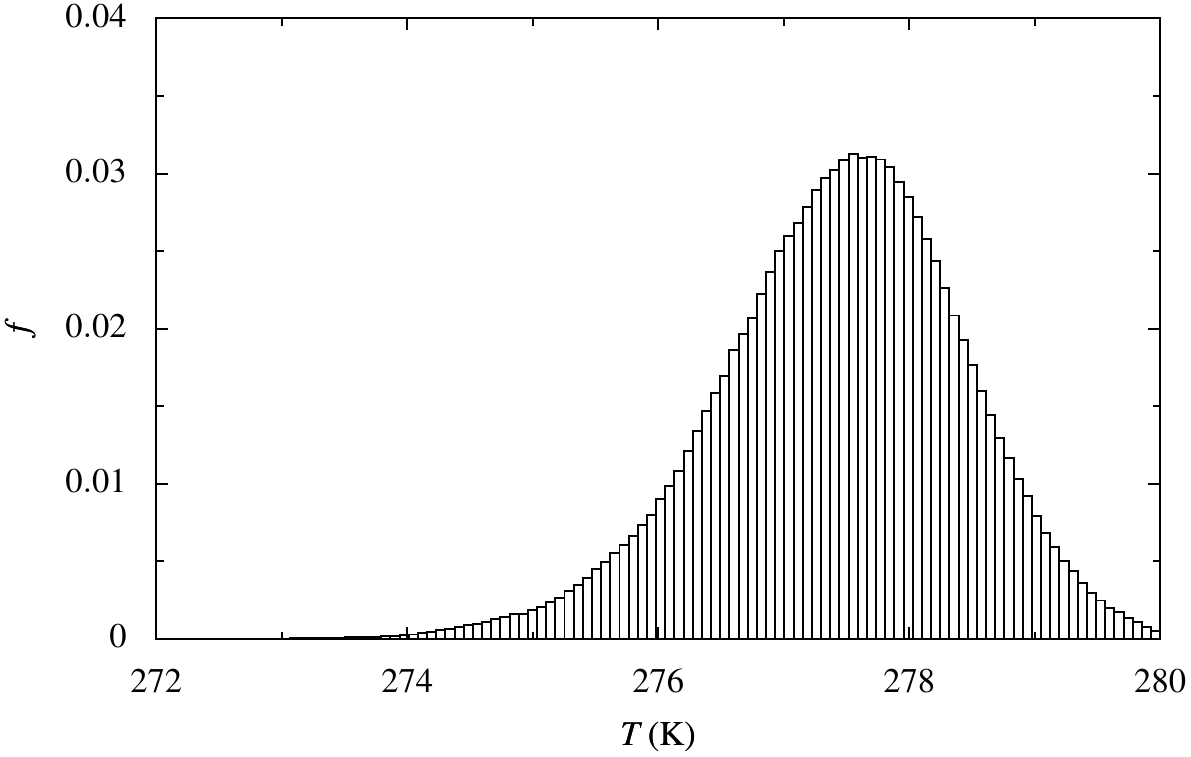}&
    \includegraphics[clip,width=0.33\linewidth]{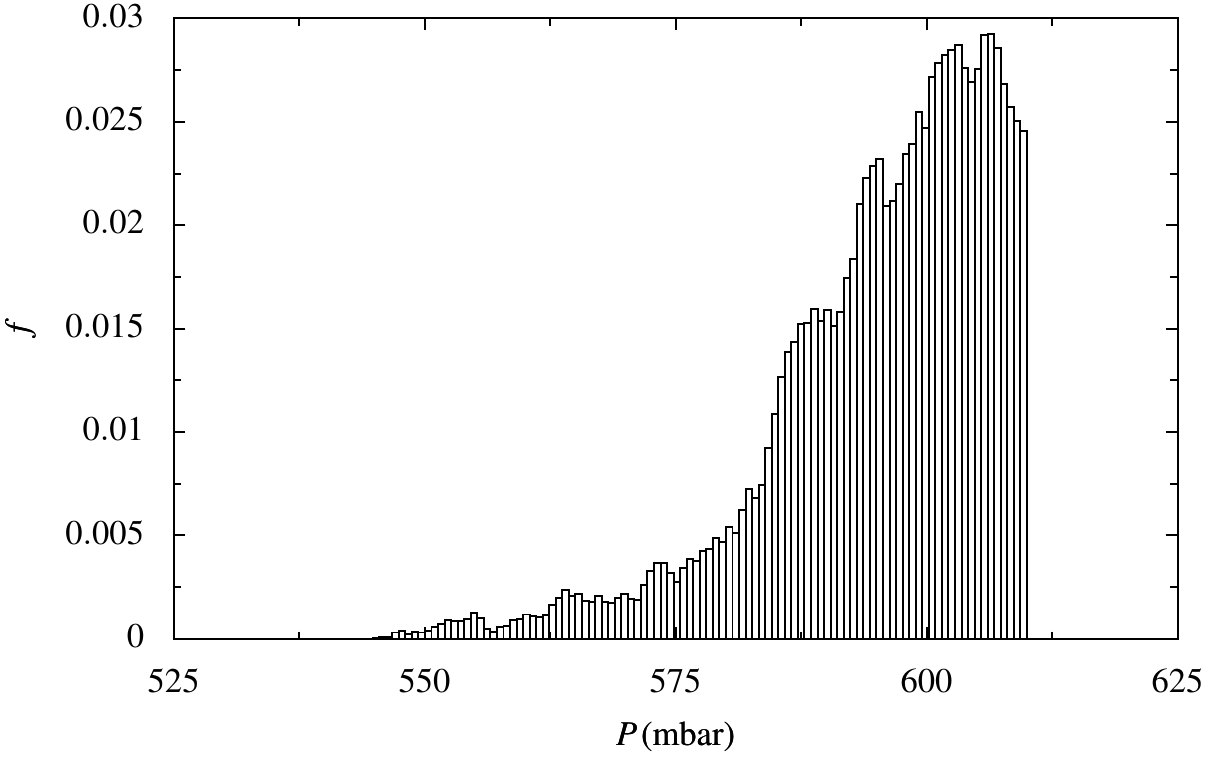}\\
    \includegraphics[clip,width=0.33\linewidth]{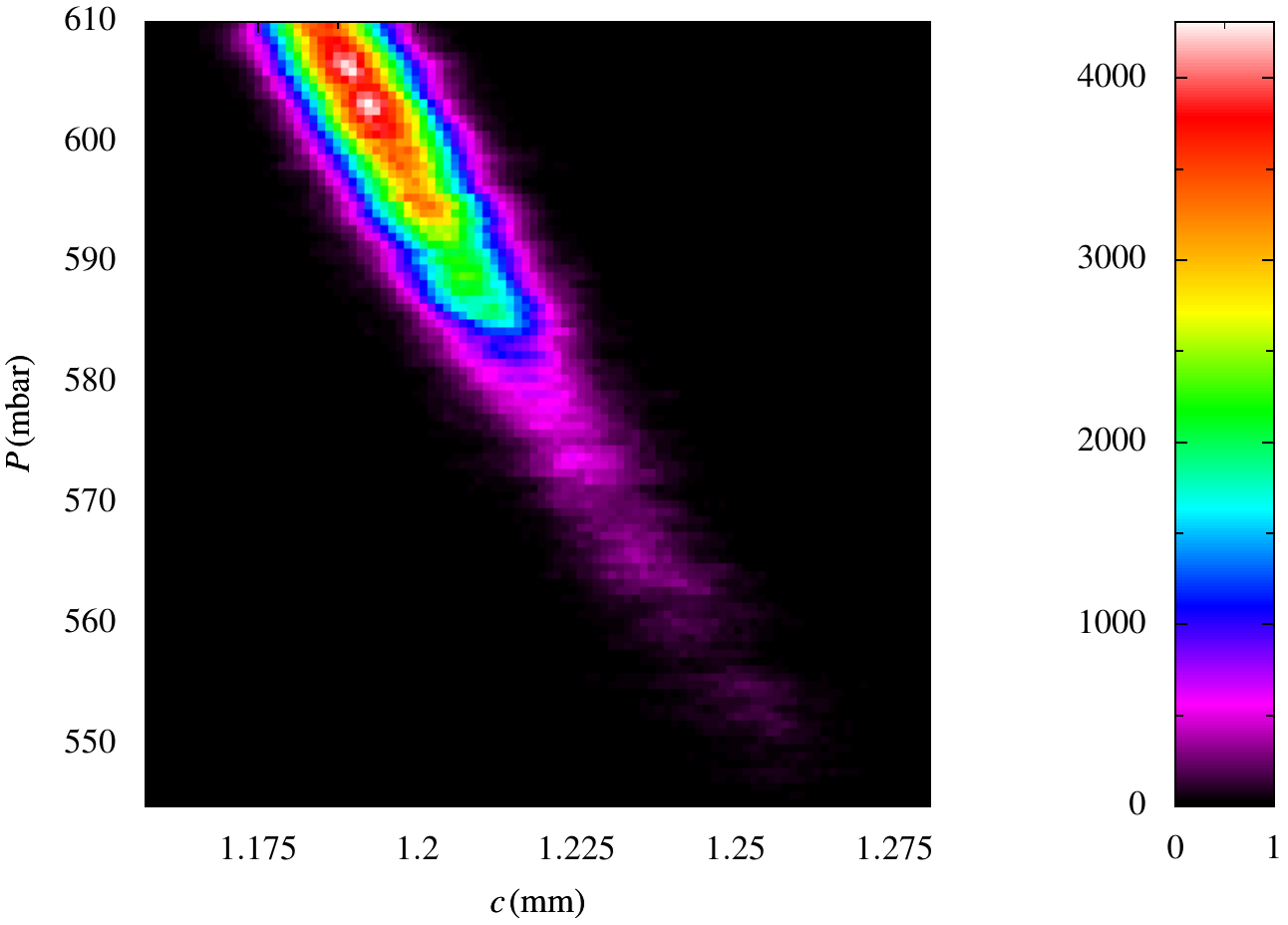}&
    \includegraphics[clip,width=0.33\linewidth]{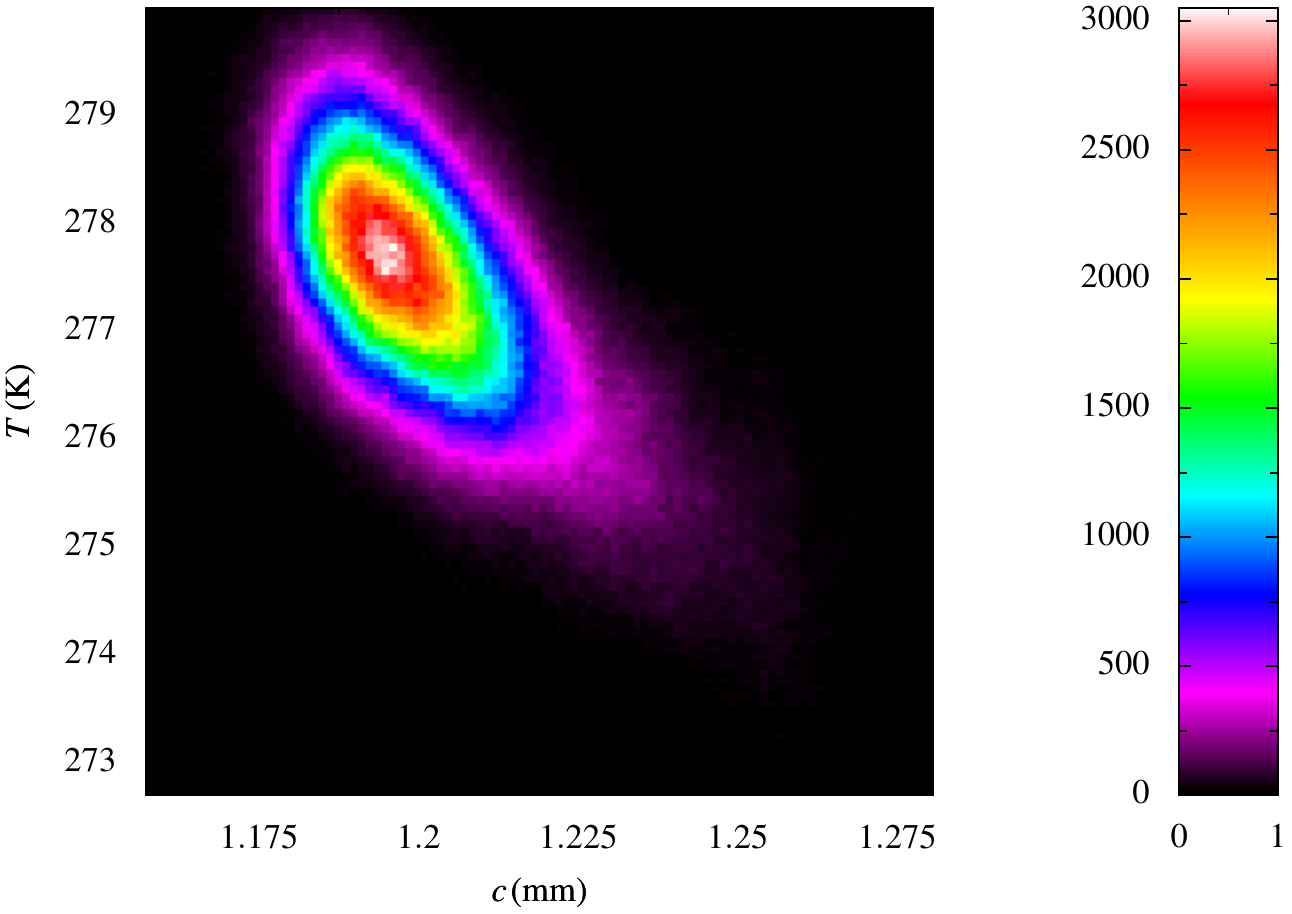}&
    \includegraphics[clip,width=0.33\linewidth]{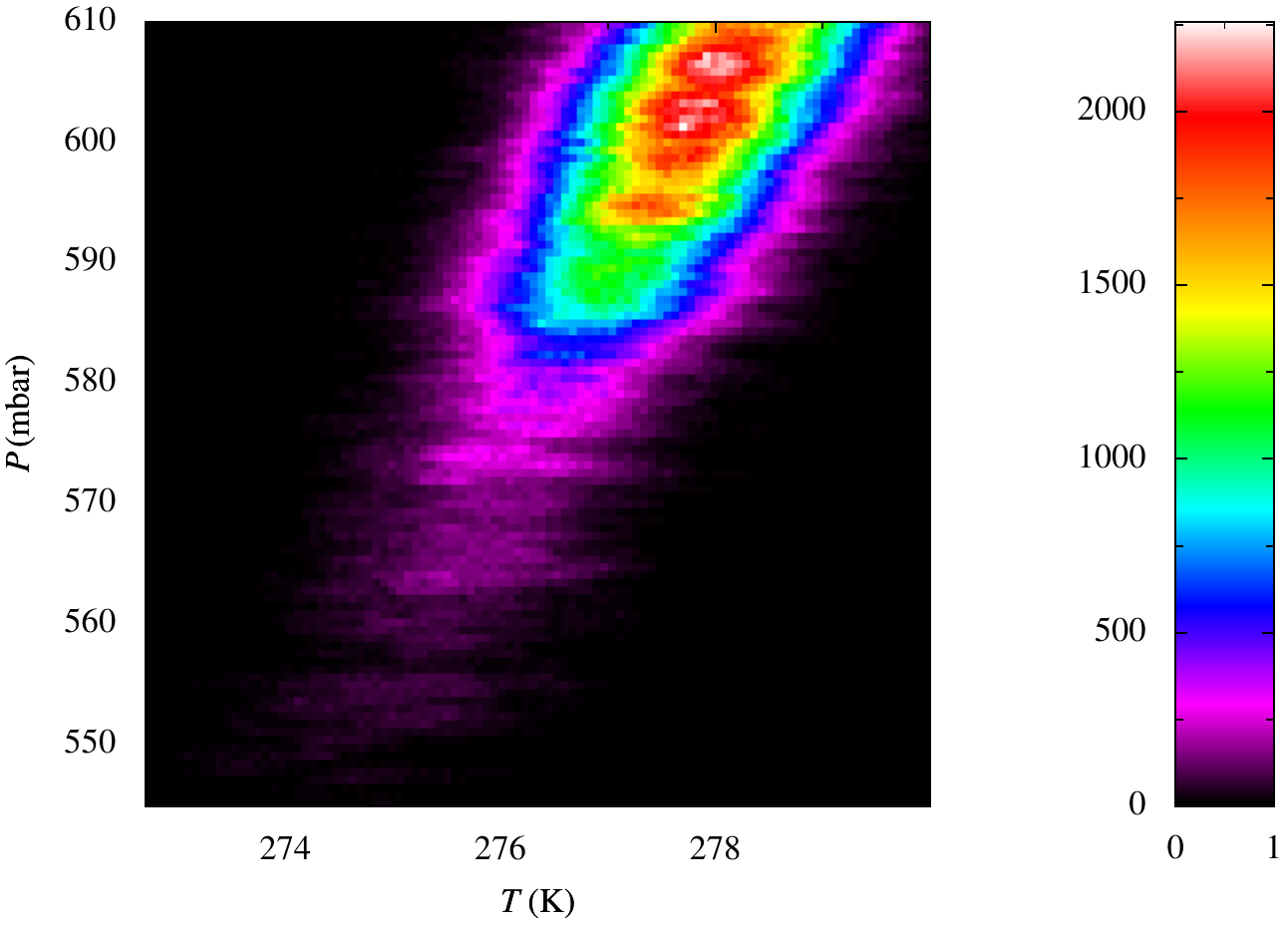}
  \end{tabular}
  \caption{Example posterior distribution of model parameters, from a
    real observation at the SMA at a low elevation (28 degrees) and
    with priors. Note that $c$ represents the zenith water vapour
    column.}
  \label{fig:absfeb17mid}
\end{figure*}

A more detailed analysis of the retrieval from one set of sky
brightness measurements in the middle of the observation is presented
in Figure~\ref{fig:absfeb17mid} in the same format as the previous
plots of the posterior distributions of model parameters. As the line
of sight water vapour during this observations is significantly higher
than the simulations shown in Section~\ref{sec:results}, some
qualitatively differently different results are seen.

The first noticeable feature is that in this retrieval the temperature
of the water vapour layer is in fact quite well constrained, to a
range of around 4 Kelvin, in contrast the results at 1\,mm water
vapour line of sight column where the posterior distribution filled
the entire prior range of 20\,K. This is a direct consequence of the
almost complete saturation of the innermost channel of the radiometer
which means that it is in effect measuring the temperature of the
water vapour rather than line of sight column.

The saturation of the inner channel, however, now also causes a
degeneracy between the pressure and the water vapour column parameter
leading to a non-Gaussian posterior distribution for $c$. The reason
for this is that at lower pressure, the line opacity becomes highly
peaked, but this can not be detected because of the saturation. The
outcome therefore is that at this point in parameter space the water
vapour column is somewhat less well constrained and the temperature is
better constrained. It should again be noted that the inference errors
we present are derived from the model itself and therefore do not
account for inaccuracies of the model.

\begin{figure*}
  \begin{tabular}{cc}
    \includegraphics[clip,width=0.45\linewidth]{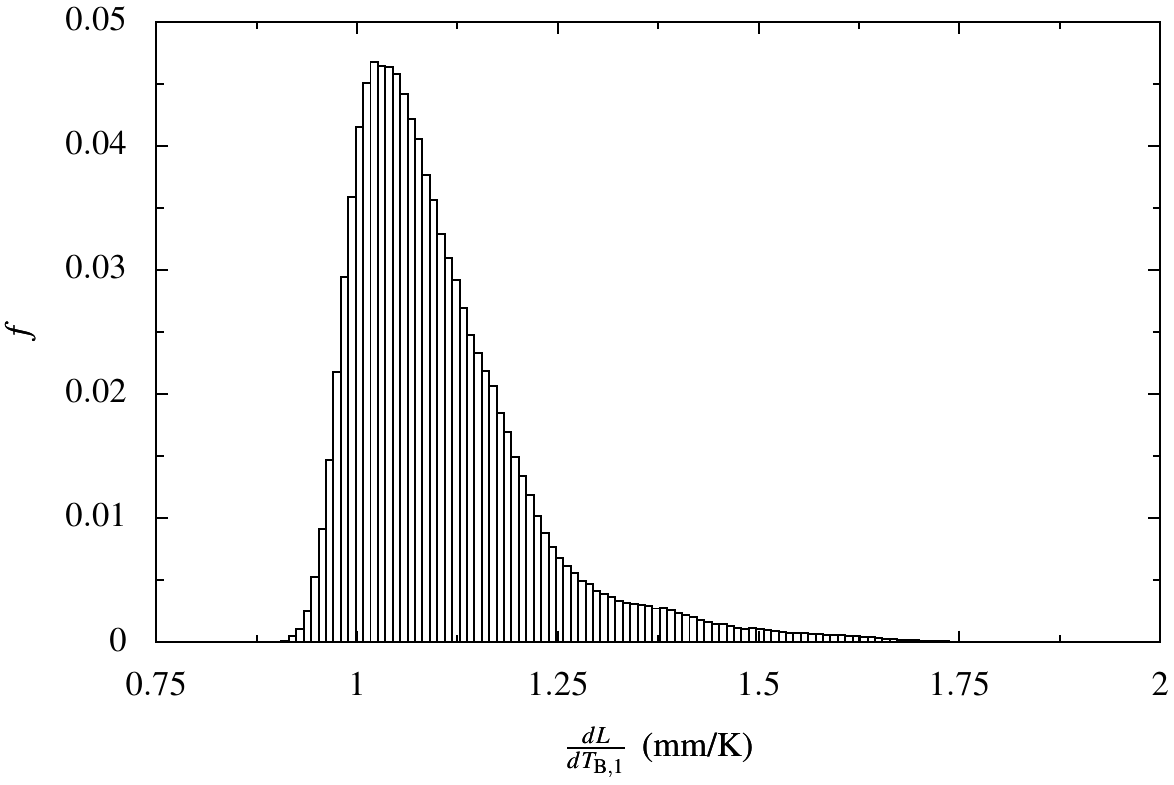}&
    \includegraphics[clip,width=0.45\linewidth]{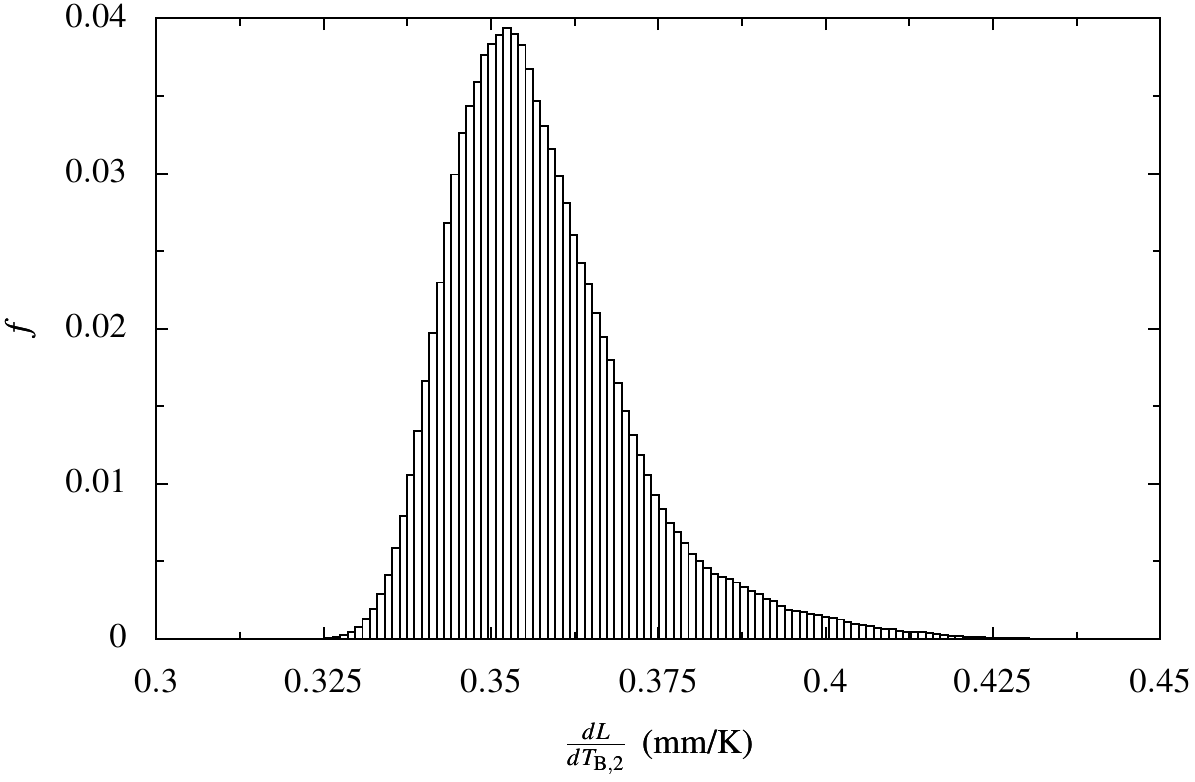}\\
    \includegraphics[clip,width=0.45\linewidth]{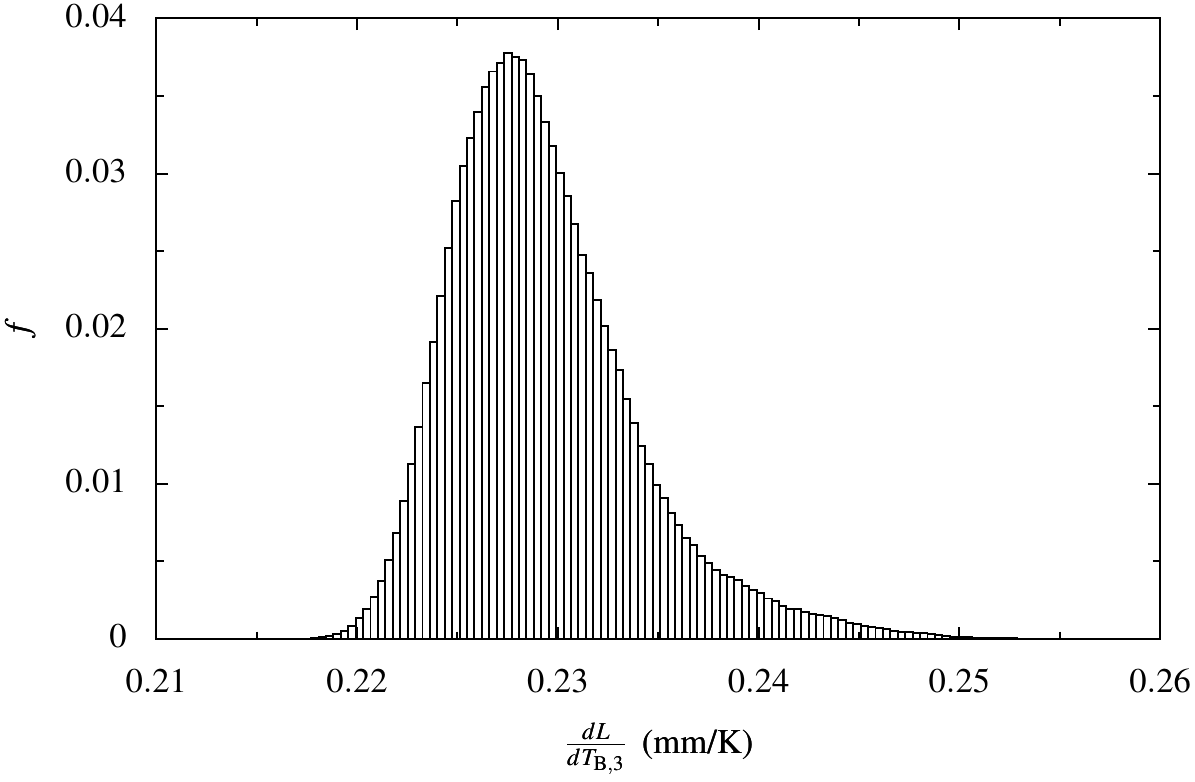}&
    \includegraphics[clip,width=0.45\linewidth]{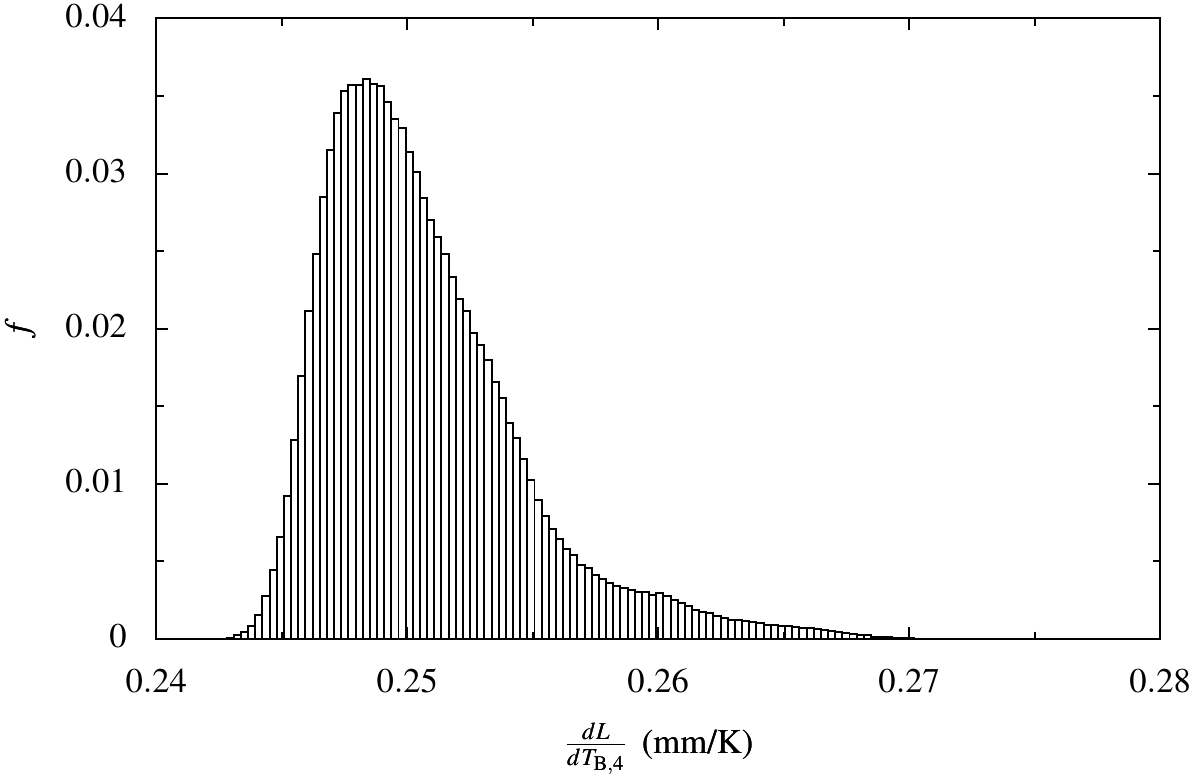}
  \end{tabular}
  \caption{The posterior distribution of ${\rm d}L/{\rm d}
      T_{{\rm B},i}$ corresponding to Figure~\ref{fig:absfeb17mid}.}
  \label{fig:absfeb17mid-dldt}
\end{figure*}

We next consider the posterior distribution of the phase correction
coefficients for this detailed retrieval, as shown in
Figure~\ref{fig:absfeb17mid-dldt}. Because of the high line of sight
water vapour column, it can be seen that the coefficient for channel 1
is very high, i.e., around 1.1 mm/K. This means that thermal noise of
0.1\,K within the radiometer would be sufficient to produce a high
random path fluctuation of 110\,\micron. Furthermore, other effects
which are not modelled are likely to cause large errors in the path
fluctuations calculated from this channel. Therefore, in this case we
do not expect this channel to provide useful data for the phase
correction itself.

It can also be seen in Figure \ref{fig:absfeb17mid-dldt} that although
the posterior distributions of the phase correction coefficients are
reasonable centrally peaked, they do all have a tail to higher
values. This tail is a result of the non-Guassian inference of the
total water vapour column, as seen in the top-left panel of
Figure~\ref{fig:absfeb17mid}. The errors on these coefficients are not
dominated by the error on inferred water vapour temperature because,
as discussed above, the saturated inner channel in allows for this
error to be minimised in this case. Finally it should be noted that
the confidence interval of the inferred phase correction coefficients
is about 10 to 15\% in this case. 

\begin{figure*}
\begin{tabular}{cccc}
  Channel \#1 & Channel \#2 & Channel \#3 & Channel \#4\\
  \includegraphics[clip,width=0.25\linewidth]{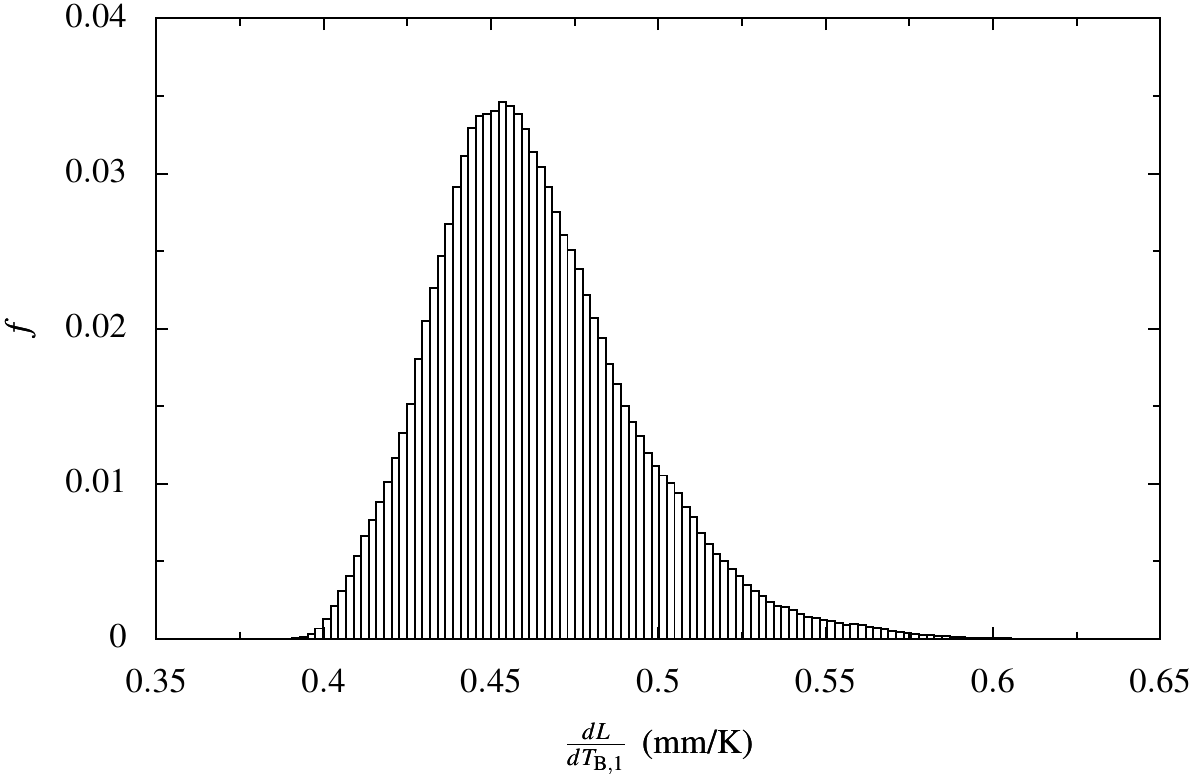}&
  \includegraphics[clip,width=0.25\linewidth]{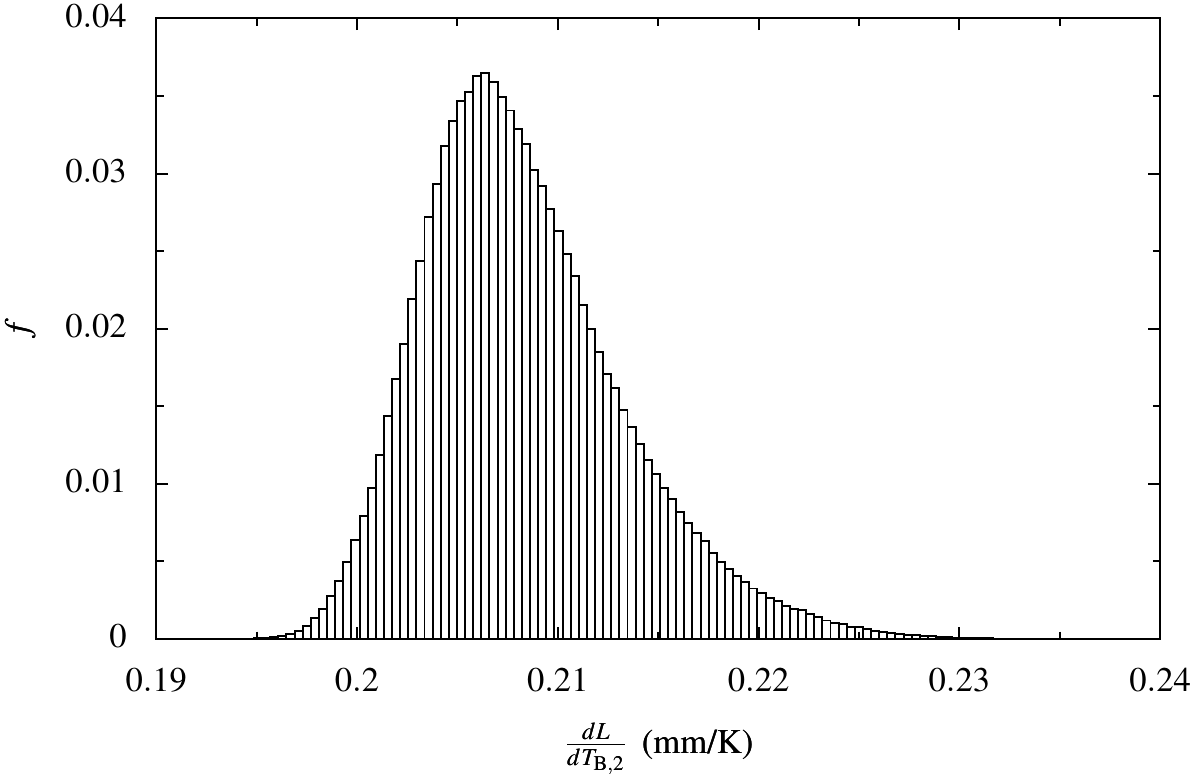}&
  \includegraphics[clip,width=0.25\linewidth]{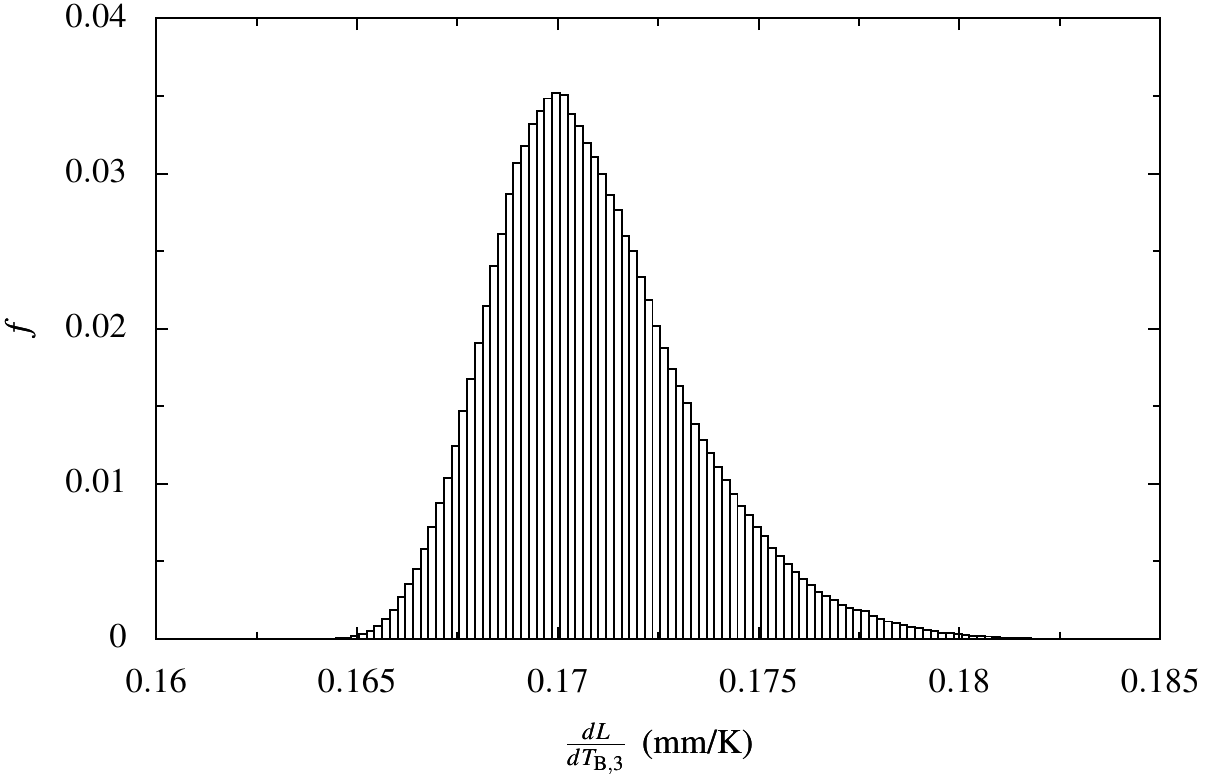}&
  \includegraphics[clip,width=0.25\linewidth]{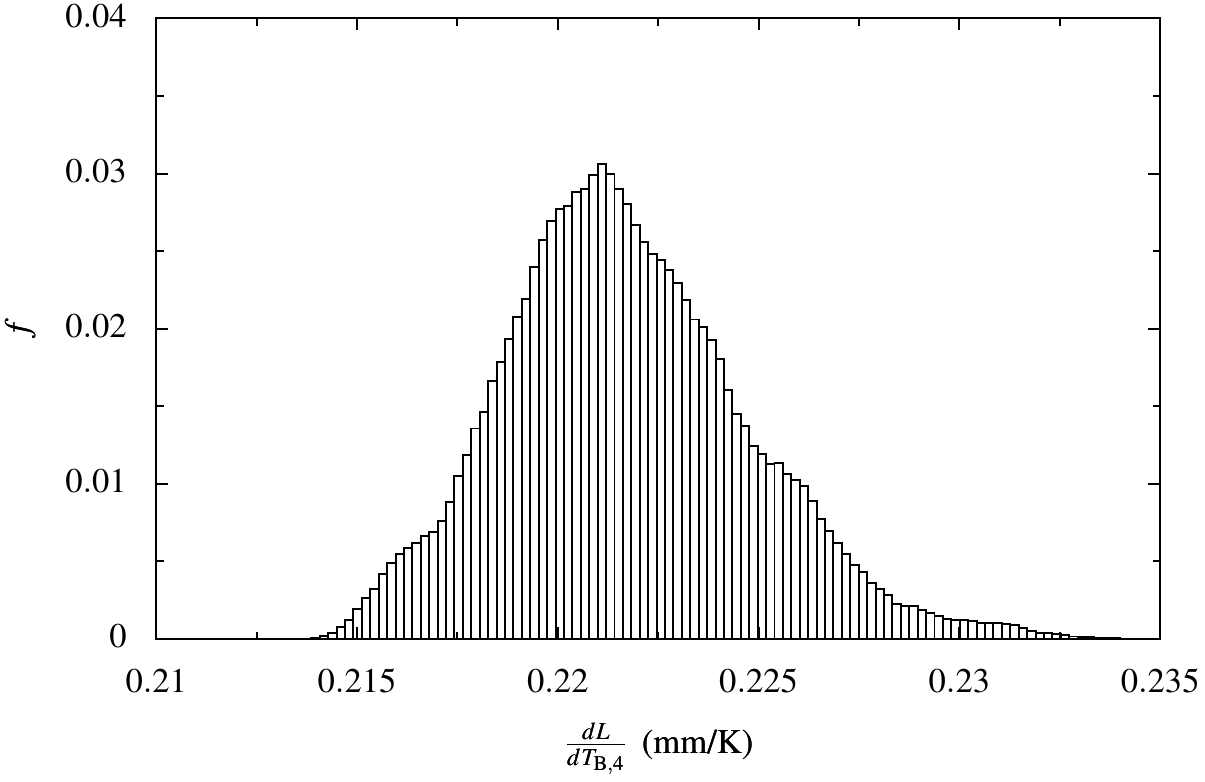}\\
  \includegraphics[clip,width=0.25\linewidth]{figs/feb17/f17midhist-dLdT1}&
  \includegraphics[clip,width=0.25\linewidth]{figs/feb17/f17midhist-dLdT2}&
  \includegraphics[clip,width=0.25\linewidth]{figs/feb17/f17midhist-dLdT3}&
  \includegraphics[clip,width=0.25\linewidth]{figs/feb17/f17midhist-dLdT4}\\
  \includegraphics[clip,width=0.25\linewidth]{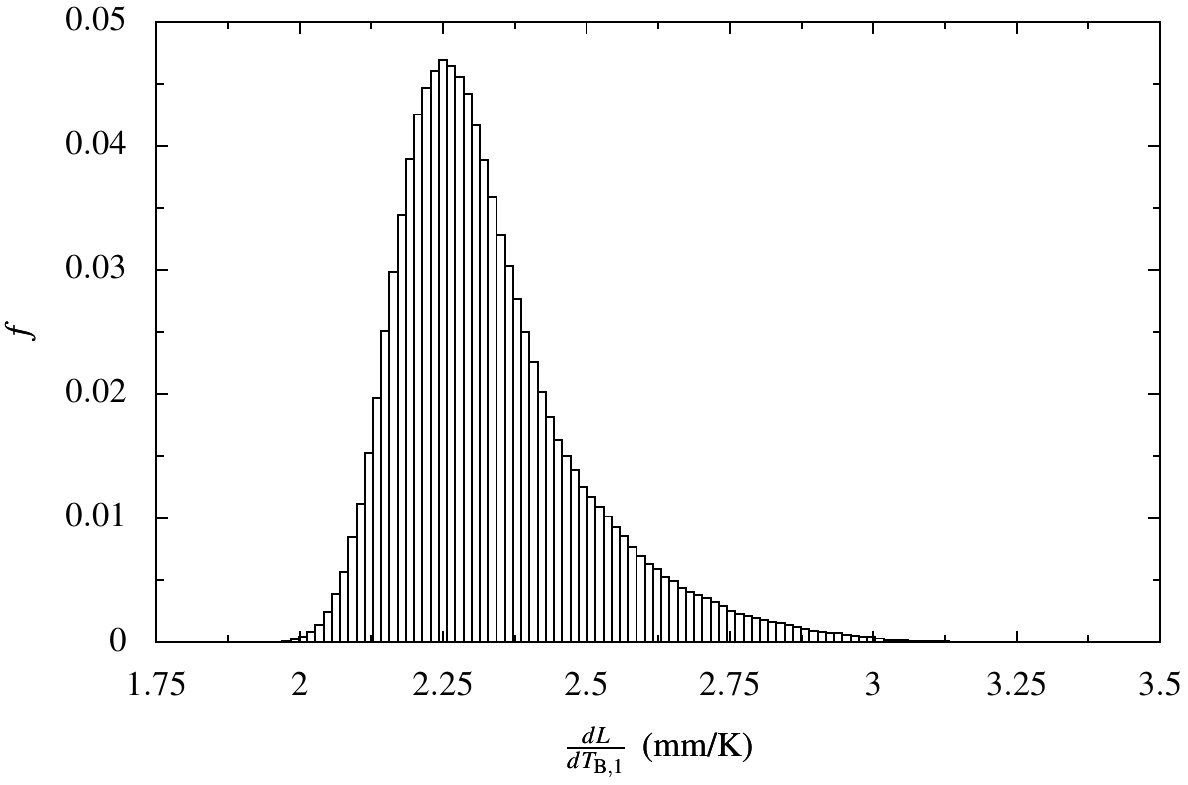}&
  \includegraphics[clip,width=0.25\linewidth]{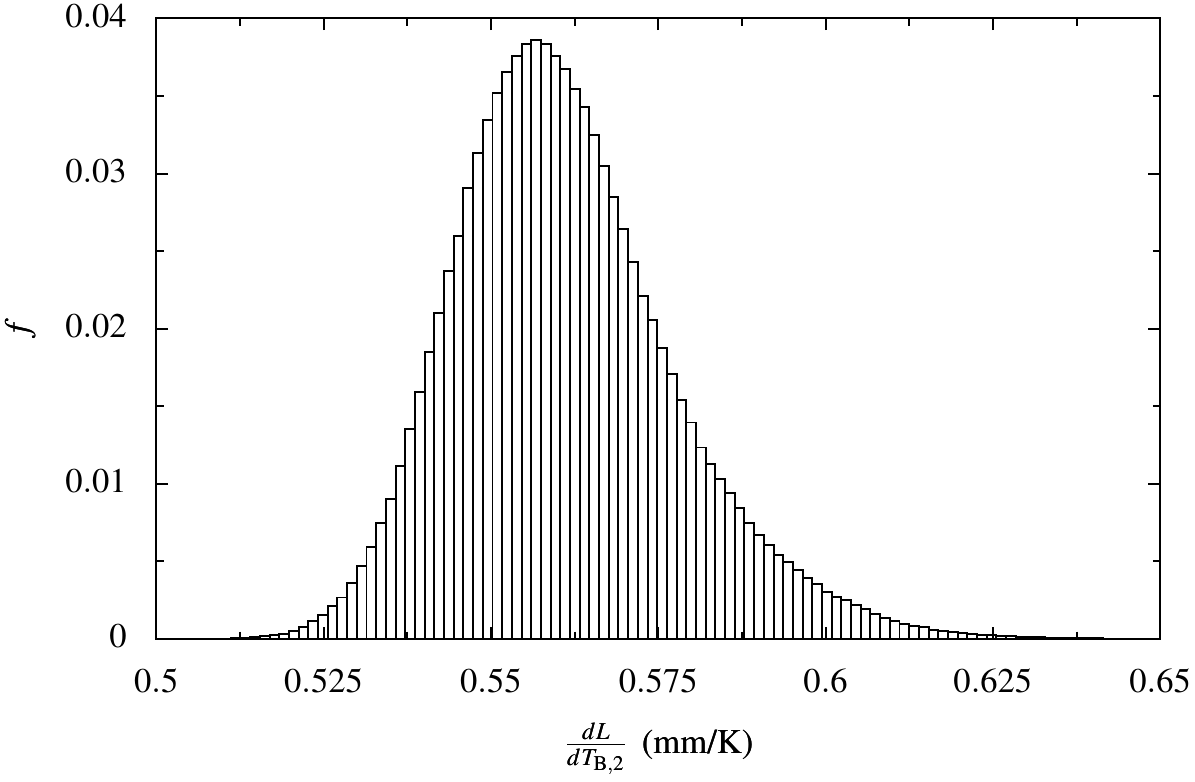}&
  \includegraphics[clip,width=0.25\linewidth]{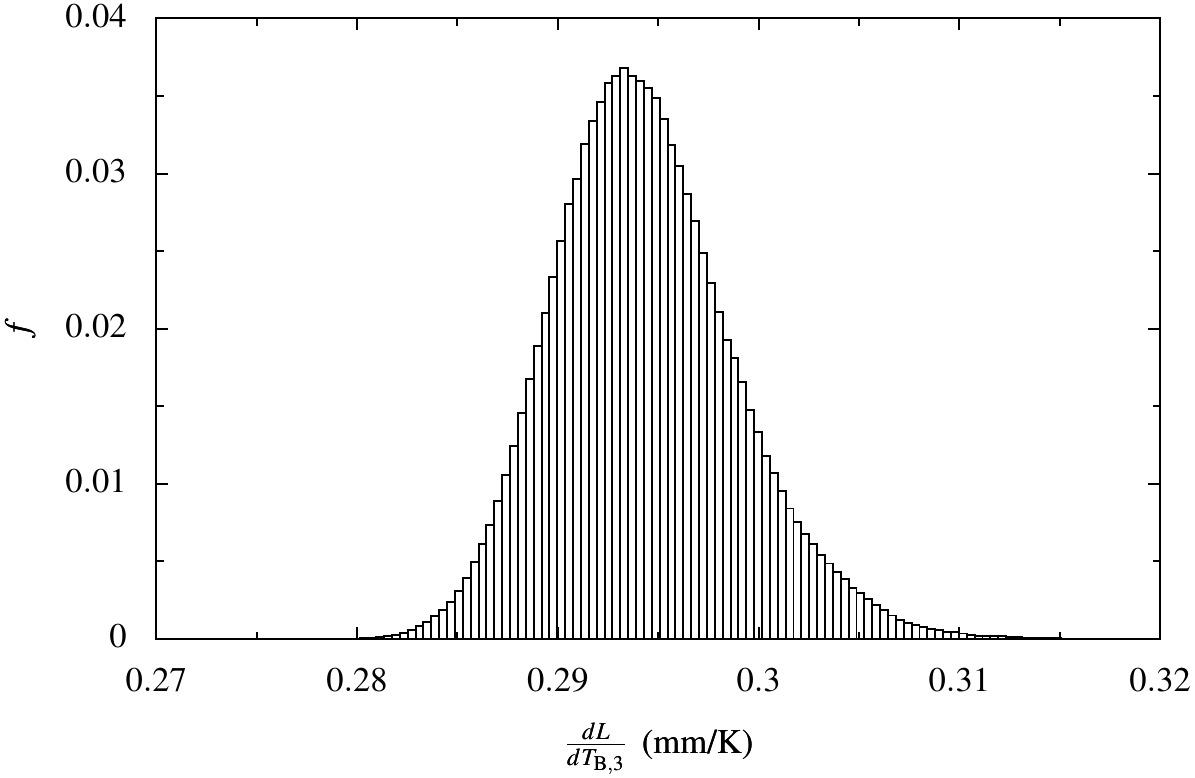}&
  \includegraphics[clip,width=0.25\linewidth]{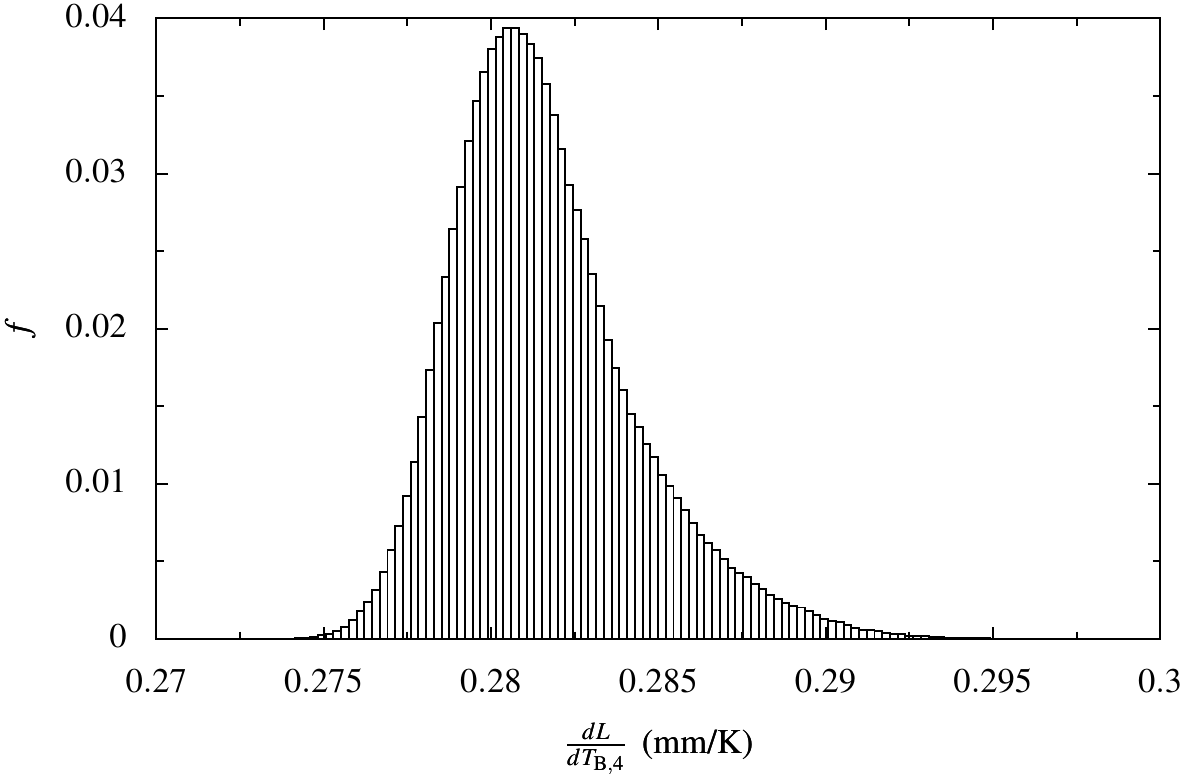}\\
\end{tabular}
\caption{Change in the posterior distribution of the four
  phase-correction coefficients during the one-hour long scan tracking
  a quasar. Top row is beg ginning of the scan (UT 17.1), middle row
  is the middle of the scan (UT 17.5) and bottom row is the end of the
  scan (UT 17.8).}
\label{fig:dldt-threetimes}
\end{figure*}

The final part set of inferences we made was to repeat the prediction
of the phase correction coefficients at the beginning and at the end
observation so that we can examine how they change during the course
of the observation. The results of this analysis are shown in
Figure~\ref{fig:dldt-threetimes}, with each retrieval on a separate
row. 

It can be seen in that the values of all of the coefficients increase
as the observation progresses. This is of course due to the increase
in airmass. The greatest increase is seen in channel 1, which starts
at a median value of about 0.45 and finishes at a median value of
about 2.25.

\subsection{Phase correction}

We next examine how well we would have been able to do phase
correction at the SMA during this observations if we had used the
phase correction coefficients retrievals as shown previously.

\begin{figure*}
  \begin{tabular}{cc}
    \subfloat[Channel 1 (innermost channel)]
    {\includegraphics[clip,width=0.45\linewidth]{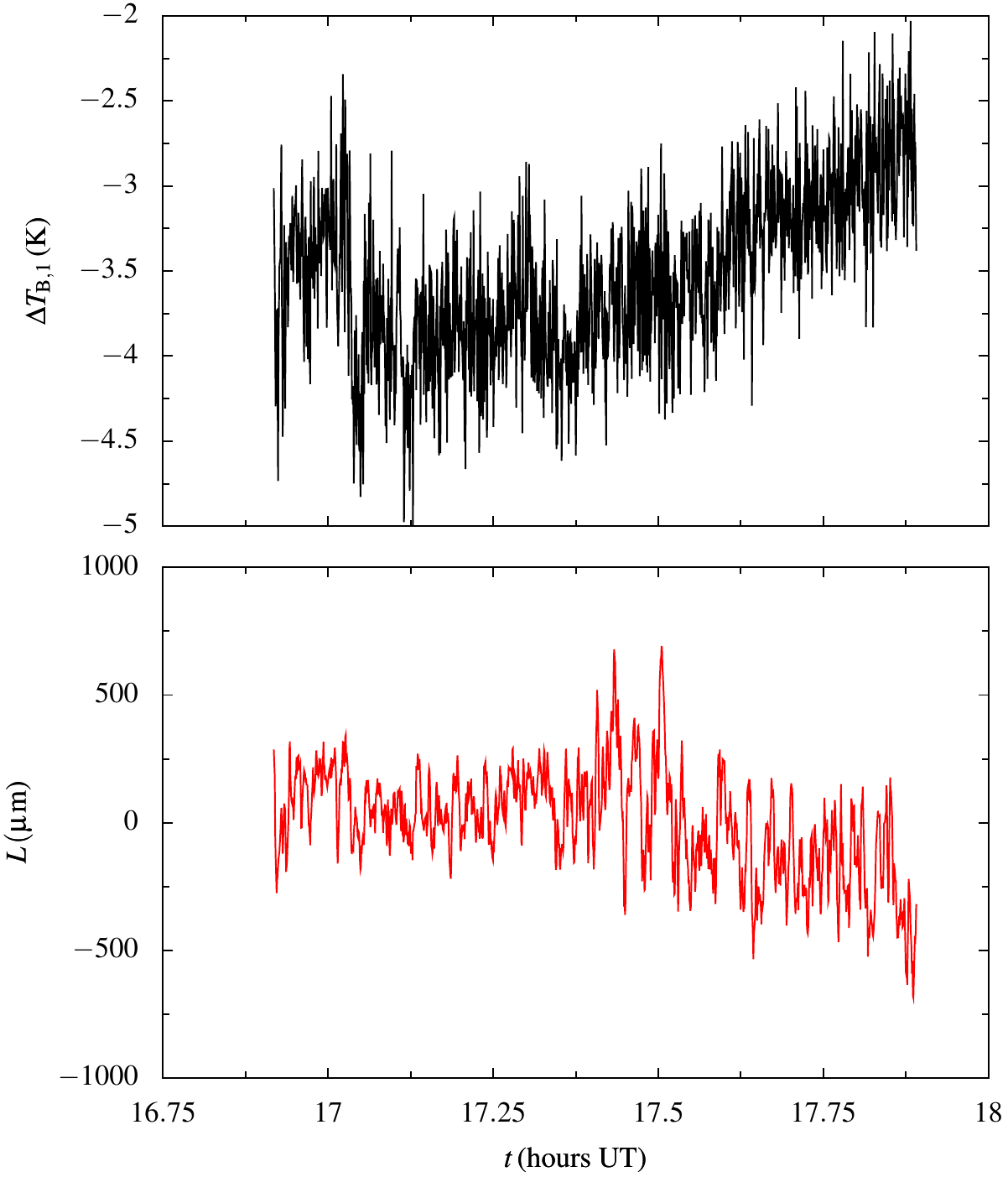}}&
    \subfloat[Channel 2]{
    \includegraphics[clip,width=0.45\linewidth]{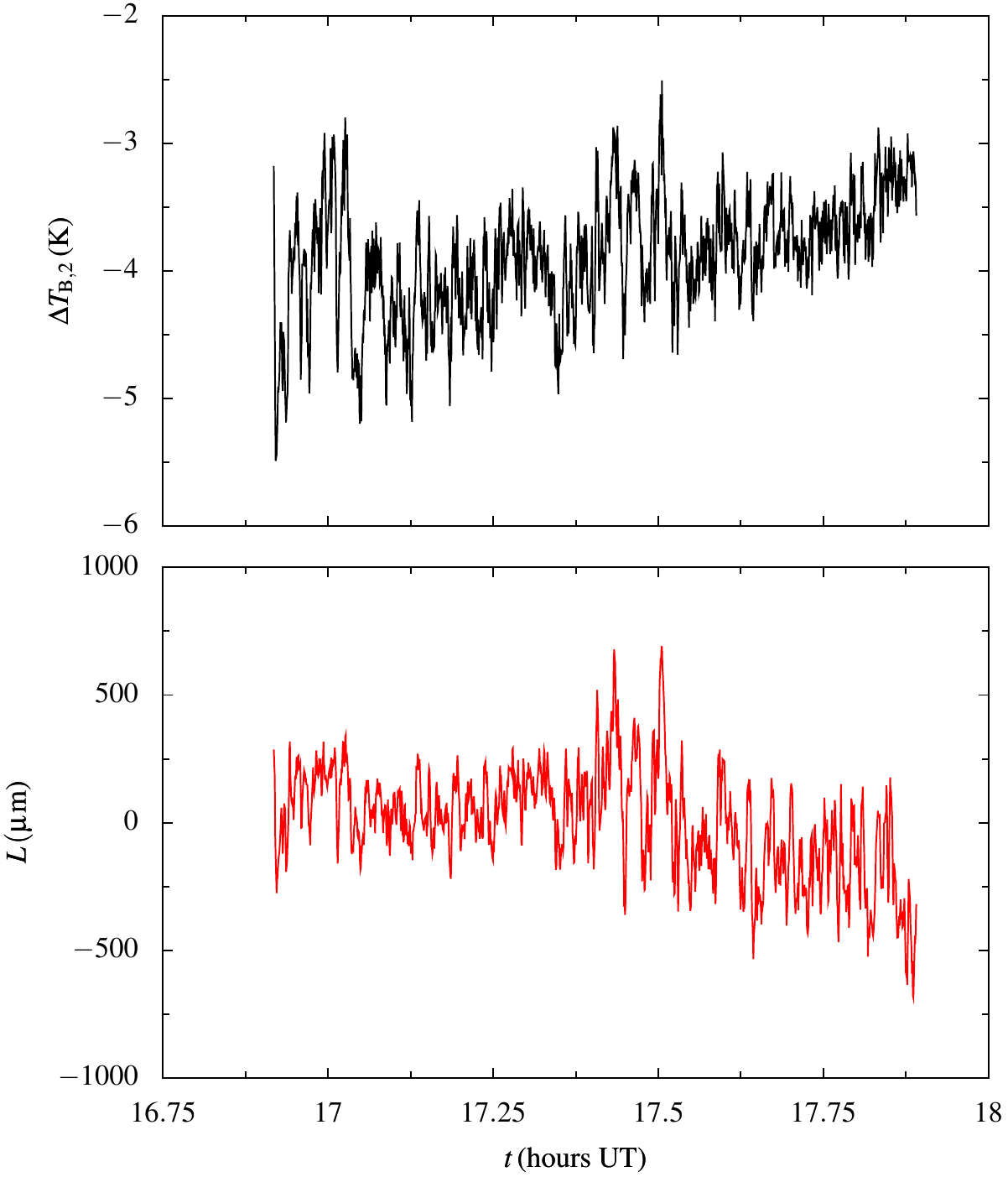}}\\
    \subfloat[Channel 3]{\includegraphics[clip,width=0.45\linewidth]{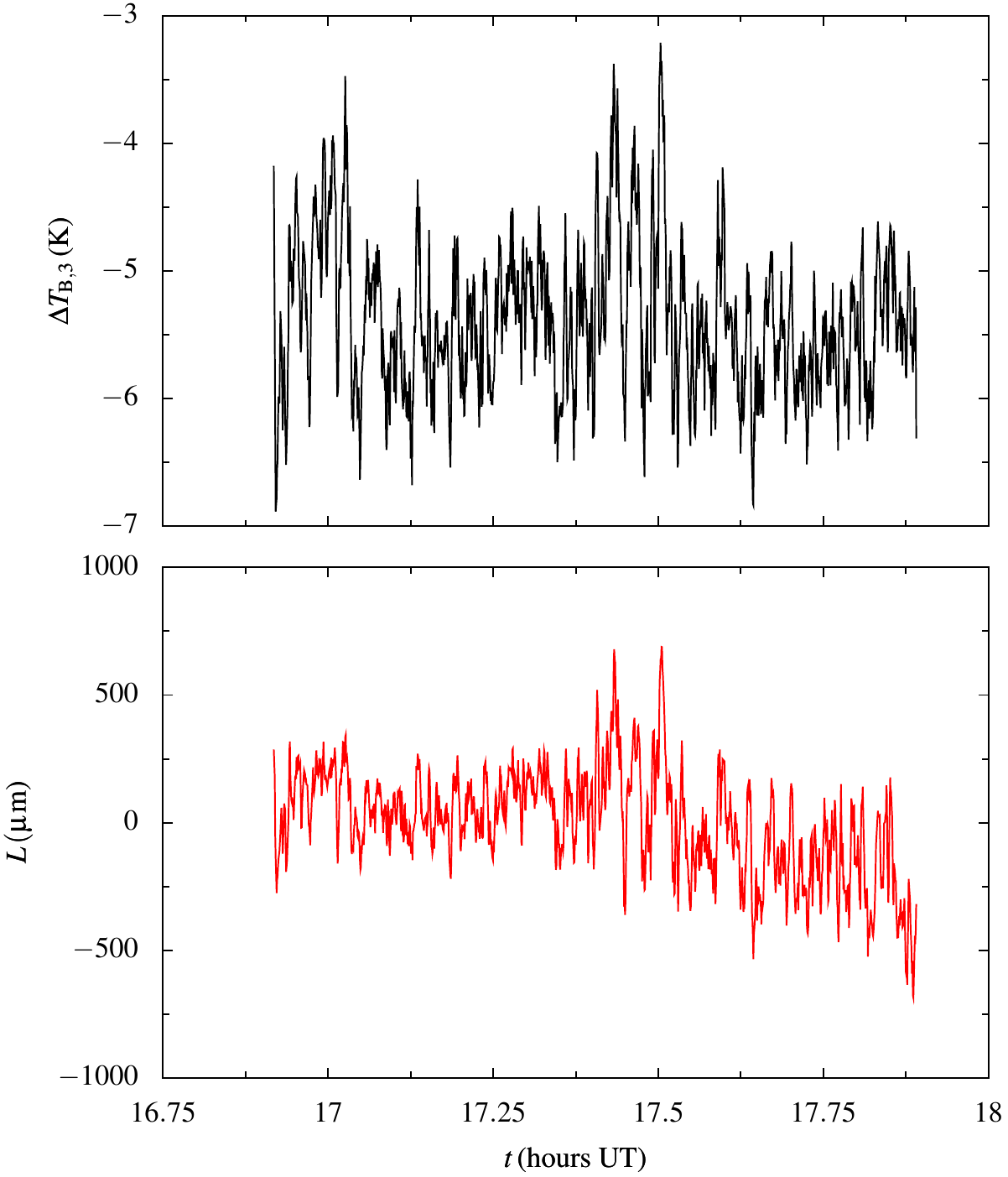}}&
    \subfloat[Channel 4 (outermost channel)]{\includegraphics[clip,width=0.45\linewidth]{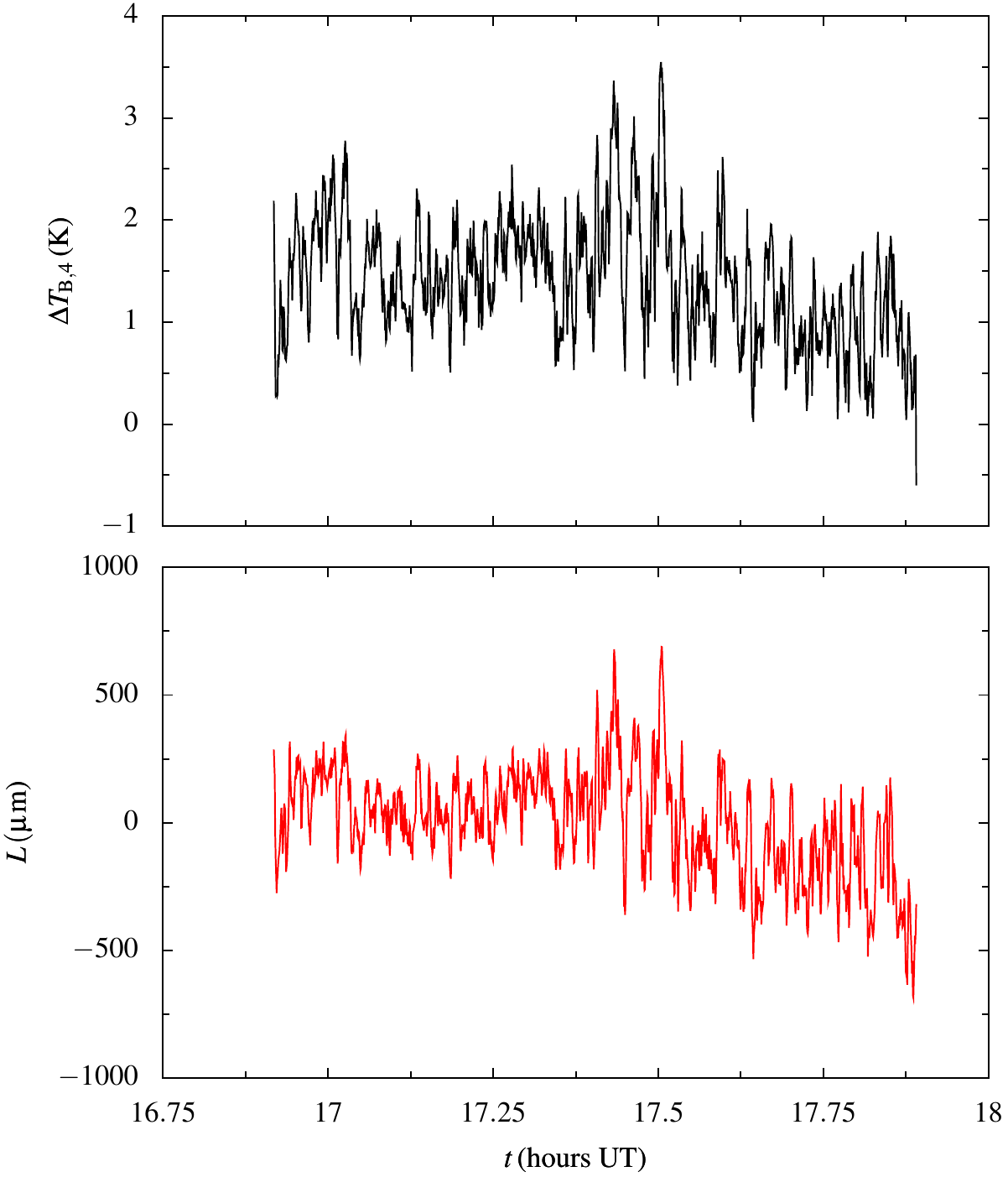}}
  \end{tabular}
  \caption{Comparison of the differences between each of the channels
    (parts a-d) on the two radiometers at SMA test with the
    fluctuation of path observed by the interferometer.}
  \label{fig:smaradiointfdata}
\end{figure*}

As mentioned previously the SMA interferometer made it possible to
infer the fluctuation of path between the two antennas of the array
that the WVRs mounted on them. We can also difference the signals
between two radiometers to calculate the differential fluctuation of
sky brightness as seen from the two antennas. Those two signals are
compared (for each of the four channels of the WVRs) in Figure
\ref{fig:smaradiointfdata}. Of course we hope that when multiplied by
the phase correction coefficient, the sky brightness fluctuations will
look very similar to the path fluctuation. 

The comparison shown in Figure~\ref{fig:smaradiointfdata} can
immediately confirm that channel 1 is unlikely to yield a useful phase
correction, while channel 2 show some correlation with path, and
channels 3 and 4 show very high correlation.

\begin{table*}

\subfloat[Part I]{
\begin{tabular}{ccccccccccccccccccccccc}
UT Range & $\sigma_{\rm raw}$ 
& $\hat{\frac{{\rm d}L}{{\rm d} T_{{\rm B},1}}}$
& $\sigma_{\rm r,1}$ 
& ${\rm Opt}({\frac{{\rm d}L}{{\rm d} T_{{\rm B},1}}})$
& $\sigma_{\rm opt,1}$
& $\hat{\frac{{\rm d}L}{{\rm d} T_{{\rm B},2}}}$
& $\sigma_{\rm r,2}$ 
& ${\rm Opt}({\frac{{\rm d}L}{{\rm d} T_{{\rm B},2}}})$
& $\sigma_{\rm opt,2}$
\\(hours)&
(\micron) & (\micron/K) & (\micron) & (\micron/K) & (\micron) & (\micron/K) & (\micron) & (\micron/K) & (\micron)
\\\hline
17.1--17.9&
157 & 
1260 & 420 & 64 & 156&
383 & 95 & 404 & 95
\\
17.1--17.4&
101 &
499 & 175 & 64 & 99 &
221 &  61 & 274& 59 
\\
17.4--17.7&
197 &
1260 & 421 & 96 & 195&
383  & 99  & 469 & 94
\\
17.7--17.9&
162&
2415 & 783 & 20 & 162 &
588  & 119 & 552 & 119
\\
\end{tabular}}

\subfloat[Part II]{
\begin{tabular}{ccccccccccccccccccccccc}
UT Range & $\sigma_{\rm raw}$ 
& $\hat{\frac{{\rm d}L}{{\rm d} T_{{\rm B},3}}}$
& $\sigma_{\rm r,3}$ 
& ${\rm Opt}({\frac{{\rm d}L}{{\rm d} T_{{\rm B},3}}})$
& $\sigma_{\rm opt,3}$
& $\hat{\frac{{\rm d}L}{{\rm d} T_{{\rm B},4}}}$
& $\sigma_{\rm r,4}$ 
& ${\rm Opt}({\frac{{\rm d}L}{{\rm d} T_{{\rm B},4}}})$
& $\sigma_{\rm opt,4}$
\\(hours)&
(\micron) & (\micron/K) & (\micron) & (\micron/K) & (\micron) & (\micron/K) & (\micron) & (\micron/K) & (\micron)
\\\hline
17.1--17.9&
157 & 
242 & 74  & 286 & 71 &
268 & 64  & 304 & 61 
\\
17.1--17.4&
101 &
179 & 53 & 220 & 51 &
234 & 47 & 259 & 46
\\
17.4--17.7&
197 &
242 & 77 & 290 & 70 &
268 & 64 & 303 & 61 &
\\
17.7--17.9&
162&
310 & 86 & 403 & 79 &
297 & 79 & 361 & 75
\\
\end{tabular}
}
\caption{Results of analysis of SMA data from February 17th. The part
  I of the table shows results for channels 1 and 2; the second part
  for channels 3 and 4. The UT range column indicates the time range
  of the data analysed: the first row corresponds to the whole set and
  the other three rows to respective thirds of the data. The column
  $\sigma_{\rm raw}$ is the RMS of phase fluctuations without
  correction. The column $\hat{\frac{{\rm d}L}{{\rm d} T_{{\rm
          B},i}}}$ is the best estimate from retrieval of the
  correction coefficient for the $i$th channel and for that section in
  time (see also Figure). The column $\sigma_{\rm r,i}$ is the RMS
  phase fluctuation after correction using the estimated
  coefficient. The column ${\rm Opt}({\frac{{\rm d}L}{{\rm d} T_{{\rm
          B},i}}})$ is the coefficient which would give the minimum
  RMS give the radiometer and interferometer data. The column
  $\sigma_{\rm opt,i}$ is that optimum RMS. }
\label{tab:phasecorrresult}
\end{table*}

We next quantify the performance that the phase correction technique
would have produced based on the phase correction coefficients
inferred in this section. For this we consider both the observation as
a whole and three sub-sets of the observation, namely the first, the
middle and the last thirds. The reason for considering the subsets is
to investigate if the change in the phase correction coefficients
inferred during the observation (Figure~\ref{fig:dldt-threetimes}) is
reflected in improved correction performance.

Before processing the data further we remove from both the
interferometric path and the sky temperature measurements a
three-minute running mean. This trend removal should approximately
model the effect of the further offset calibration scheme that ALMA
will use in practice. Although we do not present them here, the data
without the running mean removal lead to generally the same
conclusions as below.

We next asses the potential performance of the phase correction. For
these tests we consider phase correction using the fluctuations
observed in only one channel at the time. In principle, using a
combinations of channels rather than a single one should give better
performance because of the reduced effect of thermal noise and
possibly the averaging out of errors in the inferred phase correction
coefficients. We do pursue this multiple-channel approach further in
this paper but we expect it will be used in practice with ALMA.

We calculated the corrected path fluctuation using the inferred phase
correction coefficients and we also find the `best-fit' phase
coefficient, i.e., the coefficient which would have produced smallest
corrected path fluctuation given the data that we have.  The results
are present in Table~\ref{tab:phasecorrresult} with following columns:
\begin{enumerate}
  \item The root-mean-square of the uncorrected path fluctuation as
    seen by the interferometer, $\sigma_{\rm raw}$ 
  \item The approximate median inferences for the phase correction
    coefficients for each channel and for each observation subset
    $\hat{\frac{{\rm d}L}{{\rm d} T_{{\rm B},i}}}$ as shown in
    Figure~\ref{fig:dldt-threetimes}
  \item The root-mean-square path fluctuation after correction using
    radiometric data from each channel, $\sigma_{\rm r,i}$
  \item The phase correction coefficient that would have given the
    optimum reduction of path fluctuation, ${\rm Opt}({\frac{{\rm
          d}L}{{\rm d} T_{{\rm B},i}}})$
  \item The resultant root-mean-square path fluctuation that would
    have been obtained if the optimum correction coefficient were
    used, $\sigma_{\rm opt,i}$
\end{enumerate}

As can be seen from the table, channel 1 can not be used in these
conditions used for phase correction, as expected from the results
already presented. This is evident in phase correction with the
`optimum', i.e., best-fit, phase correction coefficient which makes a
negligible improvement to the path fluctuation, reducing it from
157\,micron to 156\,\micron. If the phase correction were done with
the inferred coefficient of 1.25\,mm/K (this was calculated for the
middle of the observation) then the path fluctuation would have been
drastically \emph{increased}. This is due to the high effect of
thermal noise in this channel when the line is essentially saturated
and other effects such as temperature variations of the water vapour.

The results for the other three channels show that significant
improvements in corrected versus uncorrected path fluctuations can be
made based on the radiometer data. Furthermore, it can be seen path
corrected with our inferred coefficients $\widehat{{\rm d}L/{\rm d}
    T_{{\rm B},i}}$ is close to the path corrected with the best-fit
coefficient, i.e., is fairly close to the limit that be achieved with
the present data. The difference between using the inferred and
best-fit coefficients results in changes of as little as
1\,\micron\ rms in path for some of the data sections up to
7\,\micron\ path rms in the worst combination of section and filter.

Some trends can however be noticed in the data presented in the
table. Firstly, excluding channel 1 which is saturated, the inferred
phase correction coefficients are systematically underestimating the
best-fit coefficients. The only exception is for channel 2 in the last
third of the observation, which may however be also affected by
saturation. Secondly there is evidence that at lower elevations the
gap between correction using inferred and best-fit coefficients
widens, suggesting that refinements to the model may be necessary.

\subsection{Discussion}

The analysis presented in Table~\ref{tab:phasecorrresult} shows that
the Bayesian inference approach that we described in
Section~\ref{sec:method} can produce very useful phase correction
coefficients even with a very simple atmospheric model. We have shown
that the inferred coefficients give a corrected path fluctuation to
within about five percent of what the optimal coefficient would have
given.

There are however a number of interesting points that these data (and
the simulations) raise. Firstly, the confidence interval for the
inference from our model analysis (Figure~\ref{fig:dldt-threetimes})
show a range of about 10\% and furthermore the optimal coefficients
found empirically are about 15\% higher than the median inferred
values. This needs to be compared against the specification for ALMA
phase correction which is:
\begin{equation}
  \delta L_{\rm corrected} \leq
  \left(1+\frac{c}{1\,\unit{mm}}\right)\,10\,\unit{\micron} +
  0.02 \times \delta L_{\rm raw}. 
\end{equation}
If we interpret the first term on the right hand side as the budget
for the thermal noise contribution to corrected phase fluctuations
then the second term, which is proportional to the raw phase
fluctuation, among other effects \cite[e.g.,][]{ALMANikolic573} must
account for the imperfect phase correction due to errors on the
inferred phase correction coefficients. Therefore we need to be able
to infer phase correction coefficients to better to 2\% accuracy, but
the present analysis of the model and the data suggest that we will do
significantly worse than that. 

There are a number of improvements that we are planning that will
allow better inference of the phase correction coefficients.  One of
these is the use of the \emph{observed\/} correlation between the
change in path and sky temperature which is likely to allow a
significant improvement of the inference accuracy and will be
presented in one of the next papers in this series.  Secondly, it
should be noted that in this case one of the limiting factors on the
inference is the absolute calibration of the radiometers. Although we
do not think that the \emph{a-priori\/} calibration of the radiometers
when mounted on the ALMA telescopes will be better than what we
assumed here, it will be probably the case that the major sources of
calibration inaccuracy will be fixed in time. Therefore we may be able
to improve the accuracy of these devices over time by empirical
corrections to their calibration. 

We have not discussed in detail in this paper the dispersive path
delay, in particular how it depends on the parameters other than the
water vapour column. This can be investigated by more fully featured
atmospheric models along the lines of the simple calculations
performed in Appendix~\ref{sec:dispersive}. We have also not
considered the `dry' path fluctuations, that is path fluctuation due
to changes in refractive index of dry air. It is not presently clear
how significant in practice these will be but data to be collected
during the commissioning of ALMA should provide constraints on this.

\section{Conclusions}

We have described an approach for calculation of the phase correction
coefficients based on:
\begin{itemize}
  \item A very simple, essentially \emph{minimal\/}, model of the
    atmosphere with three parameters: water vapour column, pressure
    and temperature
  \item Observables which are the four absolute sky brightness
    temperatures measured by the water vapour radiometers
  \item Priors on the model parameters based either on basic
    considerations or ancillary measurements
  \item Bayesian inference 
\end{itemize}
The approach produces the full probability distributions of the phase
correction coefficients making it easier to asses in a timely way the
accuracy of phase correction that can be expected.

The simulations we presented in Section~\ref{sec:results} show that
although there are inherent degeneracies in retrieval of atmospheric
paramours from sky brightness only, with the appropriate use of prior
information fairly good retrievals may be obtained. 

We have then applied this approach to one test observation taken
during the testing of the ALMA prototype radiometers at the SMA. We
found that the inferred water vapour column during this observation is
stable and its fluctuations are not greater than expected given the
observed path fluctuations (Figure~\ref{fig:retseq}). In tests of
applying the phase correction, we found that good phase corrections
can be made by taking representative inferences of the phase
correction coefficients and applying them to correct the observed path
fluctuation: in fact the typical corrected path fluctuation is within
about 5\% of the best correction that could be achieved with these
data, assuming only one channel is used at a time. In order for it to
be possible to meet the very demanding ALMA specifications, it is
clear to we will need to make use of further data and possibly more
complicated models. The approach we are currently working on is to
incorporate information from the occasional phase calibration
observations of the observed correlation between path and sky
brightness fluctuations into the inference process.

The source code for all of the algorithms presented in this paper is
available for public download under the GNU Public License at
\url{http://www.mrao.cam.ac.uk/~bn204/alma/memo-infer.html}. The
observational data from the SMA is also available at the same internet
address.

\section*{Acknowledgements}

This work is a part of the ALMA Enhancement programme funded by the
European Union Framework Programme 6. The Cambridge effort within this
programme has been lead by R.~E. Hills and subsequently
J.~S. Richer. 

We thank the team involved in construction and the testing of the ALMA
prototype water vapour radiometers: P.~G. Anathasubramanian,
R.~E. Hills, K.~G.Isaak, M. Owen, J.~S. Richer, H. Smith,
A.~J. Stirling, R.~Williamson, V.~Belitsky, R. Booth, M. Hagstr\"om,
L. Helldner, M. Pantaleev, L.~E.~Pettersson, T.~R. Hunter, S. Paine,
A. Peck, M.~A. Reid, A.~Schinckel and K.~Young.

We would like to thank S. Paine for making the code for {\tt am}
publicly available, which allowed us to match our model of the
183\,GHz line to his. We would also like to thank J. Pardo for making
the code for {\tt ATM} publicly available. We used {\tt ATM} for
computation of the dispersive phase adjustment.

We thank Robert Laing for a careful reading of the manuscript and
helpful comments.

\bibliographystyle{mn2e} 
\bibliography{wvrretrieval.bib}

\begin{thebibliography}{15}
\expandafter\ifx\csname natexlab\endcsname\relax\def\natexlab#1{#1}\fi

\bibitem[{Abramowitz \& Stegun(1964)}]{abramowitz+stegun}
Abramowitz M., Stegun I.~A., 1964, Handbook of Mathematical Functions with
  Formulas, Graphs, and Mathematical Tables, ninth dover printing, tenth gpo
  printing edn. Dover, New York

\bibitem[{Asaki {et~al.}(2005)Asaki, Saito, Kawabe, Morita, Tamura, \&
  Vila-Vilaro}]{ALMAAsaki535}
Asaki A., Saito M., Kawabe R., Morita K., Tamura Y., Vila-Vilaro B., 2005,
  Simulation series of a phase calibration scheme with water vapor radiometers
  for the atacama compact array. ALMA Memo Series 535, The ALMA Project

\bibitem[{Hills {et~al.}(2001)Hills, Gibson, Richer, Smith, Belitsky, Booth, \&
  Urbain}]{ALMAHills352}
Hills R.~E., Gibson H., Richer J.~S., Smith H., Belitsky V., Booth R., Urbain
  D., 2001, Design and development of 183ghz water vapour radiometers. ALMA
  Memo Series 352, The ALMA Project

\bibitem[{Jaynes(2003)}]{Jaynes:PTLS}
Jaynes E.~T., 2003, Probability Theory : The Logic of Science. Cambridge
  University Press

\bibitem[{MacKay(2003)}]{MackayBayesain}
MacKay D. J.~C., 2003, Information Theory, Inference, and Learning Algorithms.
  Cambridge University Press

\bibitem[{{Metropolis} {et~al.}(1953){Metropolis}, {Rosenbluth}, {Rosenbluth},
  \& {Teller}}]{1953Metropolis}
{Metropolis} N., {Rosenbluth} A.~W., {Rosenbluth} M.~N., {Teller} A.~H., 1953,
  The Journal of Chemical Physics, 21, 1087

\bibitem[{Neal(1993)}]{NealR1993}
Neal R.~M., 1993, Probabilistic inference using markov chain monte carlo
  methods. Tech. Rep. CRG-TR-93-1, University of Toronto,
  \url{http://www.cs.toronto.edu/~radford/ftp/review.pdf}

\bibitem[{Nikolic(2009)}]{BNMin1}
Nikolic B., 2009, BNMin1 Minimisation/Inference library.
  \url{http://www.mrao.cam.ac.uk/~bn204/oof/bnmin1.html}

\bibitem[{Nikolic {et~al.}(2007)Nikolic, Hills, \& Richer}]{ALMANikolic573}
Nikolic B., Hills R.~E., Richer J.~S., 2007, Limits on phase correction
  performance due to differences between astronomical and water-vapour
  radiometer beams. ALMA Memo Series 573, The ALMA Project

\bibitem[{Nikolic {et~al.}(2008)Nikolic, Richer, \& Hills}]{ALMANikolic582}
Nikolic B., Richer J.~S., Hills R.~E., 2008, Simulating atmospheric phase
  errors, phase correction and the impact on alma science. ALMA Memo Series
  582, The ALMA Project

\bibitem[{Paine(2004)}]{PaineAmRev3}
Paine S., 2004, The am atmospheric model. Tech. rep., SMA Technical Memo,
  revision 3

\bibitem[{Pardo {et~al.}(2001)Pardo, Cernicharo, \& Serabyn}]{PARDOATM}
Pardo J.~R., Cernicharo J., Serabyn E., 2001, IEEE Trans. on Antennas and
  Propagation, 49

\bibitem[{Rothman {et~al.}(2005)}]{HITRAN04}
Rothman L., {et~al.}, 2005, JQSRT, 96

\bibitem[{Sivia \& Skilling(2006)}]{SiviaD06}
Sivia D.~S., Skilling J., 2006, Data Analysis--A Bayesian Tutorial, 2nd edn.
  Oxford Science Publications

\bibitem[{Stirling {et~al.}(2008)Stirling, Richer, Hills, \&
  Lock}]{ALMAStirling517}
Stirling A., Richer J., Hills R., Lock A., 2008, Turbulence simulations of dry
  and wet phase fluctuations at chajnantor. ALMA Memo Series 517.1, The ALMA
  Project, (Original version published 2005)

\end{thebibliography}

\appendix

\section{Calculation of atmospheric opacity}
\label{sec:atmo-opac-calc}

\subsection{The 183\,GHz water vapour line}

The water line at 183\,GHz is assumed to have the Gross line shape:
\begin{equation}
\kappa(\nu; S, \nu_0, \gamma)= \frac{4 S \gamma}{\pi} \frac{ \nu^2
}{\left(\nu^2-\nu_0^{2}\right)^2 + 4 \gamma^2 \nu^2},
\end{equation}
where $\nu$ is the frequency and the set are $\{S, \nu_0, \gamma\}$ are
the parameters of the line.

In the present model we are assuming that water vapour is a relatively
small constituent of air and so that the volume mixing ratio for water
is negligible. In this case, the line parameters can be computed as
follow:
\begin{eqnarray}
  \nu_{0} = F + P \times \Delta_{\rm air} \\
  \gamma  = \gamma_{\rm air} \times P \times \left(\frac{T_{\rm
      ref}}{T}\right)^{n_{\rm air}}\\
  S = H \times e^{\frac{E_l (T-T_{\rm ref})}{T T_{\rm ref}}}
      \frac{ 1- e^{ -h\nu/T}}{1-e^{-h\nu/T_{\rm ref}}}
      \frac{ Q(T_{\rm ref} )}{Q(T)}.
\end{eqnarray}
Here $T_{\rm ref}$ is the reference temperature of the catalogue,
which in this case is 296\,K; $T$ and $P$ are the temperature and
pressure of the air containing the water vapour, and $Q(T)$ is the
partition function of water vapour which we take from a look-up
table. The remaining parameters are taken from the HITRAN entry for
the 183.3\,GHz water vapour line and are shown in
Table~\ref{tab:hitranparams}.

\begin{table*}

\begin{tabular}{cccc}
  Parameter & Units & Value & Comment\\\hline
  $H$         & $\unitp{cm}{2}\times\unit{GHz}$  &
  $2.333884\times10^{-21}$ & Line intensity\\
  $F$         & $\unit{GHz}$ & 183.310107 & Frequency \\
  $\Delta_{\rm air}$ & $\unit{GHz}\times\unitp{mbar}{-1}$ &
  $-5.0298\times10^{-5}$ &Air-induced freq. shift\\
  $E_{\rm l}$ & $\unit{K}$ & 195.908398 & Lower state energy level\\
  $\gamma_{\rm air}$ & $\unit{GHz}\times\unitp{mbar}{-1}$ &
  $2.9114\times 10^{-3}$ & Air broadening \\
  $\gamma_{\rm self}$ & $\unit{GHz}\times\unitp{mbar}{-1}$ &
  $1.6149\times10^{-2}$ & Self-broadening \\
  $n_{\rm air}$ & none &
  0.77 & Temperature dep. of air broadening
\end{tabular}
\caption{The HITRAN parameters and values for the 183.3\,GHz line
  \citep{HITRAN04}.}
\label{tab:hitranparams}
\end{table*}

\subsection{Continuum}

The continuum cross section is :
\begin{equation}
  C(\nu; C_{0}, T_{0}, m  ) = C_{0} \left(\frac{ T_{0}}{T}\right)^{m} \times N \times \nu^{2},
\end{equation}
where N is number density of air, and the parameter set $\{C_{0},
T_{0}, m\}$ describe the continuum. For water, the values we assume
are:
\begin{tabular}{ccc}
  Param & Value & Unit\\\hline
  $C_{0}$ & $6.1\times10^{-48}$ & $\unitp{cm}{5}/\unitp{GHz}{2}$\\
  $T_{0}$ &  300.0 & K\\
  $m$     & 2.6 & none
\end{tabular}

\section{Excess path length}

\subsection{Non-dispersive path delay}
\label{sec:non-dispersive}

The non-dispersive excess path length due to water is computed using
the Smith-Weintraub equation which gives the following expression for
the excess refractive index \citep{ALMAStirling517}:
\begin{equation}
n-1 = 10^{-6} \left[ \alpha \frac{P_{\rm d}}{T} + \beta \frac{P_{\rm w}}{T} + \gamma
\frac{P_w}{T^2}\right],
\label{eq:SmithWeintraubEq}
\end{equation}
where $P_{\rm d}$ is the partial pressure of the dry air and $P_{\rm
  w}$ is the partial pressure of the water vapour and the constants
have the following values:
\begin{itemize}
\item $\alpha = 77.6\times 10^{-2}\,\unit{K}\,\unitp{Pa}{-1}$
\item $\beta = 64.8 \times 10^{-2}\,\unit{K}\,\unitp{Pa}{-1}$ 
\item $\gamma = 3.776 \times 10^{3}\,\unitp{K}{2}\,\unitp{Pa}{-1}$.
\end{itemize}
We are presently concerned with the excess path introduced with the
water vapour only, so we can drop the first term of
Equation~\ref{eq:SmithWeintraubEq}. 

Of the remaining two terms in Equation~\ref{eq:SmithWeintraubEq}, the
dominant term is the second and it can be transformed to be in terms
of water vapour density  \citep[see][]{ALMAStirling517}:
\begin{equation}
  {\rm d}n \approx  \frac{\gamma R}{\mathcal{M}_v T} {\rm d}\rho_w
\end{equation}
where $\mathcal{M}_v= 18.02\,\unit{g}\,\unitp{mol}{-1}$ is the
molecular weight of water vapour and $\rho_w$ is the density of the
water vapour.  Assuming a thin layer and by integrating along the line
of sight we get:
\begin{equation}
  {\rm d}l \approx  \frac{\gamma R}{\mathcal{M}_v T} {\rm d}c \approx
  \frac{1741\,\unit{K}}{T}{\rm d}c.
  \label{eq:nondisp-simple}
\end{equation}

\subsection{Dispersive path delay}
\label{sec:dispersive}

The dispersive delay adjustment was calculated using the {\tt ATM}
code \cite{PARDOATM}. The calculation was made using a simple program
{\tt dispersive} that takes a frequency and ground-level relative
humidity and computes the dispersive and non-dispersive delay due to
the water vapour and the dry atmosphere at the supplied frequency.

The program {\tt dispersive}, and the code for the library {\tt ATM},
are available at
\url{http://www.mrao.cam.ac.uk/~bn204/alma/atmomodel.html}, packaged
as {\tt aatm}. All of the code is made available under the GPL
license. For this paper we used {\tt aatm-0.06}, i.e., version 0.06.

\label{lastpage}
\end{document}